\documentclass[a4paper,fleqn,usenatbib]{mnras}

\usepackage{newtxtext,newtxmath}

\usepackage[T1]{fontenc}
\usepackage{ae,aecompl}

\usepackage{graphicx}	
\usepackage{amsmath}	
\usepackage{amssymb}	

\title[Star formation and nuclear activity in local LIRGs]{Star formation and AGN activity in a sample of local Luminous Infrared Galaxies through multi-wavelength characterization }

\author[Herrero-Illana et al.]{Rub\'en Herrero-Illana,$^{1,2}$\thanks{E-mail:
rherrero@eso.org} Miguel \'A. P\'erez-Torres,$^{1,3}$ Zara Randriamanakoto,$^{4}$
\newauthor Antxon Alberdi,$^{1}$ Andreas Efstathiou,$^{5}$ Petri V\"ais\"anen,$^{6,7}$, Erkki Kankare,$^{8}$
\newauthor Erik Kool,$^{9,10}$ Seppo Mattila,$^{11,12}$ Rajin Ramphul,$^{4,6}$ and Stuart Ryder.$^{9}$ 
\\
$^{1}$Instituto de Astrof\'isica de Andaluc\'ia (IAA-CSIC). Glorieta de la Astronom\'ia s/n, 18008, Granada, Spain \\
$^{2}$European Southern Observatory (ESO), Alonso de C\'ordova 3107, Vitacura, Casilla 19001, Santiago de Chile, Chile\\
$^{3}$Visiting Scientist: Departamento de F\'isica Te\'orica, Facultad de Ciencias, Universidad de Zaragoza, Spain\\
$^{4}$University of Cape Town, Astronomy Department, Private Bag X3, Rondebosch 7701, South Africa \\
$^{5}$School of Sciences, European University Cyprus, Diogenes Street, Engomi, 1516 Nicosia, Cyprus \\
$^{6}$South African Astronomical Observatory, P.O. Box 9 7935, South Africa\\
$^{7}$Southern African Large Telescope, PO Box 9, Observatory 7935, Cape Town, South Africa\\
$^{8}$Astrophysics Research Centre, School of Mathematics and Physics, Queen's University Belfast, Belfast BT7 1NN, UK\\
$^{9}$Australian Astronomical Observatory, 105 Delhi Rd, North Ryde, NSW 2113, Australia\\
$^{10}$Department of Physics and Astronomy, Macquarie University, Sydney, NSW 2109, Australia\\
$^{11}$Tuorla Observatory, Department of Physics and Astronomy, University of Turku, V\"ais\"al\"antie 20, FI-21500 Piikki\"o, Finland\\
$^{12}$Finnish Centre for Astronomy with ESO (FINCA), University of Turku, V\"ais\"al\"antie 20, FI-21500 Piikki\"o, Finland\\
}

\date{Accepted XXX. Received YYY; in original form ZZZ}

\pubyear{2015}

\begin{document}
\label{firstpage}
\pagerange{\pageref{firstpage}--\pageref{lastpage}}
\maketitle

\begin{abstract}
Nuclear starbursts and AGN activity are the main heating processes in luminous infrared galaxies (LIRGs) and their relationship is fundamental to understand galaxy evolution. In this paper,  we study the star-formation and AGN activity of a sample of 11 local LIRGs imaged with subarcsecond angular resolution at radio (8.4\,GHz) and near-infrared ($2.2\,\mu$m) wavelengths. 
This allows us to characterize the central kpc of these galaxies with a spatial resolution of $\simeq100$\,pc. In general, we find a good spatial correlation between the radio and the near-IR emission, although radio emission tends to be more concentrated in the nuclear regions. Additionally, we use an MCMC code to model their multi-wavelength spectral energy distribution (SED) using template libraries of starburst, AGN and spheroidal/cirrus models, determining the luminosity contribution of each component, and finding that all sources in our sample are starburst-dominated, except for NGC\,6926 with an AGN contribution of $\simeq64$\%. Our sources show high star formation rates ($40$ to $167\,M_\odot\,\mathrm{yr}^{-1}$), supernova rates (0.4 to $2.0\,\mathrm{SN}\,\mathrm{yr}^{-1}$), and similar starburst ages (13 to $29\,\mathrm{Myr}$), except for the young starburst (9\,Myr) in NGC\,6926. A comparison of our derived star-forming parameters with estimates obtained from different IR and radio tracers shows an overall consistency among the different star formation tracers. AGN tracers based on mid-IR, high-ionization line ratios also show an overall agreement with our SED model fit estimates for the AGN. Finally, we  use our wide-band VLA observations to determine pixel-by-pixel radio spectral indices for all galaxies in our sample, finding a typical median value ($\alpha\simeq-0.8$) for synchrotron-powered LIRGs. 
\end{abstract}

\begin{keywords}
galaxies: interactions -- galaxies: nuclei -- galaxies: starburst -- infrared: galaxies -- radio continuum: galaxies   
\end{keywords}

\clearpage

\section{Introduction}\label{sec:intro}

Luminous Infrared Galaxies (LIRGs) are defined as those galaxies with an infrared luminosity $L_\mathrm{IR}$[8--1000\,$\mu\mathrm{m}] >10^{11}L_\odot$. The majority of these sources, discovered in the early 1980s by the \emph{IRAS} satellite, are actually galaxies undergoing a merging process.

While most LIRGs are dominated by violent episodes of nuclear star formation \citep[starbursts; see e.g.,][]{petric11,stierwalt13}, many of them also contain an active galactic nucleus (AGN). The existence and link between these two processes \citep[e.g.,][]{shao10} make LIRGs ideal laboratories in which to study the AGN-starburst connection.

The parameters describing the star formation in a galaxy can be obtained by means of a number of indirect tracers and prescriptions based on different bands of the electromagnetic spectrum, including UV and far-IR continuum, as well as several recombination or forbidden lines  \citep[see review by][]{kennicutt98}. Recently, near-IR observations have also proven useful to characterize the star formation properties of LIRGs through the study of super star clusters \citep[SSCs;][]{portegies-zwart10, randriamanakoto13b, randriamanakoto13a}, which are young ($\simeq10$\,Myr) massive star clusters that preferentially form whenever there is strong ongoing starburst activity. Other tracers are used to infer the AGN type, its luminosity and its relative contribution to the bolometric luminosity of galaxies, such as X-ray emission \citep[][]{treister09, mullaney11}, mid-IR continuum \citep{asmus14} or optical \citep[e.g.][]{bassani99, heckman05} and IR spectral lines \citep[][]{genzel98, pereira-santaella10, diamond-stanic09}.

However, the large amounts of dust present in LIRGs, whose reprocessing of ultraviolet photons from massive stars is responsible for the high IR luminosities in these systems, impose a limitation to optical and near-IR tracers due to dust obscuration \citep{mattila01}. For this reason, observations at radio wavelengths, unaffected by dust extinction, are an alternative and powerful tool to trace starburst (SB) and AGN processes in the innermost regions of these systems \citep[][]{condon92,parra07,perez-torres09b,perez-torres10,murphy11,romero-canizales12a}.

There is a well-known tight linear correlation between the far-IR and radio (1.4\,GHz) luminosities \citep{vanderkruit73, dejong85, helou85, condon92}, with no evident dependence with redshift \citep{ivison10, pannella15}. To quantify this correlation, the so called $q$-factor \citep{helou85} was defined as:
\begin{equation}\label{eq:q}
q=\log\left(\frac{\mathrm{FIR}/3.72\times10^{12}\,\mathrm{Hz}}{S_{1.49\,\mathrm{GHz}}}\right),
\end{equation}
where $S_{1.49\,\mathrm{GHz}}$ is the flux density at 1.49\,GHz in units of W\,m$^{-2}$\,Hz$^{-1}$ and FIR is defined as
\begin{equation}
\mathrm{FIR}=1.26\times10^{-14}(2.58 S_{60\,\mu\mathrm{m}} + S_{100\,\mu\mathrm{m}}),
\end{equation}
with $S_{60\,\mu\mathrm{m}}$ and $S_{100\,\mu\mathrm{m}}$ being the \emph{IRAS} fluxes in Jy at 60 and 100\,$\mu$m, respectively. The mean value of the $q$-factor in the \emph{IRAS} Bright Galaxy Sample \citep{condon90} is $\left<q\right>=2.34$ \citep{condon91b, yun01}. 
While a radio excess ($q<2.34$) is typically associated with a strong AGN contribution \citep[e.g.][]{roy97, donley05,ivison10,delmoro13} and a FIR excess ($q>2.34$) is suggestive of intense star formation, this is not always the case \citep[e.g., the obscured AGN in the most FIR-excess galaxy, NGC\,1377;][]{costagliola16}. In general, the $q$-factor cannot be used as a standalone tool to separate AGN from starburst galaxies \citep{moric10, padovani11}.

The underlying physics for the FIR-radio correlation is usually associated with a star formation origin: dust reprocesses massive-star UV radiation into far infrared photons, while the explosions of those same stars as supernovae (SNe) accelerate the cosmic ray electrons, responsible for the radio non-thermal synchrotron radiation \citep[e.g.,][]{voelk89,lisenfeld96, lacki10a, lacki10b}. Thermal bremsstrahlung arising from  H\,{\sc ii} regions (free-free emission) is also linked to this correlation \citep[e.g., ][]{murphy11}. Therefore, both thermal and non-thermal emission are correlated with the far-IR and are good tracers of star-formation.
Correlations between radio emission and windows in the IR regime other than the far-IR exist \citep[e.g., $3.3\mu$m and $8.7\mu$m; as used for NGC\,1614 in][]{alonso-herrero01, vaisanen12, herrero-illana14}, but are not as clear nor that well studied.

A problem of many of the starburst indicators mentioned above is the frequent contamination of the tracers by the effects of an AGN. Notable examples are the contamination of the radio continuum by putative jets \citep{cleary07} and the significant contribution to the far-IR through the heating of the narrow line region by the AGN \citep{tadhunter07}. A way to overcome this problem is to fit a multi-wavelength spectral energy distribution (SED) combining templates of both starbursts and AGN \citep[e.g.,][]{efstathiou00, netzer07, mullaney11,calistro-rivera16}.

In this paper, we model the SED of a sample of 11 local LIRGs and compare our derived starburst and AGN properties with other models. We also present radio X-band (8.4\,GHz) and near-IR K$_\mathrm{S}$-band ($2.2\,\mu$m; hereafter referred to as K-band) observations of this sample, comparing them and analyzing possible near-IR/radio correlations. The paper is structured as follows: in section~\ref{sec:sample} we present the sample, together with a concise individual description of the sources, and in 
 section~\ref{sec:observations} we present our observations and details on the data reduction. Through the discussion (section~\ref{sec:discussion}) we describe our SED modeling in section~\ref{sec:sed}, and compare it with other tracers of star formation and AGN activity. We then compare our radio and near-IR observations (section~\ref{sec:radiovsir}), analyze the special case of IRAS\,16516-0948 (section~\ref{sec:offnuclear}) and discuss the radio spectral index of our sources (section~\ref{sec:spix}). We summarize our results in section~\ref{sec:summary}.

Throughout this paper we adopt a cosmology with $H_{\rm 0} = 75$\,km\,s$^{-1}$\,Mpc$^{-1}$, $\Omega_\Lambda = 0.7$ and $\Omega_m = 0.3$.

\section{The sample} \label{sec:sample}

The sources in our LIRG sample are taken from the \emph{IRAS} Revised Bright Galaxy Sample \citep[IRBGS;][]{sanders03}, and fulfill the following criteria:  $D<110$\,Mpc, luminosity $\log(L_\mathrm{IR}/L_\odot)>11.20$, and declination $\delta>-35^\circ$. We chose those criteria so that our 8.4\,GHz Karl G. Jansky Very Large Array (VLA) observations in A-configuration (angular resolution of $\simeq 0.3^{\prime\prime}$)
could image the central kpc region of all galaxies with spatial resolutions of $\simeq 70-150$ pc, which would allow us to disentangle the compact radio emission from a putative AGN from the diffuse, extended radio emission linked to a starburst, as well as potentially allowing  the detection of other compact sources, e.g., individual supernovae or supernova remnants (SNR).
Furthermore, we excluded \emph{warm} LIRGs with \emph{IRAS} color $f_{25}/f_{60}>0.2$ to prevent contamination from obscured AGN activity \citep[e.g., ][]{farrah07}, except for Arp\,299 ($f_{25}/f_{60}=0.22$), which is borderline, but was included in the study since it is one of the most luminous local LIRGs. A total of 54 out of 629 sources in the IRBGS satisfy the conditions above, from which ours is a representative sample. Finally, we also excluded galaxies with no nearby reference star to guide the adaptive optics (AO) observations with the VLT or the  Gemini telescopes.

\begin{figure}
\includegraphics[width=\columnwidth]{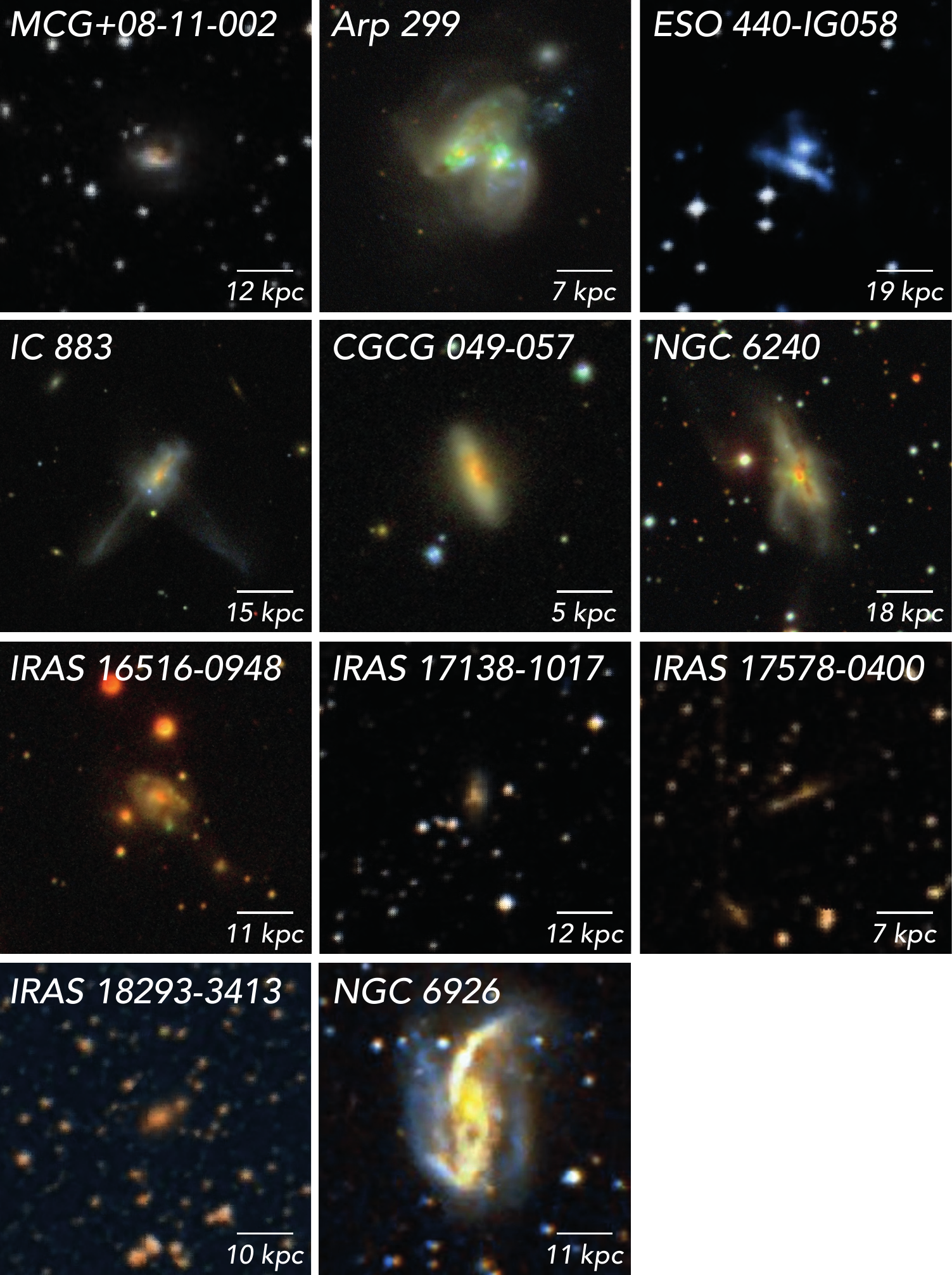}
\caption{Optical color composite images of the galaxies in our LIRG sample. Images are from the Sloan Digital Sky Survey (SDSS) DR9 \citep{ahn12} when available, or from the STScI Digitized Sky Survey (DSS) otherwise.}
 \label{dss.pdf}
\end{figure}

\begin{figure}
\includegraphics[width=\columnwidth]{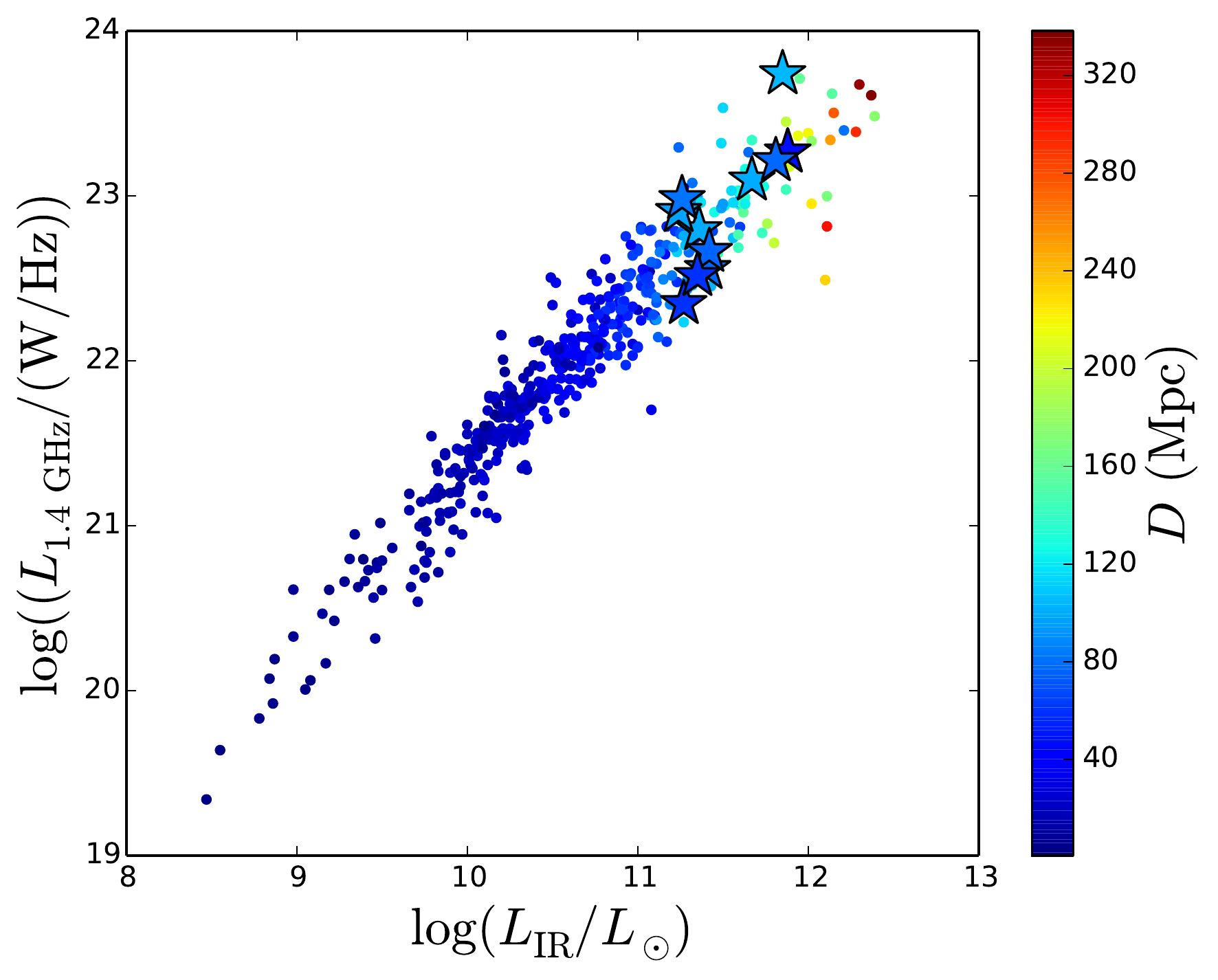}
 \caption{Radio (1.4\,GHz) vs. infrared ($8-1000\,\mu$m) luminosity plotted for a sample of \emph{IRAS} galaxies (dots) and our galaxy sample (stars), which falls into the LIRG regime. The most radio-powerful galaxy in our sample, NGC\,6240, is also the one that deviates most from the correlation. 
 }
 \label{radiovsir.pdf}
\end{figure}

In Table~\ref{tab:galaxysample} we show our final sample, which consists of 11 LIRGs (although the two components of one of them, Arp~299, are treated separately), all of them included in the GOALS sample \citep{armus09}, together with their IR luminosity and luminosity distance. We also include the merger stage as derived from IRAC 3.6\,$\mu$m morphology \citep{stierwalt13}, and the $q$-factor values \citep{helou85}, obtained using the \emph{IRAS} fluxes from the IRBGS and the 1.4\,GHz fluxes from the NRAO VLA Sky Survey \citep[NVSS;][]{condon98}. An optical image of each source in our sample is shown in Figure~\ref{dss.pdf}. In Figure~\ref{radiovsir.pdf} we show the correlation between $L_\mathrm{IR}$ and the radio luminosity at 1.4\,GHz for the sources in our sample (plotted with a star symbol) compared with the resulting sources of a cross-match between the IRBGS and the New VLA Sky Survey by \citet{yun01}.

\begin{table*}
\caption{Galaxy sample properties.}
\begin{tabular}{rcccccccc}
  	\hline
	Name & RA & Dec. &  $D_\mathrm{L}$ & Merger stage$^2$ &$\log(L_\mathrm{IR})$  & $\log(L_\mathrm{1.4\,GHz})$ &  $q$-factor$^3$ & X-ray\\
   	 & (J2000) & (J2000) & (Mpc) & &($L_\odot$) & (erg\,s$^{-1}$\,Hz$^{-1}$)   & & classification$^{4}$\\
	\hline
MCG+08-11-002&           05 40 43.7 &            +49 41 41&            77.2 &        d &            11.41&          29.56&      2.61 & $\cdots$ \\
Arp299$^1$ &        11 28 29.8 &             +58 33 43 &           47.7 &        c &            11.88 &    30.27 &      2.30 & AGN \\
ESO\,440-IG058 &         12 06 51.9 &            $-$31 56 54 &         100.5 &       b &            11.36 &         29.80&      2.32 & $\cdots$\\
IC\,883&                 13 20 35.3 &            +34 08 22 &           100.0 &       d &            11.67 &         30.10&      2.34 & AGN \\
CGCG\,049-057&           15 13 13.1 &            +07 13 32 &           59.1 &        N &            11.27&          29.35&      2.74 & No AGN \\
NGC\,6240 &              16 52 58.9 &            +02 24 03 &           103.9 &       d &            11.85 &         30.74&      1.83  & AGN \\
IRAS\,16516-0948&        16 54 24.0&            $-$09 53 21 &          96.9 &        d&             11.24&          29.91&      2.08 & $\cdots$ \\
IRAS\,17138-1017&        17 16 35.8 &            $-$10 20 39  &        75.9 &        d &           11.42 &          29.66&      2.47 & $\cdots$ \\
IRAS\,17578-0400&        18 00 31.9 &            $-$04 00 53  &        58.6 &        b &           11.35 &          29.52&      2.64 & No AGN \\
IRAS\,18293-3413&        18 32 41.1 &            $-$34 11 27 &         77.8 &        c &            11.81 &         30.21&      2.33 &  No AGN \\
NGC\,6926 &              20 33 06.1 &            $-$02 01 39 &         81.9 &        d &            11.26 &         29.98&      1.97 & AGN \\
	\hline
\end{tabular}

\medskip

IR luminosities and luminosity distances were obtained from \citet{sanders03}, while radio 1.4\,GHz luminosities come from \citet{condon98}.\\
$^1$ {Including both NGC\,3690W and NGC\,3690E.} \\
$^2$ From the visual inspection of the \emph{Spitzer}-IRAC 3.6\,$\mu$m images \citep{stierwalt13}. The code is as follows: (N) non merger; (a) pre-merger; (b) early-stage merger; (c) mid-stage merger; (d) late stage merger.\\
$^3$ See equation~\ref{eq:q} for the definition of $q$-factor \citep{helou85}.\\
$^4$ When not stated, X-ray data are not available or inconclusive. See individual description of the galaxies for the corresponding references.
\label{tab:galaxysample}
\end{table*}

A short individual description of each source, some of which have been barely studied, is shown below.

\subsection{MCG\,+08-11-002}
This galaxy, also named IRAS\,05368+4940, is a barred spiral galaxy (type SBab) with a complex morphology, in a late stage of merging \citep{stierwalt13,davies16}. Its mid-IR extended emission is clearly silicate dominated \citep{diaz-santos11}. \citet{davies16} found evidence of a possible preceding starburst episode, likely linked to the previous encounter of the galaxy nuclei.

\subsection{Arp\,299}
\begin{figure}
\includegraphics[width=\columnwidth]{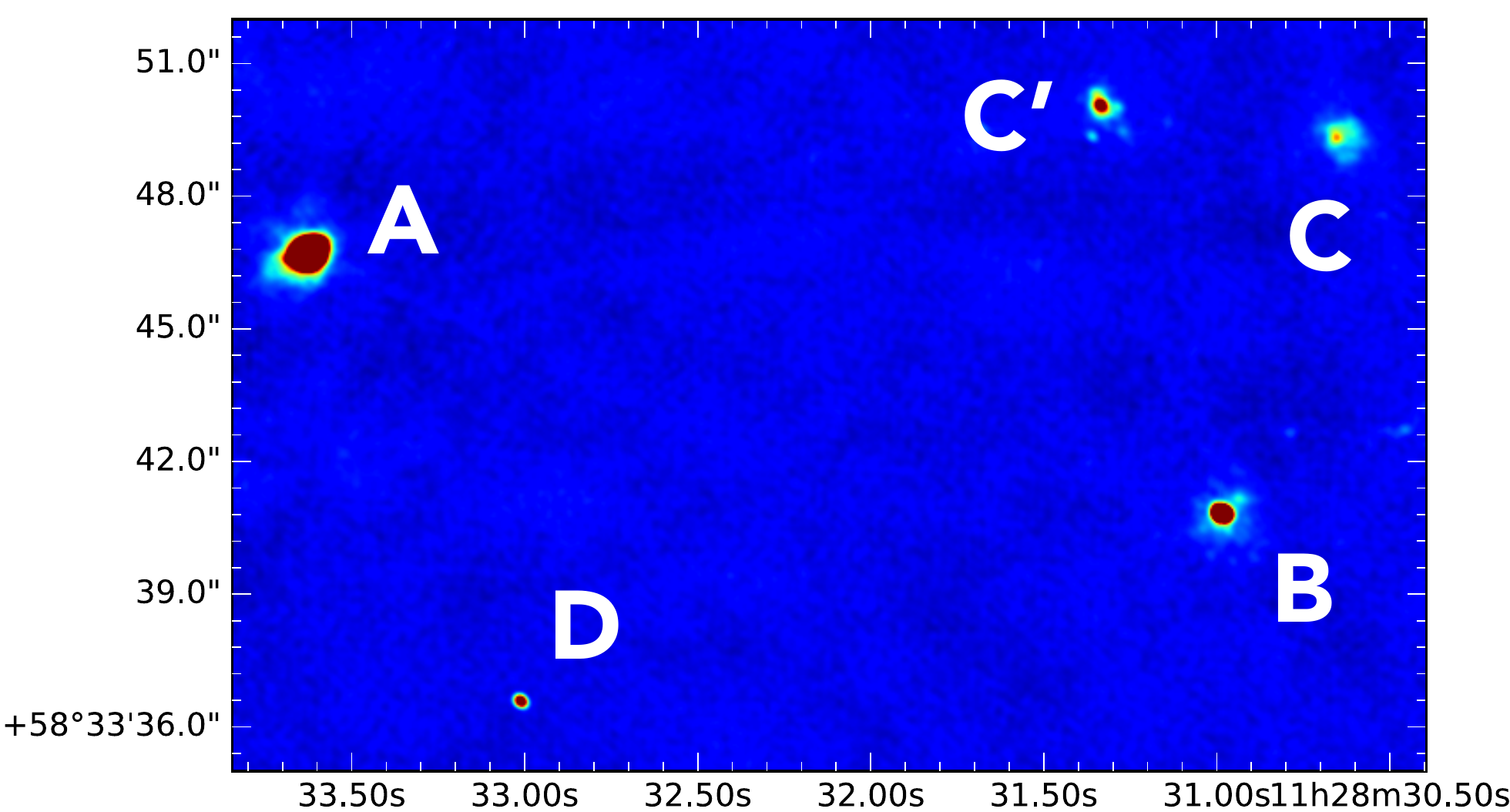}
 \caption{VLA 8.4\,GHz image showing the radio sources in the Arp~299 system. Nuclei A and B are the main components of NGC\,3690 East and NGC\,3690 West, respectively. Further details on the image can be found in section~\ref{sec:radio}}.
 \label{fig:arp299all}
\end{figure}

Arp\,299 is one of the most luminous LIRGs in the local Universe, and for that reason one of the most studied systems. It is in an early merger stage according to \citet{keel95} or in a mid-stage according to \citet{stierwalt13} and \citet{larson16}. Arp\,299 is formed by two galaxies and exhibits two clear radio nuclei \citep{gehrz83}, A and B, and two secondary components, C and C$^\prime$. The remaining compact structure, D, is believed to be a background quasar, unrelated to the system \citep{ulvestad09}. Each radio source is identified in Figure~\ref{fig:arp299all}. Several supernovae have been recently discovered in the inter nuclear region \citep{ryder10,mattila10b, herrero-illana12b, romero-canizales14,kankare14} and in nucleus B \citep[][P\'erez-Torres et al., in prep.]{ulvestad09, romero-canizales11}. However, it is the A nucleus that hosts a very rich supernova factory \citep{neff04, perez-torres09b,ulvestad09, bondi12,herrero-illana12a}, as well as a low luminosity AGN, suggested from X-ray observations \citep{dellaceca02, zezas03,ballo04}, and found by means of VLBI observations \citep{perez-torres10}. SED modeling for this system yielded a star formation rate (SFR) of 90 and 56\,$M_\odot\,\mathrm{yr}^{-1}$ for the east and west components, respectively \citep{mattila12}.

While there are different nomenclatures to name the two galactic components of Arp\,299, some have produced confusion in the literature \citep{corwin04}. To avoid that, we use the unequivocal designations NGC\,3690 East and NGC\,3690 West for these components, which are treated individually throughout this paper.

\subsection{ESO\,440-IG058}
This LIRG, also known as IRAS\,12042-3140, consists of two merging galaxies separated by $\simeq6\,$kpc. The northern component is very compact and has been classified as a LINER \citep{corbett03}. While the northern galaxy is dominated by star formation, the southern emission appears to be dominated by shocks \citep[$v=100-150\,$km\,s$^{-1}$;][]{monreal-ibero10}.

Based on the IR luminosity, \citet{miluzio13} estimated that ESO\,440-IG058 has a star formation rate of $36\,M_\odot\,\mathrm{yr}^{-1}$ and an expected SN rate of 0.4\,SN\,yr$^{-1}$. \citet{rodriguez-zaurin11} derived an age of the stellar population of $t\lesssim6.5\,$Myr for the currently star forming stellar population.

\subsection{IC\,883}
IC\,883, also known as UGC\,8387, is a late-merger LIRG at a distance of 100\,Mpc, showing a peculiar morphology, with extended perpendicular tidal tails visible in the optical and near-IR \citep{smith95, scoville00, modica12}.
This source was classified as an AGN/SB composite \citep{yuan10}, which has been confirmed by the direct detection of a number of radio componentes that are consistent with an AGN and with SNe/SNRs \citep{romero-canizales12b} and with the detection, in the near-IR, of two SNe  \citep{kankare12} within the innermost nuclear region of the galaxy.
Through SED model fitting, \citet{romero-canizales12b} estimated a core collapse supernova (CCSN) rate of $1.1\,$SN\,yr$^{-1}$ and a SFR of 185\,$M_\odot$\,yr$^{-1}$. 
IC\,883 also presents strong PAH emission, silicate absorption and a steep spectrum beyond 20\,$\mu$m \citep{vega08}. Recently, using radio VLBI and X-ray data, \citet{romero-canizales17} have reported unequivocal
evidence of AGN activity, with the nucleus showing a core-jet structure and the jet having subluminal proper-motion.

\subsection{CGCG\,049-057}
Also known as IRAS\,15107+0724, this is the only LIRG in our sample classified as isolated. However, despite its apparent isolation \citep{larson16}, it has a complex and dusty nuclear morphology, so some form of past interaction cannot be ruled out. It hosts an OH megamaser \citep{bottinelli86, baan87}. It is optically classified as a pure starburst, supported by \emph{Chandra} X-ray observations \citep{lehmer10}. However multi-band radio observations show evidence of a buried AGN within the SB \citep{baan06}.

\subsection{NGC\,6240}
This bright LIRG is a well-studied late-stage merger \citep{stierwalt13}. It hosts one of the few binary AGN detected so far using \emph{Chandra} hard X-ray observations \citep{komossa03}, with a projected distance of $\simeq1$\,kpc. This was later supported by the detection of two compact unresolved sources at radio wavelengths with inverted spectral indices \citep[$\alpha=+1.0$ and $\alpha=+3.6$ for the north and south component respectively; see][]{gallimore04}, finding also a third component with a spectral index consistent with a radio supernova. Close to the southern nucleus, a water-vapor megamaser was found \citep{nakai02, sato05}.

\subsection{IRAS\,16516-0948}
This unexplored LIRG was optically identified as a star forming galaxy \citep{mauch07}, and classified as a late merger from its infrared morphology \citep{stierwalt13}. IRAS\,16516-0948 was part of the COLA project, for which high-resolution radio imaging did not detect any compact core \citep{corbett02}. Despite the scarce information available for this galaxy, we found it to be an interesting source, as discussed in section~\ref{sec:offnuclear}.

\subsection{IRAS\,17138-1017}
This LIRG is a highly obscured starburst galaxy \citep{depoy88} in a late stage of interaction. An extremely extinguished supernova ($A_V=15.7\pm0.8\,$mag) was discovered in IRAS\,17138-1017 using infrared K-band observations \citep[SN2008cs;][]{kankare08b} Two more supernovae were also found: SN2002bw \citep{li02, matheson02} and SN2004iq \citep{kankare08a}.

\subsection{IRAS\,17578-0400}

This is a galaxy pair in an early stage of merging \citep{stierwalt13}. Ultra-hard X-ray (14--195\,keV) observations with the \emph{Swift} Burst Alert Telescope (BAT) searching for AGN did not detect any compact emission in this source \citep{koss13}. 
There is abundant archival data in the X-ray, optical, IR, and millimeter bands for this source \citep[see, e.g.,][]{lu14,stierwalt14,rich15}, although no significant results for the purposes of this paper have been reported in the literature.

\subsection{IRAS\,18293-3413}

This source was classified as an H\,{\sc ii} galaxy based on its optical spectrum \citep{veilleux95}. It was detected in hard X-ray data (2--10\,keV), obtained with \textit{ASCA}, at a 5$\sigma$ level by \citet{risaliti00}, finding no evidence of any AGN contribution to the X-ray spectrum. There are discrepancies on the merger stage determination. \citet{stierwalt13} classified it as a mid-stage merger from the visual inspection of \emph{Spitzer}-IRAC 3.6\,$\mu$m images; on the other hand, \citet{haan11} classified it as a very early merger, with canonical disks and no tidal tails based on its \emph{HST} morphology. However, \citet{vaisanen08b}, using high-resolution near-IR K-band adaptive optics imaging with VLT/NACO, showed the galaxy to have a very complex morphology, strongly suggesting a late stage interaction. The system includes a rare un-evolved elliptical companion as well. Using SED modeling, \citet{mattila07b} estimated a CCSN rate of $1$\;SN\;yr$^{-1}$ for IRAS\,18293-3413.  Supernovae SN2004ip \citep{mattila07a, perez-torres07} and AT2013if (Kool et al. in prep.) were discovered within the nuclear regions of this galaxy.

\subsection{NGC\,6926}
This relatively low luminosity LIRG is a spiral galaxy in a very early phase of interaction with the dwarf elliptical NGC\,6929, located $4^\prime$ to the east. Optically identified as a Seyfert~2 \citep{veilleux95}, NGC\,6926 has a powerful water-vapor megamaser \citep{greenhill03, sato05}, typically found in heavily obscured AGN \citep[e.g.][]{greenhill03,masini16}. Its X-ray hardness ratio \citep{terashima15} is also suggestive of an AGN. This is the only source for which we lack near-IR observations.

\section{Observations and data reduction} \label{sec:observations}
The observational data used in this study comes from (1) multi-wavelength archival data obtained through the VizieR photometry tool (used for the SED model fits) and (2) from our own observations at both radio and near-IR wavelengths. Table~\ref{tab:beams} shows a summary of these observations. The images are shown in Figure~\ref{fig:radioirratios} and discussed in section~\ref{sec:radiovsir}.

\begin{figure*}
		{\includegraphics[width=0.8\textwidth]{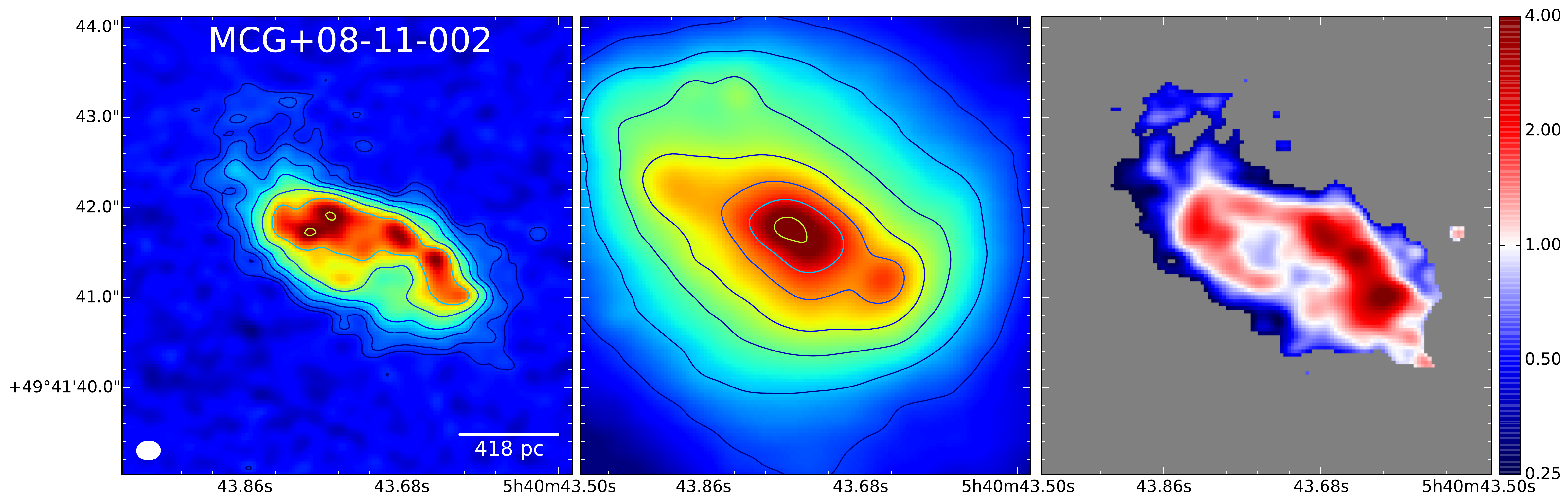}}
		\includegraphics[width=0.8\textwidth]{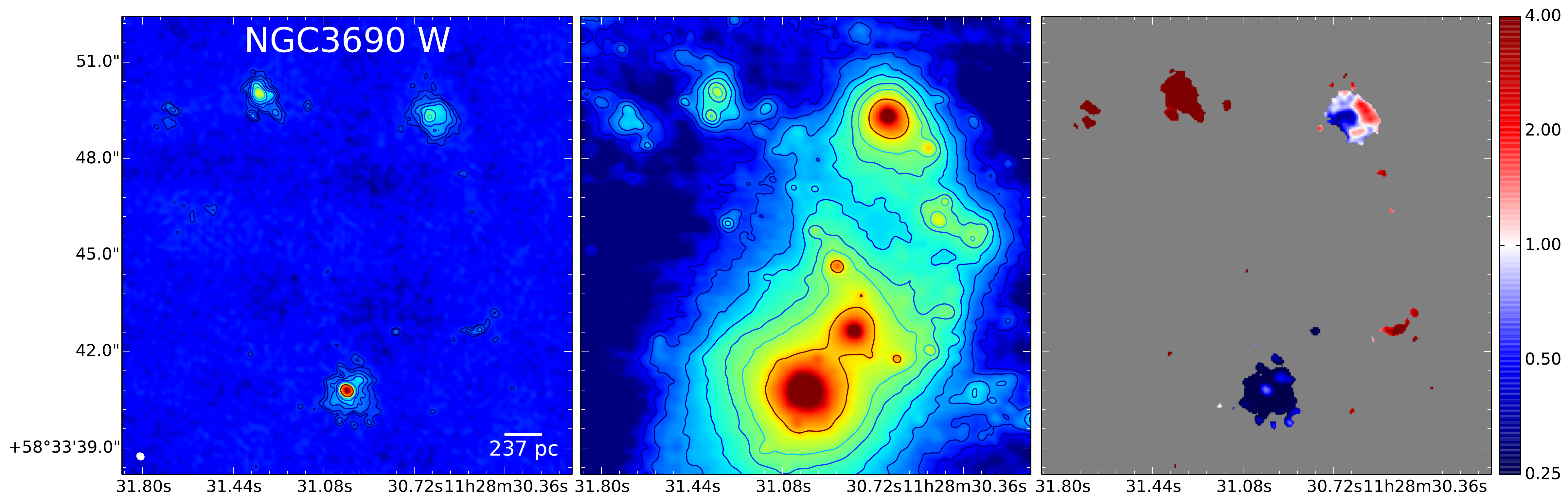} 
		\includegraphics[width=0.8\textwidth]{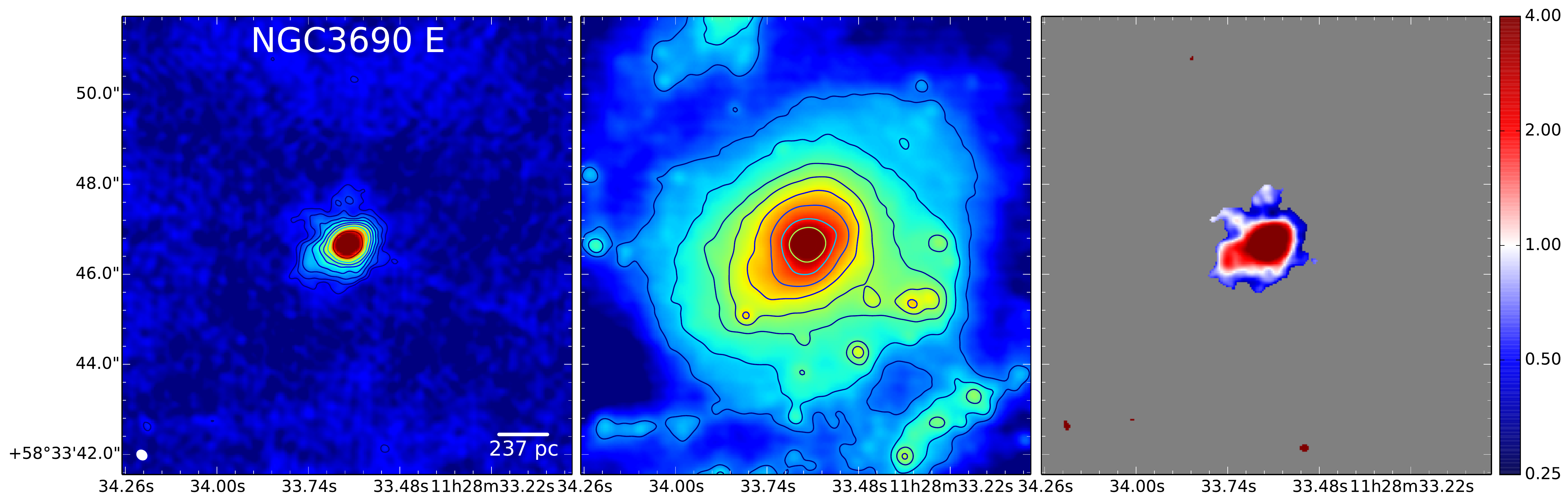} 
		\includegraphics[width=0.8\textwidth]{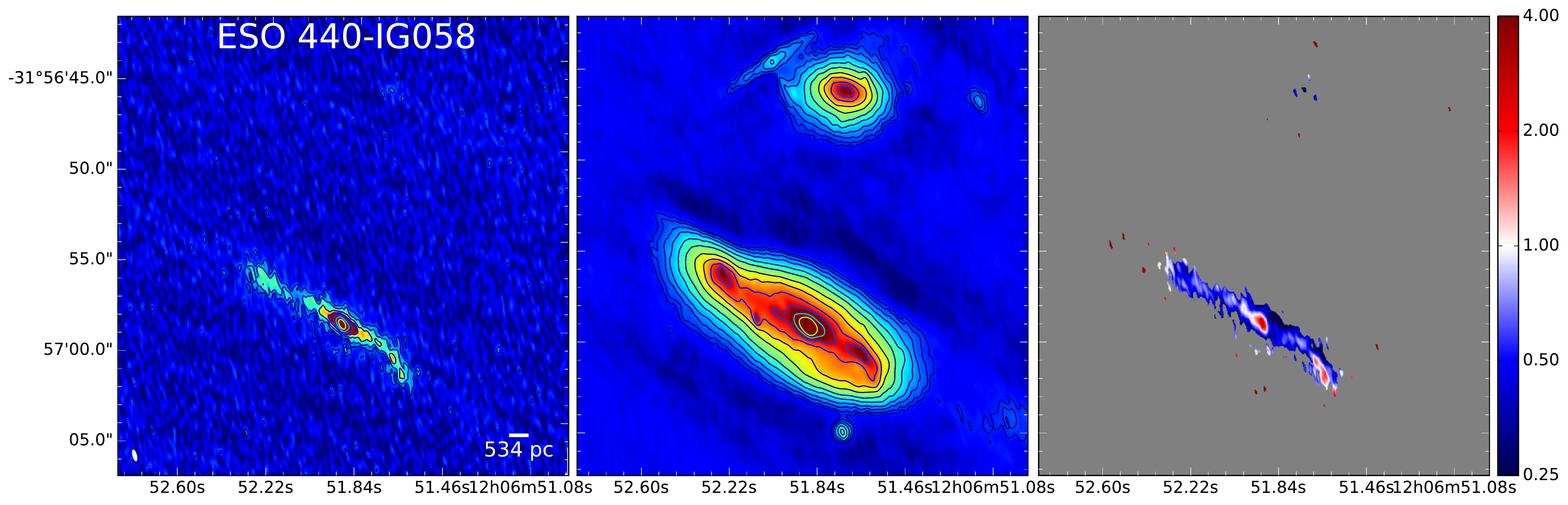} 
		\includegraphics[width=0.8\textwidth]{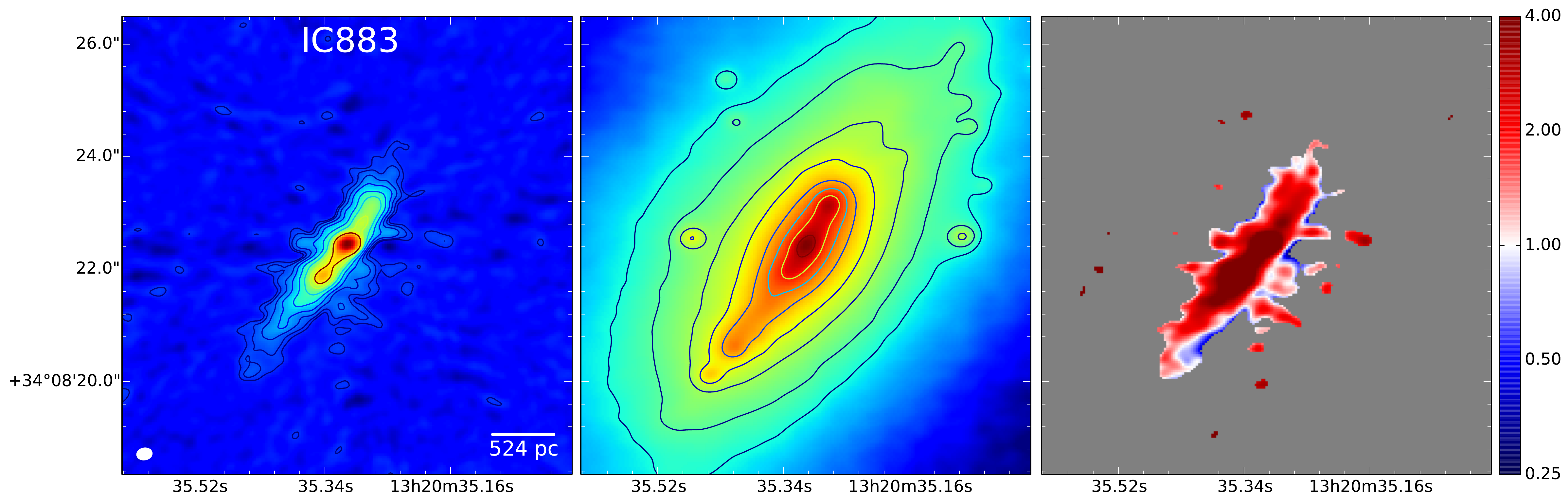} 
	\caption{Radio and near-IR comparison of our sample. The image shows the 8.4\;GHz (3.6\,cm) radio maps (left), the 2.2\,$\mu$m IR maps (middle) and the ratio between radio and IR (right), where redder colors imply radio dominated regions in contrast with bluer colors. Contours are drawn every $3\left(\sqrt{3}\right)^n\times\mathrm{rms}$ level, with $n=0,1,2,3...$ A direct comparison in a single plot between radio and near-IR is shown in Figure~\ref{fig:app:radioir}.}
	\label{fig:radioirratios}
\end{figure*}

\begin{figure*}
		\includegraphics[width=0.8\textwidth]{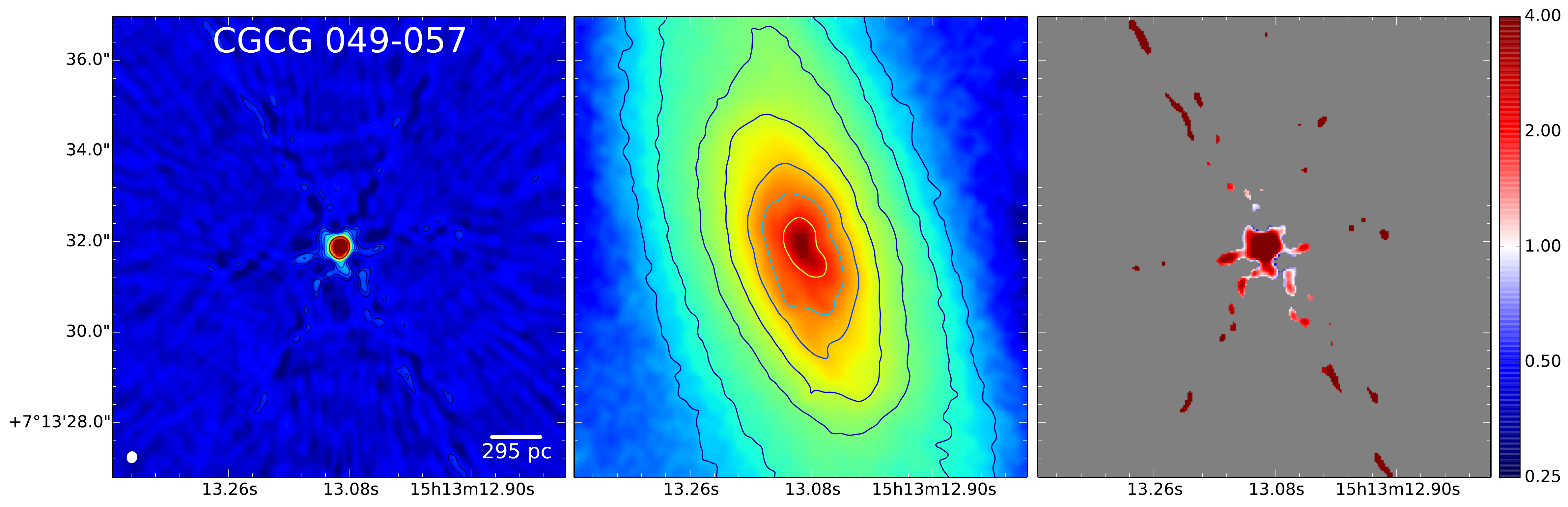}
		\includegraphics[width=0.8\textwidth]{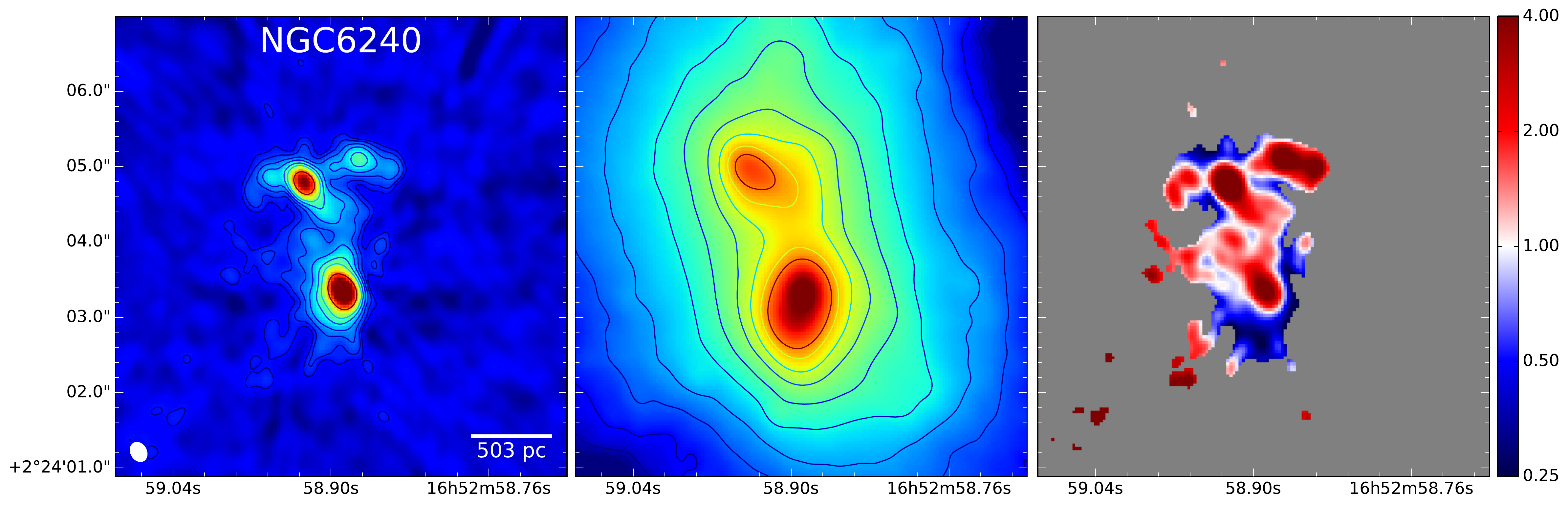} 
		\includegraphics[width=0.8\textwidth]{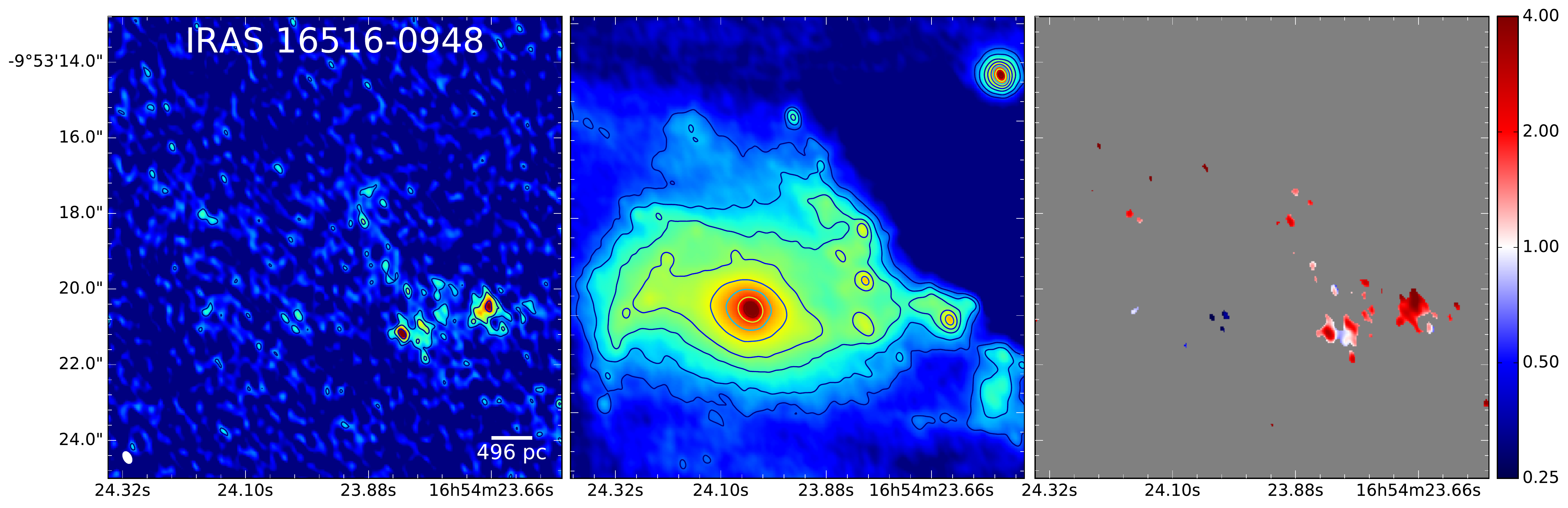}
		\includegraphics[width=0.8\textwidth]{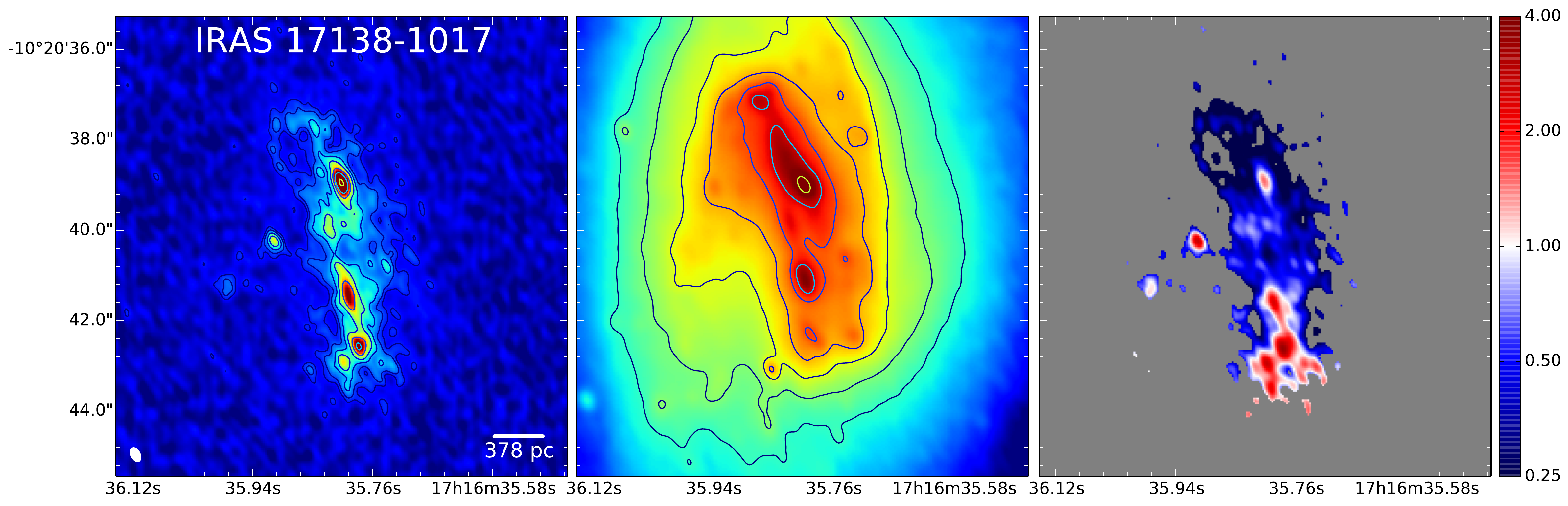} 
		\includegraphics[width=0.8\textwidth]{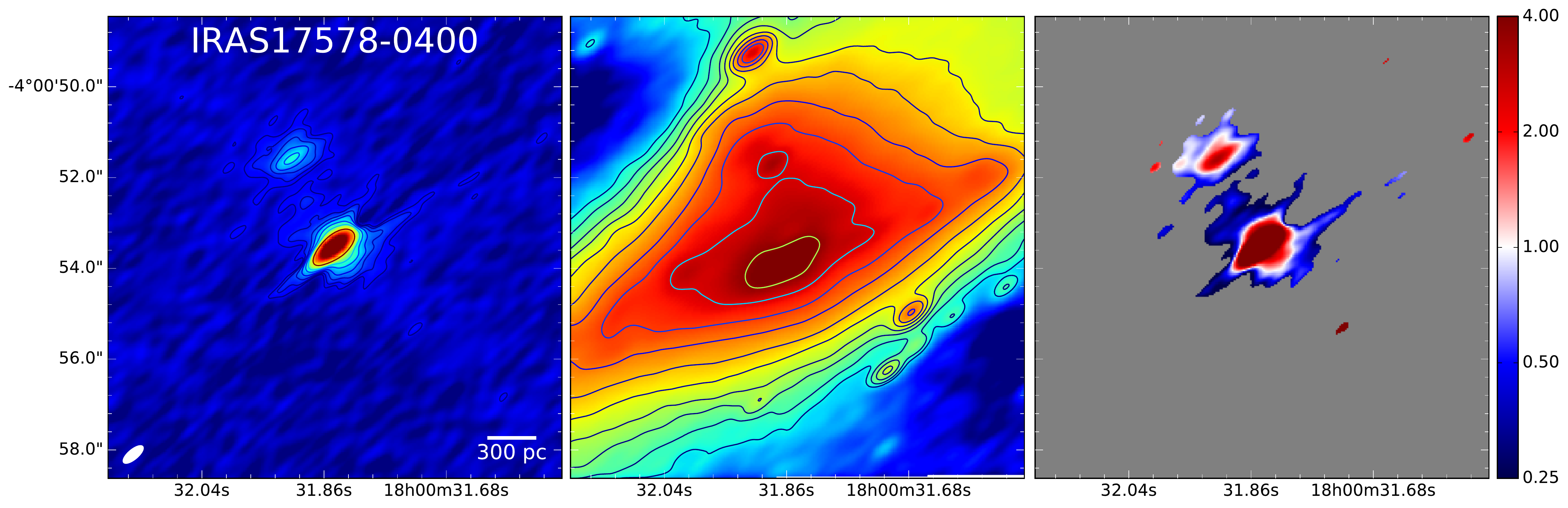} 

\contcaption{}
\end{figure*}

\begin{figure*}
		{\includegraphics[width=0.8\textwidth]{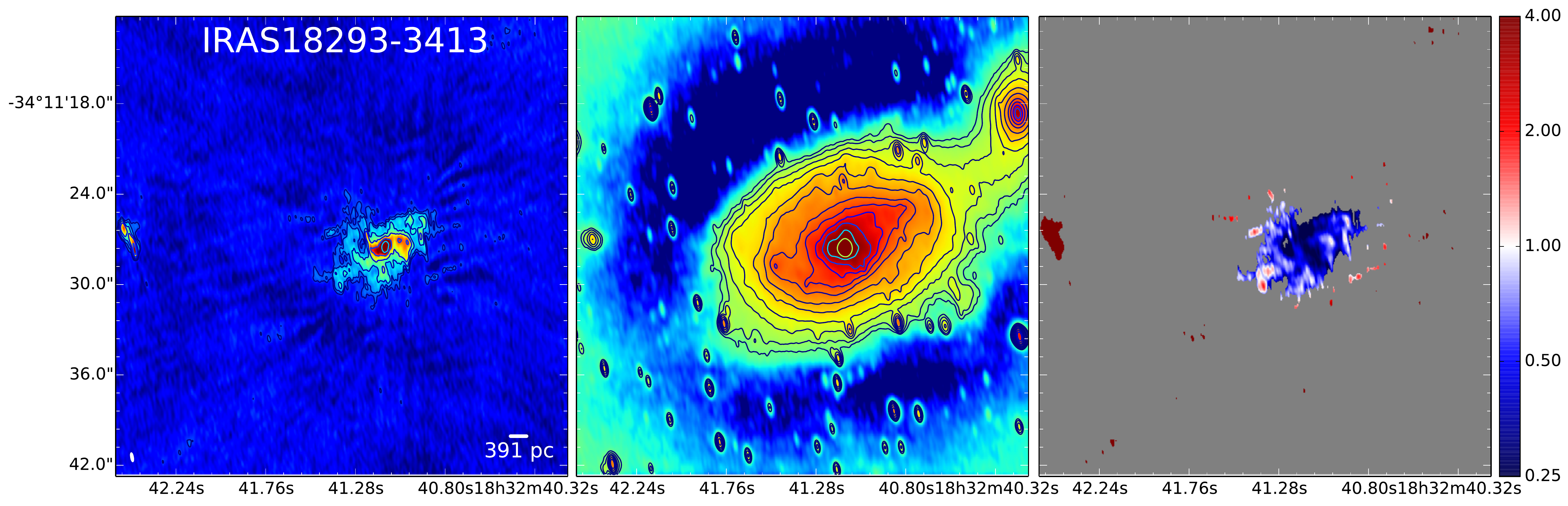}
		\phantom{wwwwwwwwwi}\includegraphics[width=0.285\textwidth]{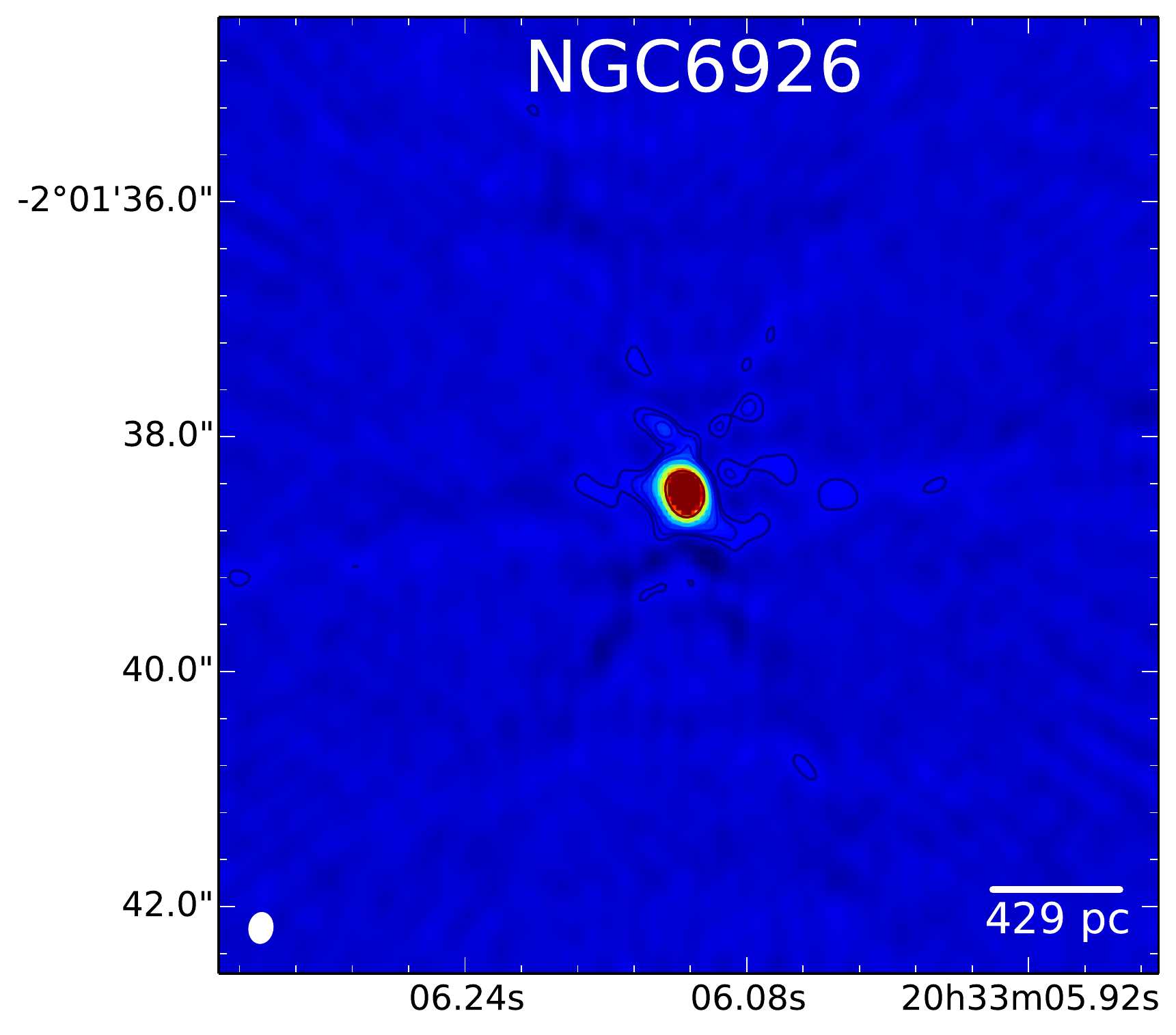}\hfill}

\contcaption{}
\end{figure*}

\begin{table*}
\caption{Observations summary.}
\begin{tabular}{rcccccccccc}
  	\hline
 & \multicolumn{4}{c}{Radio (8.4\,GHz or 3.6\,cm)} &  & \multicolumn{4}{c}{Near-IR (2.2\,$\mu$m)}   \\
\cline{2-5} \cline{7-10} \\
Name & FWHM & rms & Peak & Int. flux &   &  Pixel size & rms & Peak  & Int. flux \\
 & (arcsec) & ($\mu$Jy\,beam$^{-1}$) & (mJy\,beam$^{-1}$) & (mJy) &  &  (arcsec) & ($\mu$Jy) & (mJy) &   (mJy) \\
	\hline
MCG+08-11-002 	& $0.26\times0.21$   & $8.64$ & $0.73$ & 16.68 & & 0.022 & 0.06 & 0.006   &  21.27 \\
NGC\,3690W 		& $0.24\times0.20$   & $12.59$ & $8.85$ & 82.20 & &  0.022 & 0.01 & 0.404  & 31.61 \\
NGC\,3690E 		& $0.24\times0.20$   & $16.91$ & $24.85$ & 22.13 & &  0.022 & 0.03 & 0.017  & 93.90 \\
ESO\,440-IG058	& $0.60\times0.18$   & $8.12$ & $1.07$ & 6.39 & &  0.02 & 0.007 & 0.006   & 21.07  \\
IC\,883 			& $0.27\times0.21$   & $13.52$ & $10.51$ & 34.06 & &  0.022 & 0.02 & 0.006  & 17.09 \\
CGCG\,049-057 	& $0.24\times0.21$   & $23.77$ & $19.73$ & 26.94 & &  0.022 & 0.02 & 0.002  &  17.89 \\
NGC\,6240 		& $0.27\times0.20$   & $29.26$ & $19.70$ & 52.86 & &0.027&  0.008 & 0.078 & 36.67 \\
IRAS\,16516-0948 & $0.33\times0.20$   & $$8.28 & $0.09$ & 0.27 & &  0.022 & {0.006} & {0.005}  & 10.84 \\
IRAS\,17138-1017 & $0.33\times0.20$   & $7.61$ & $0.68$ & 8.79 & & 0.022 & 0.03 & 0.004  & 47.42  \\ 
IRAS\,17578-0400 & $0.55\times0.21$   & $9.66$ & $17.21$ & 23.24  & & 0.022 & 0.005 & 0.002  & 22.17 \\
IRAS\,18293-3413  & $0.60\times0.17$   & $11.64$ & $0.83$ & 25.30 & &0.027 & 0.001 & 0.039  & 101.27 \\
NGC\,6926 		& $0.25\times0.19$    & $7.26$ & $3.99$ & 6.00 & &  $\cdots$ & $\cdots$ & $\cdots$  & $\cdots$ \\
	\hline
\end{tabular}

\medskip

The angular resolution for the near-IR images is between $0.07^{\prime\prime}$ (diffraction-limited FWHM at K-band) and $0.10^{\prime\prime}$. {ESO\,440-IG058 was observed with GeMS/GSAOI, NGC\,6240 and IRAS\,18293-3413 with NACO, and the remaining sources with ALTAIR/NIRI. Integrated fluxes are quoted above $5\sigma$.}
\label{tab:beams}
\end{table*}

\subsection{Radio} \label{sec:radio}

The radio data used in this paper were X-band (8.4\,GHz, 3.6\,cm) observations in full polarization mode and with a total bandwidth of 2048\,MHz (project 12B-105; PI: M. \'A. P\'erez-Torres) taken between 5 October and 26 December 2012 using the new VLA capabilities. The total on-source time for each target was approximately $30$\,min. For every source we used 3C286 as flux and bandpass calibrator, while we observed for each case a nearby bright point-like source for phase calibration purposes. We used the \emph{Common Astronomy Software Applications} \citep[\emph{CASA}, ][]{mcmullin07} package for data reduction purposes, which consisted of standard amplitude and phase calibration. We imaged the sources using a Briggs weighting scheme with \texttt{ROBUST = 0.5} to obtain the best compromise between sensitivity and resolution. Table~\ref{tab:beams} shows the beam size, noise achieved and peak flux density for each image. To compare these images with our near-IR data, we converted the flux density units from Jy\,beam$^{-1}$ to Jy\,px$^{-1}$.

Although our data were taken in a single band, the large bandwidth (2\,GHz) VLA X-band receivers allowed us to estimate the spectral index, $\alpha$ ($S_{\nu} \propto \nu^{\alpha}$), through this radio band. To this end, we imaged our sources using multiscale multifrequency synthesis \citep[\texttt{mode = mfs} within \emph{CASA}; see][]{rau11}, which models the wide-band sky brightness as a linear combination of Gaussian-like functions whose amplitudes follow a Taylor-polynomial in frequency. The task performs a Taylor expansion to the second order (\texttt{nterms = 2}) of the function:
\begin{equation}
I_\nu^\mathrm{sky}=I_{\nu_0}^\mathrm{sky}\left(\frac{\nu}{\nu_0}\right)^\alpha,
\label{eq:spix}
\end{equation}
where $I^\mathrm{sky}$ is the multiscale image, $\nu_0$ is the reference frequency (in our case $\nu_0=8.459\,$GHz), and $\alpha$ is the spectral index.

The two output images are the first two coefficients of the expansion, i.e.:
\begin{equation}
I_0 = {I_{\nu_0}}\;\;\;\;\;\;\;\;\;\;\;\;\;\;\;\; I_1=I_{\nu_0}\alpha.
\end{equation}

We then obtained the radio spectral index map of each source by simply using the ratio $I_1/I_0$. These maps are discussed in section~\ref{sec:spix}.

\subsection{Near-IR data}
We used ALTAIR/NIRI and GeMS/GSAOI on the Gemini North and South telescopes, respectively (PI: S. Ryder), and  NACO on the VLT (PIs: S. Mattila) to obtain near-IR K-band ($2.2\,\mu$m) laser guide star adaptive optics (AO) images. These instruments have a pixel scale of $0.022^{\prime\prime}\,$px$^{-1}$ (ALTAIR/NIRI), $0.02^{\prime\prime}\,$px$^{-1}$ (GeMS/GSAOI), and $0.027^{\prime\prime}\,$px$^{-1}$ (NACO, with camera S27). IRAS\,18293-3413 and NGC\,6240 were observed using NACO, ESO\,440-IG058  using GeMS/GSAOI, and the remaining targets were observed with ALTAIR/NIRI. Data from ALTAIR/NIRI come from a multi-epoch survey intended for the detection and study of nuclear core-collapse supernovae (CCSNe). The ALTAIR/NIRI images used in this paper are a combination of data from 2008 to 2012. NACO observations were performed on 13 September 2004 (IRAS\,18293-3414)  and 31 May 2011 (NGC\,6240), and GeMS/GSAOI observations for ESO\,440-IG058 were performed on 05 March 2015. The total integration time spans from 990\,s (for NGC\,6240) to 2192\,s (for NGC\,3690W). No near-IR data are available for NGC\,6926.
Spatial resolution was typically $\simeq0.1^{\prime\prime}$ in the AO corrected images.

We reduced the near-IR data using standard \emph{IRAF}-based tasks, including flat-fielding and sky subtraction, and created the final images by average-combining individual frames from different epochs after having shifted them to a common reference. A detailed description of the reduction process of the ALTAIR/NIRI and NACO/VLT data is presented in \citet{randriamanakoto13a}, while GeMS/GSAOI data are part of the SUNBIRD survey (Kool et al. 2016, in prep.).  In multi-epoch data, we found no evidence of variability among the different individual epochs
above 2-$\sigma$.

\subsection{Astrometry calibration and image convolution}
Since the Field of View (FOV) of the near-IR images was small (between 22 and 54\,arcsec on a side), we first calibrated larger FOV archival data from NOT or \emph{HST}/ACS using stars from Guide Star Catalogue-2 or 2MASS K$_\mathrm{S}$-band catalogue. We then added the WCS information in the FITS header of our targets using these intermediate images \citep[see][for details]{randriamanakoto13a}. We estimate the astrometry calibration uncertainty to be $\simeq0.15^{\prime\prime}$.

Given the significantly different resolution of the radio and near-IR images (see Table~\ref{tab:beams}), we convolved the near-IR images with a Gaussian with the size of the radio beam for each case, and then rebinned the near-IR images to the pixel size used in our radio images ($0.04^{\prime\prime}$). 

\section{Results and discussion}\label{sec:discussion}

\subsection{SED modeling} \label{sec:sed}
{There is a wide variety of methods, based on data at different wavelengths, to determine the starburst and AGN properties of galaxies. These are not always consistent and often depend on the individual environment and star forming history of each galaxy. A more reliable method is to take advantage of multi-wavelength observations and fit the SED to combined templates of starbursts and AGN to derive the relevant parameters (AGN contribution, starburst age, star formation rate and supernova rate). In this section, we obtain these parameters and compare them with those obtained with traditional tracers.}

We modeled the multi-wavelength SED for the sources in our sample. To this end, we combined libraries of starburst models \citep{efstathiou00, efstathiou09}, AGN torus models \citep{efstathiou95, efstathiou13} and models of the spheroidal/cirrus component (Efstathiou et al., \emph{in prep.}). The latter is based on cirrus models by \citet{efstathiou03} that rely on calculations of the radiative transfer problem in a medium where dust and stars are mixed in a spheroidal distribution. We used a Markov Chain Monte Carlo (MCMC) fitting code  \citep[SATMC; ][]{johnson13} to obtain realistic uncertainties on the fitted parameters. {SATMC is based on Bayesian statistics, resulting in more reliable results and error determination than traditional least-square fitting. For details on the SATMC fitting process, we refer the reader to \citet{johnson13}.}

{The SFR of the starburst and its errors were derived self-consistently from the radiative transfer models which incorporate the stellar population synthesis model of \citet{bruzual93, bruzual03}. The libraries for a Salpeter initial mass function (IMF), solar metallicity and stars in the range 0.1--125\,$M_\odot$ were used. The models of Efstathiou et al. predict the SED of a starburst at different ages $t_*$ assuming the star formation rate declines exponentially with an e-folding time $\tau$. From the SATMC fit we know both the best fit values of $t_*$ and $\tau$ and their $1\sigma$ uncertainties. Additionally, we know the fitted starburst luminosity $L_\mathrm{SB}$ and the associated uncertainties. The combination of $L_\mathrm{SB}$, $t_*$, $\tau$ and the tables of \citet{bruzual93,bruzual03} uniquely determined the total mass of stars formed in the starburst episode for each combination. The stellar mass is divided by 50\,Myr to give the mean SFR that we provide.
}

\begin{figure*}
\includegraphics[width=1\textwidth]{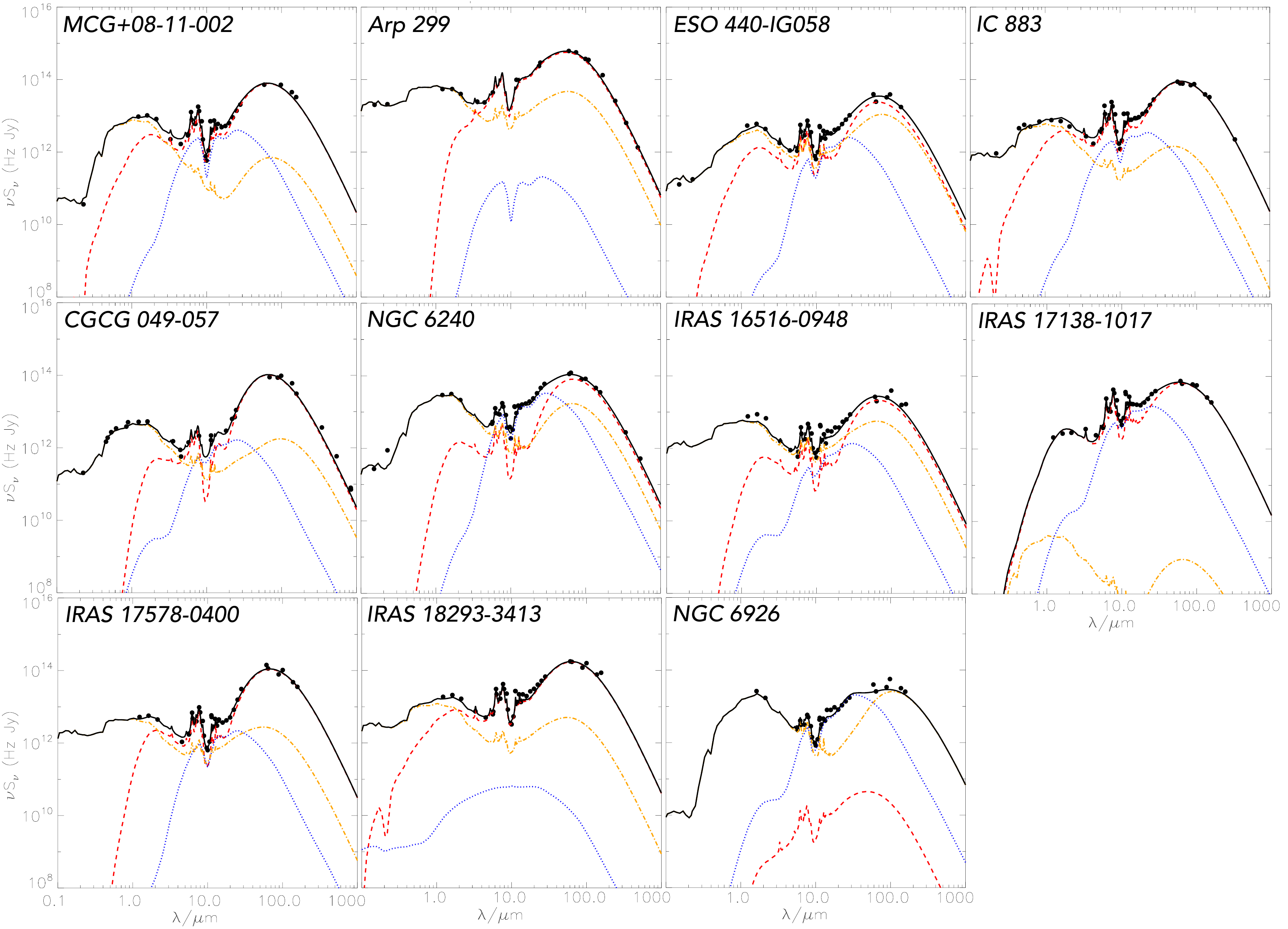}
 \caption{Best SED fitting model for all the sources in our sample. Black data points correspond to fitted data points (IRS spectral data are not plotted at their full resolution for the sake of clarity). The lines correspond to the starburst contribution (dotted red), the AGN torus (dotted blue), and the spheroidal/cirrus component (dash-dotted {orange}). The solid black lines show the total (starburst+AGN+spheroidal/cirrus) fit. 
See main text for details.
}
\label{SEDselection.pdf}
\end{figure*}

We obtained the SED data points from public data through the VizieR photometry tool, which included near- to far-IR photometry {(2MASS, IRAC, WISE, AKARI, IRAS)}, as well as UV {(GALEX)}, optical {(SDSS)}, and (sub-)mm {(SCUBA)} data for some of the sources (see photometric data points in Figure~\ref{SEDselection.pdf}). {All the photometric data we are using refer to the whole galaxy, either because the PSF was big enough (far-IR data) or because we used integrated fluxes.} In addition, we included the \emph{Spitzer} IRS spectra {when available}. {For these, and to minimize aperture effects, we scaled up their fluxes in such a way that the edges of each spectrum match the flux of the adjacent photometric data points.
The spectral resolution of the IRS data is reduced so that they are better matched to the resolution of the radiative transfer models. The IRS data included in the fitting with SATMC have a wavelength grid which is separated in steps of 0.05 in the log of rest wavelength. We additionally added more points around the 9.7$\mu m$ silicate feature and the PAH features.
There was no spectrum available for CGCG049-057, and we did not use IRS data for the fitting of Arp299 either, since we are fitting both components together.}

We show in Table~\ref{tab:andreas} the best fit parameters, and in Figure~\ref{SEDselection.pdf} show the fitted SED for ten sources. The CCSN rates are estimated by convolving the SN rate at a given time with the star formation history of the starburst  \citep[see details in][]{mattila12}. {For consistency we modeled Arp\,299 (NGC\,3690E + NGC\,3690W) as a unique galaxy but we note that  a similar, yet not as complete model was published in \citet{mattila12} for the individual components.} We note a discrepancy in the SB ages of NGC\,3690E {(45\,Myr)} and NGC\,3690W {(55\,Myr)} obtained by \citet{mattila12} and those derived {here and also in the modeling} by \citet{alonso-herrero00} using the evolutionary synthesis models by \citet{rieke93} and \citet{engelbracht96}, which yield significantly lower ages ($6-8$\,Myr and $4.5-7$\,Myr for NGC\,3690E and NGC\,3690W, respectively).
However, we find the models not comparable since they are considering different IMF, different star formation history, and, specially, different apertures: the whole {system in this study, versus each} component in \citet{mattila12} or the innermost $<5\,$arcsec region in \citet{alonso-herrero00} {We also note that the higher SFRs quoted in \citet{mattila12} were averaged over the whole duration of the starburst, while in this study we are averaging it over 50\,Myr. Nonetheless, the SN rate between both studies are compatible.}


\begin{table*}
\caption{SED model fitting parameters.}
\renewcommand{\arraystretch}{1.2}
\begin{tabular}{rccccccc}
  	\hline
Name & $L_\mathrm{sph}$ & $L_\mathrm{SB}$ & $L_\mathrm{AGN}$ & AGN contribution & $\mathrm{SFR}_\mathrm{mean}$ & $\mathrm{Age}_\mathrm{SB}$  & $\nu_\mathrm{CCSN}$ \\
 &$(10^{10}L_\odot)$ & $(10^{10}L_\odot)$ & $(10^{10}L_\odot)$ & $($\%$)$ & $(M_\odot$\,yr$^{-1}$) & (Myr)  & (SN\,yr$^{-1}$) \\
(1) &(2)& (3) & (4) & (5) & (6)  & (7)  \\
	\hline
{MCG+08-11-002} 	 & $2.36_{-1.00}^{+1.26}$  & $21.52_{-0.58}^{+0.64}$ &  $2.86_{-1.72}^{+5.99}$ & $10.7_{-6.1}^{+16.5}$ & $97.2_{-6.2}^{+2.3}$  & $29.1_{-0.8}^{+0.4}$  & $1.07_{-0.05}^{+0.02}$   \\
{Arp\,299}                     &  $15.31_{-2.83}^{+1.80}$  & $63.28_{-1.85}^{+0.96}$ & $0.06_{-0.05}^{+0.45}$ & $0.1_{-0.1}^{+0.6}$ & $86.1_{-18.4}^{+7.4}$  & $12.6_{-3.1}^{+2.6}$ & $1.45_{-0.22}^{+0.10}$ \\
ESO\,440-IG058	 & $6.38_{-2.22}^{+1.39}$  & $9.65_{-1.25}^{+2.02}$ &  $1.78_{-1.46}^{+1.89}$ &  $10.0_{-8.1}^{+8.9}$  & $48.3_{-8.0}^{+11.1}$   & $27.9_{-3.4}^{+1.6}$  & $0.51_{-0.06}^{+0.11}$   \\
IC\,883 		 & $3.57_{-0.65}^{+7.93}$  & $37.02_{-1.63}^{+1.10}$ &  $2.31_{-0.69}^{+1.21}$ & $5.4_{-1.5}^{+2.6}$  & $157.6_{-8.7}^{+8.6}$  & $28.6_{-1.8}^{+0.8}$  & $1.78_{-0.07}^{+0.07}$   \\
CGCG\,049-057 	 & $0.88_{-0.22}^{+0.57}$  &  $11.21_{-0.35}^{+0.13}$ &   $0.31_{-0.24}^{+0.30}$ &  $2.5_{-1.9}^{+2.3}$  & $43.0_{-10.8}^{+4.5}$   & $24.8_{-5.0}^{+1.0}$  & $0.52_{-0.11}^{+0.03}$   \\
NGC\,6240 		 & $21.13_{-4.20}^{+7.70}$  & $33.93_{-8.03}^{+2.86}$ & $28.75_{-5.74}^{+11.97}$ & $34.3_{-4.5}^{+8.2}$  & $166.2_{-64.7}^{+22.6}$  & $27.6_{-3.7}^{+1.2}$  & $1.80_{-0.54}^{+0.18}$   \\
IRAS\,16516-0948 & $5.55_{-0.89}^{+2.82}$  &  $7.84_{-2.89}^{+0.78}$ &  $1.35_{-1.35}^{+1.12}$ & $9.2_{-9.2}^{+7.0}$  & $40.3_{-18.8}^{+2.1}$   & $28.5_{-4.1}^{+0.3}$  & $0.42_{-0.17}^{+0.02}$   \\
IRAS\,17138-1017 & $<0.05$  & $16.45_{-0.87}^{+0.87}$ &  $4.54_{-1.27}^{+2.67}$ & $21.6_{-5.3}^{+9.7}$  & $58.8_{-3.2}^{+2.3}$   & $23.0_{-2.0}^{+1.3}$  & $0.75_{-0.04}^{+0.01}$   \\
IRAS\,17578-0400 & $1.52_{-0.49}^{+0.38}$  & $14.84_{-0.24}^{+0.42}$ &  $0.46_{-0.24}^{+0.42}$ &  $2.7_{-1.4}^{+2.4}$ & $73.8_{-6.5}^{+5.4}$   & $27.7_{-1.5}^{+1.8}$  & $0.79_{-0.05}^{+0.04}$   \\
IRAS\,18293-3413 & $5.44_{-2.18}^{+3.34}$  & $45.62_{-1.88}^{+0.86}$ & $0.02_{-0.01}^{+0.11}$ & $<0.2$  & $167.1_{-14.4}^{+11.8}$  & $27.5_{-0.5}^{+2.0}$  & $1.97_{-0.17}^{+0.07}$   \\
NGC\,6926 		 & $12.41_{-1.73}^{+2.42}$  &  $0.01_{-0.01}^{+0.8}$ &  $21.67_{-10.80}^{+6.41}$ & $63.6_{-19.5}^{+7.1}$  & $<0.5$    & $9.1_{-8.1}^{+5.2}$  & $<0.01$   \\
	\hline
\end{tabular}

\medskip

(1) Source name; (2) Bolometric luminosity of the spheroid; (3) Starburst luminosity; (4) AGN luminosity \citep[anisotropy corrected; see][]{efstathiou06}; (5) Percentage of AGN luminosity with respect to total luminosity; (6) Star formation rate, averaged over the past 50\,My; (7) SB age; (8) Core-collapse supernova rate.

$^\dagger$ NGC\,3690W and NGC\,3690E were modeled by \citet{mattila12}. The starburst parameters from their SED modeling are quoted in the table. 
\label{tab:andreas}
\end{table*}

Our modeling allowed us to quantify the luminosity contribution of each component: the starburst, the AGN, and the spheroid, i.e., the underlying host galaxy stellar population minus the current starburst. 
We find that the SED of all sources in our sample is starburst-dominated except for NGC\,6926 (AGN contribution of $\simeq64\%$), with two additional sources having a significant AGN contribution ($\gtrsim20\%$).
Of the seven sources with an X-ray classification (see Table~\ref{tab:galaxysample}), four of them are catalogued as AGN: NGC\,6926 and NGC\,6240 show an important AGN contribution in our analysis, although Arp299 and IC883 do not. However, we note that these two sources are known to host an AGN.
We also derive star formation rates ($40$ to $167\,M_\odot\,\mathrm{yr}^{-1}$), supernova rates (0.3 to $2.0\,\mathrm{SN}\,\mathrm{yr}^{-1}$), and starburst ages (23 to $55\,\mathrm{Myr}$) that are consistent with their LIRG nature, with the exception of NGC\,6926, a young starburst (9\,Myr) where supernova events have not yet been triggered. 
The oldest starburst of our sample corresponds to {MCG+08-11-002 at $\simeq29$\,Myr}, while the youngest one is NGC\,6926 at just $\simeq9\,$Myr. 

Three of our sources show a remarkably high CCSN rate (which depends on both SFR and age) {above 1.5\,SN\,yr$^{-1}$}: IRAS\,18293-3413, with 1.97\,SN\,yr$^{-1}$, NGC\,6240 with 1.80\,SN\,yr$^{-1}$, and IC\,883 with 1.78\,SN\,yr$^{-1}$, the latter consistent with previous studies \citep[$\nu_\mathrm{CCSN}=1.1^{+1.3}_{-0.6}$,][]{romero-canizales12b}. The extreme CCSN rate and SFR of IRAS\,18293-3413, together with the controversy on its merger stage and its extremely steep spectrum ($\alpha=-1.73\pm0.70$, see section~\ref{sec:spix}), turns it into an interesting case, likely hosting a very rich starburst. This is supported by the detection of hundreds of {super star clusters (SSCs)} in the field of IRAS\,18293-3413 \citep{randriamanakoto13a}. We note however, the lower SN rate derived with a previous simpler model \cite[1\;SN\;yr$^{-1}$;][]{mattila07b}. By contrast, our fit for NGC\,6926 is modeled by such a young starburst (9\,Myr) that no SN event has yet been observed from its current star formation episode. {The ratio of the SB luminosity to $L_\mathrm{IR}$ from Table~\ref{tab:galaxysample} \citep{sanders03} is only $5\%$ for NGC\,6926, in agreement with the above, while it ranges between 42 and 89\% for the rest of the sources.}

In the following sections we compare the results from our SED models described above with other proxies and indicators of both SFR and AGN.

\subsubsection{Comparison with SFR indicators} \label{sec:sfrindicators}
\begin{figure}
\includegraphics[width=\columnwidth]{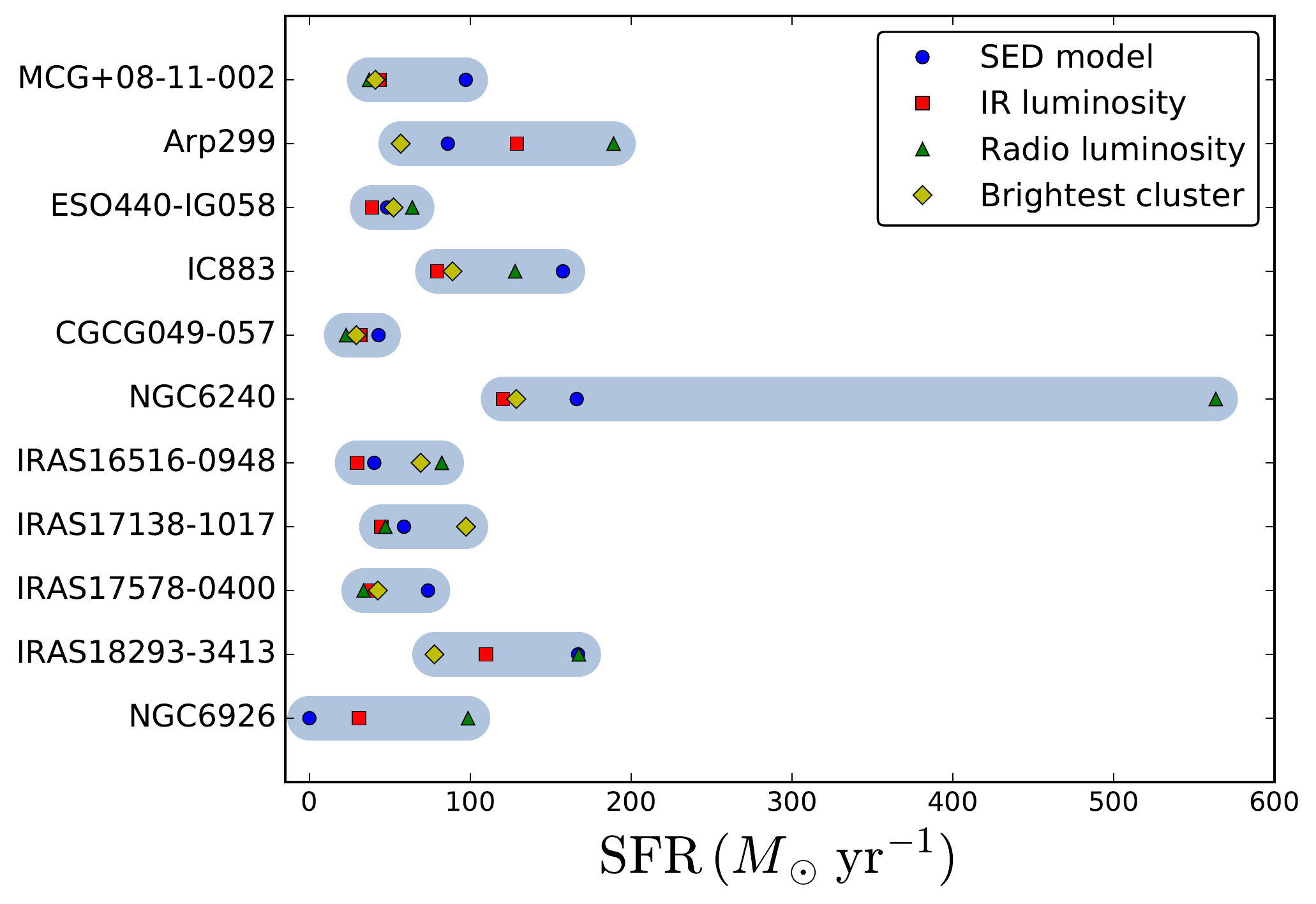}
\caption{Star formation rate SED models (blue circle) compared with three extinction-free prescriptions, based on the IR luminosity \citep[][red squares]{kennicutt98}, the radio luminosity \citep[][green triangles]{murphy11}, and the magnitude of the brightest super star cluster in the galaxy \citep[][yellow diamonds]{randriamanakoto13b}. The blue shaded area simply indicates the range between the lowest and the highest estimate for visual guidance. See main text for details.}
\label{fig:sfrcomparison}
\end{figure}

Table~\ref{tab:andreas} quotes the SFR obtained from the multi-wavelength SED fit of our sources, averaged  over the past 50\,Myr. However, there is a large number of methods to indirectly estimate SFR in galaxies. From these prescriptions, only those that are not strongly affected by extinction (such as IR and radio tracers) are of real use in the dusty environments of LIRGs. In Figure~\ref{fig:sfrcomparison} we show a comparison between our SED modeling and three of these methods, described below:

\begin{enumerate}

\item The outcome of massive stars heating the interstellar dust is an intense infrared flux re-emitted by the dust through black-body radiation. The following relation \citep{kennicutt98} is used to connect the infrared luminosity with the star formation rate in case of star formation $<100$\,Myr of age, which is satisfied in the cases we present in this paper: 
\begin{equation}
\mathrm{SFR}=4.5\times10^{-44}L_\mathrm{IR},
\end{equation}
with SFR in $M_\odot$\,yr$^{-1}$ and $L_\mathrm{IR}$, measured between 8 and 1000\,$\mu\mathrm{m}$, in erg\,s$^{-1}${, where we use the values quoted in Table~\ref{tab:galaxysample}}.

\item There is a tight correlation between the IR and radio luminosities (see section~\ref{sec:intro}). Thus, radio luminosity is expected to trace SFR in a similar way as $L_\mathrm{IR}$ does. A common prescription used to derive SFR from radio data (typically NVSS) is given by \citet{murphy11}, and adapting it for a Salpeter IMF (as for the rest of the SFR indicators), we have:
\begin{equation}
\mathrm{SFR}=1.02\times10^{-28}L_{1.4\,\mathrm{GHz}},
\end{equation}
where SFR is given in $M_\odot$\,yr$^{-1}$ and $L_{1.4\,\mathrm{GHz}}$ in erg\,s$^{-1}$\,Hz$^{-1}$, obtained from NVSS \citep{condon98}. The luminosity at this frequency is essentially due to non-thermal emission.

\item \citet{randriamanakoto13b} established a relation between the near-IR K-band magnitude of the brightest SSC in a galaxy and its global SFR. For the local Universe, this relation can be expressed as:
\begin{equation}
M_k^\mathrm{brightest} = -2.56\log{(\mathrm{SFR})}-13.39.
\end{equation}

\end{enumerate}

\begin{figure}
\includegraphics[width=\columnwidth]{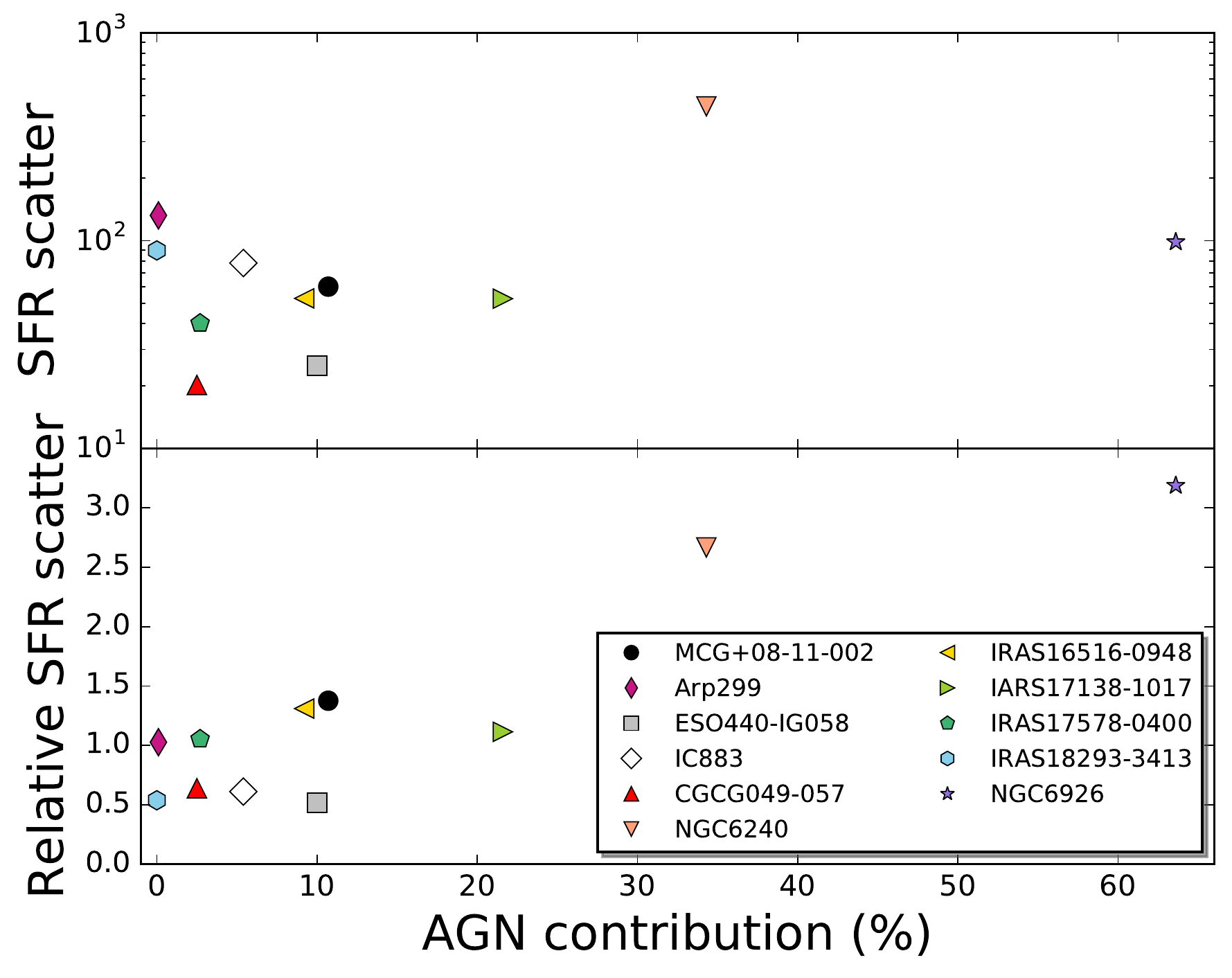}
 \caption{SFR scatter as a function of AGN contribution for each galaxy. Top panel shows the absolute scatter, given by SFR$_\mathrm{max}-$SFR$_\mathrm{min}$, while bottom panel shows the relative scatter, obtained through (SFR$_\mathrm{max}-$SFR$_\mathrm{min}$)/SFR$_\mathrm{median}$.}
\label{sfrscatter.pdf}
\end{figure}

The range between highest and lowest estimates of SFR {(i.e. $\mathrm{SFR}_\mathrm{max}-\mathrm{SFR}_\mathrm{min}$)} of the different tracers shown in Figure~\ref{fig:sfrcomparison} is approximately similar, with a median {scatter} of {78.1}\,$M_\odot$\,yr$^{-1}$, with only one source, NGC\,6240, deviating beyond the standard deviation, for which the SFR derivation through its radio luminosity is highly overestimated. {The radio luminosity tracer is indeed known to over-estimate the SFR in the presence of strong AGN activity \citep{delmoro13, bonzini15}, which is also supported by the large SFR derived from the radio luminosity in NGC\,6926, with an AGN contribution of $\simeq64\%$. In this sense, Figure~\ref{sfrscatter.pdf} shows the absolute and relative scatter of the SFR determination with respect to the relative AGN contribution. We suggest the existence of a weak trend, where more AGN dominated sources present a larger scatter in the SFR determination from different proxies.}

{For the case of NGC\,6240, \citet{gallimore04} reported that no compact sources were detected with the VLBA at 8.4\,GHz, with a $1\sigma$ upper-limit of 130\,$\mu$Jy/beam, which might lead us to think that the AGN contamination is not a satisfactory explanation for the overestimated value of the radio SFR. However, the radio AGN emission has both compact and diffuse origin, and the large difference in angular resolution between the VLA  and the VLBA can account for this apparent discrepancy. To test this, we fitted the 8.4\,GHz VLA emission from each nucleus to single Gaussian components, finding peak flux densities of 9.1\,mJy/beam and 19.0\,mJy/beam for the northern and southern components, respectively, out of a total emission of 53\,mJy (see Table~\ref{tab:beams}). Given the angular resolutions of the VLA ($0.27\times0.20$\,arcsec$^2$) and the VLBA ($2.8\times1.1$\,mas$^2$) observations, and assuming for simplicity that the total nuclear VLA emission ($28.1$\,mJy) is uniformly distributed, we would need an extreme VLBA sensitivity of $\sim0.32\,\mu$Jy/beam to detect the diffuse emission at $5\sigma$. In summary, the radio AGN contamination is compatible with the available observations.}


\subsubsection{Comparison with AGN indicators}

There are a number of AGN indicators in the mid-IR.
In particular, we focused on the comparison between our SED modeling with three of these indicators: (i) the 6.2\,$\mu$m polycyclic aromatic hydrocarbon (PAH) equivalent width (EW); (ii) the 14.3\,$\mu$m [Ne\,{\sc v}] over 12.8\,$\mu$m [Ne\,{\sc ii}] line ratio; and (iii) the 25.9\,$\mu$m [O\,{\sc iv}] over 12.8\,$\mu$m [Ne\,{\sc ii}] line ratio.

{Several authors have used the PAH features to estimate star formation rates \citep[e.g.][]{farrah07,shipley16}. However, in terms of equivalent width, a lower PAH EW is found in AGN galaxies, either because PAH molecules are destroyed by the AGN radiation field \citep[e.g.][]{voit92,sales10} or due to an increased AGN continuum contribution \citep{alonso-herrero14}}.
Several {PAH line EWs} can be used to estimate the presence and contribution of AGNs. In particular, the feature at 6.2\,$\mu$m has the advantage of not being blended with other PAH features and is less affected by silicate absorption \citep{stierwalt13}.
Pure SB galaxies have 6.2\,$\mu$m PAH EW values between 0.5 and 0.7\,$\mu$m \citep{brandl06}, while smaller values are usually indicative of an excess of hot dust due to the presence of an AGN \citep[e.g.,][]{genzel98, sturm00, wu09}.

High-ionization lines in the mid-IR (e.g., [Ne\,{\sc ii}], [Ne\,{\sc iii}], [Ne\,{\sc v}], [O\,{\sc iv}]) are also often used as tracers of AGN activity \citep[e.g.,][]{genzel98, armus07, pereira-santaella10, dixon11}, although some of these lines can be produced, with lower luminosities, in supernova remnants \citep{oliva99}.

We obtained the 6.2\,$\mu$m PAH EW value for our sample from \citet{stierwalt13}, while mid-IR line fluxes were taken from \citet{inami13}. Table~\ref{tab:agnindicators} shows these mid-IR indicators for our sample, where $1\sigma$ upper limits are quoted for non-detections. 

\begin{table*}
\caption{Mid-IR AGN properties.}
\begin{tabular}{rccccccc}
  	\hline
Name & 6.2\,$\mu$m PAH EW & 12.8\,$\mu$m [Ne\,{\sc ii}] & 14.3\,$\mu$m [Ne\,{\sc v}] & 25.9\,$\mu$m [O\,{\sc iv}] & & $L_\mathrm{AGN}^\mathrm{[Ne\,{\sc v}]}$& $L_\mathrm{AGN}$ \\
 & ($\mu$m) & ($10^{-17}\,\mathrm{W\,m}^{-2}$)  & ($10^{-17}\,\mathrm{W\,m}^{-2}$) & ($10^{-17}\,\mathrm{W\,m}^{-2}$) & & $(10^{10}\,L_\odot)$ & $(10^{10}L_\odot)$\\
\hline
MCG+08-11-002& $0.56\pm0.01$  &  $67.3\pm0.6$  &  $<2.5$  &  $1.7\pm0.4$ &&  $<3.3$                     &  $2.36_{-1.00}^{+1.26}$    \\
NGC\,3690W & $0.12\pm0.02$   & $103.8\pm2.7$ &  $<11.6$   &  $28.9\pm3.8$ &&   $<5.7$   &   --                    \\
NGC\,3690E  & $0.38\pm0.03$   & $237.4\pm2.7$ &  $<14.0$   &  $16.1\pm14.3$  &&  $<6.6$   &   --                    \\
ESO\,440-IG058 & $0.66\pm0.01$   &  $48.9\pm0.4$&  $<2.9$   &  $1.3\pm0.5$ &&  $<6.0$    &  $1.78_{-1.46}^{+1.89}$    \\
IC\,883& $0.62\pm0.01$   & $117.6\pm1.0$ &  $1.6\pm0.1$   &  $6.9\pm1.4$ &&  $9.9$    &  $2.31_{-0.69}^{+1.21}$    \\
CGCG\,049-057 & $0.51\pm0.04$   & $7.3\pm0.2$ &  $<0.4$   &  $<0.6$ &&   $<0.3$   &  $0.31_{-0.24}^{+0.30}$    \\
NGC\,6240 & $0.35\pm0.01$  &   $177.1\pm0.6$  &  $3.4\pm0.2$   &  $22.1\pm2.8$ &&  $21.6$     &  $28.75_{-5.74}^{+11.97}$  \\
IRAS\,16516-0948 & $0.69\pm0.01$   &  $27.4\pm0.3$ &  $<0.8$   &  $1.2\pm0.3$ &&  $<1.8$    &  $1.35_{-1.35}^{+1.12}$    \\
IRAS\,17138-1017&  $0.68\pm0.01$  &  $92.2\pm1.4$ &  $<3.0$   &  $<23.7$ &&  $<3.9$     &  $4.54_{-1.27}^{+2.67}$    \\
IRAS\,17578-0400& $0.68\pm0.01$   & $46.1\pm0.7$ &  $<3.2$   &  $<8.7$ &&  $<2.4$    &  $0.46_{-0.24}^{+0.42}$    \\
IRAS\,18293-3413& $0.63\pm0.01$  &  $303.5\pm2.3$ &  $<8.2$   &  $7.4\pm1.0$ &&  $<10.2 $    &  $0.02_{-0.01}^{+0.11}$    \\
NGC\,6926 &  $0.37\pm0.01$   &  $7.3\pm0.1$ &  $1.3\pm0.1$   &  $4.6\pm0.2$ &&   $5.7$   &  $21.67_{-10.80}^{+6.41}$  \\   
	\hline
\end{tabular}

\medskip

The PAH equivalent widths at 6.2\,$\mu$m were obtained from \citet{stierwalt13}, while the line fluxes come from \citet{inami13}. We quote $3\sigma$ upper limits for non-detections. The AGN bolometric luminosity obtained from the [Ne\,{\sc v}] flux ($L_\mathrm{AGN}^\mathrm{[Ne\,{\sc v}]}$) is derived from equation~\ref{eq:nev}. The last column is our SED-derived AGN luminosity (same as in Table~\ref{tab:andreas}), for comparison purposes. 
\label{tab:agnindicators}
\end{table*}

\begin{figure}
\includegraphics[width=\columnwidth]{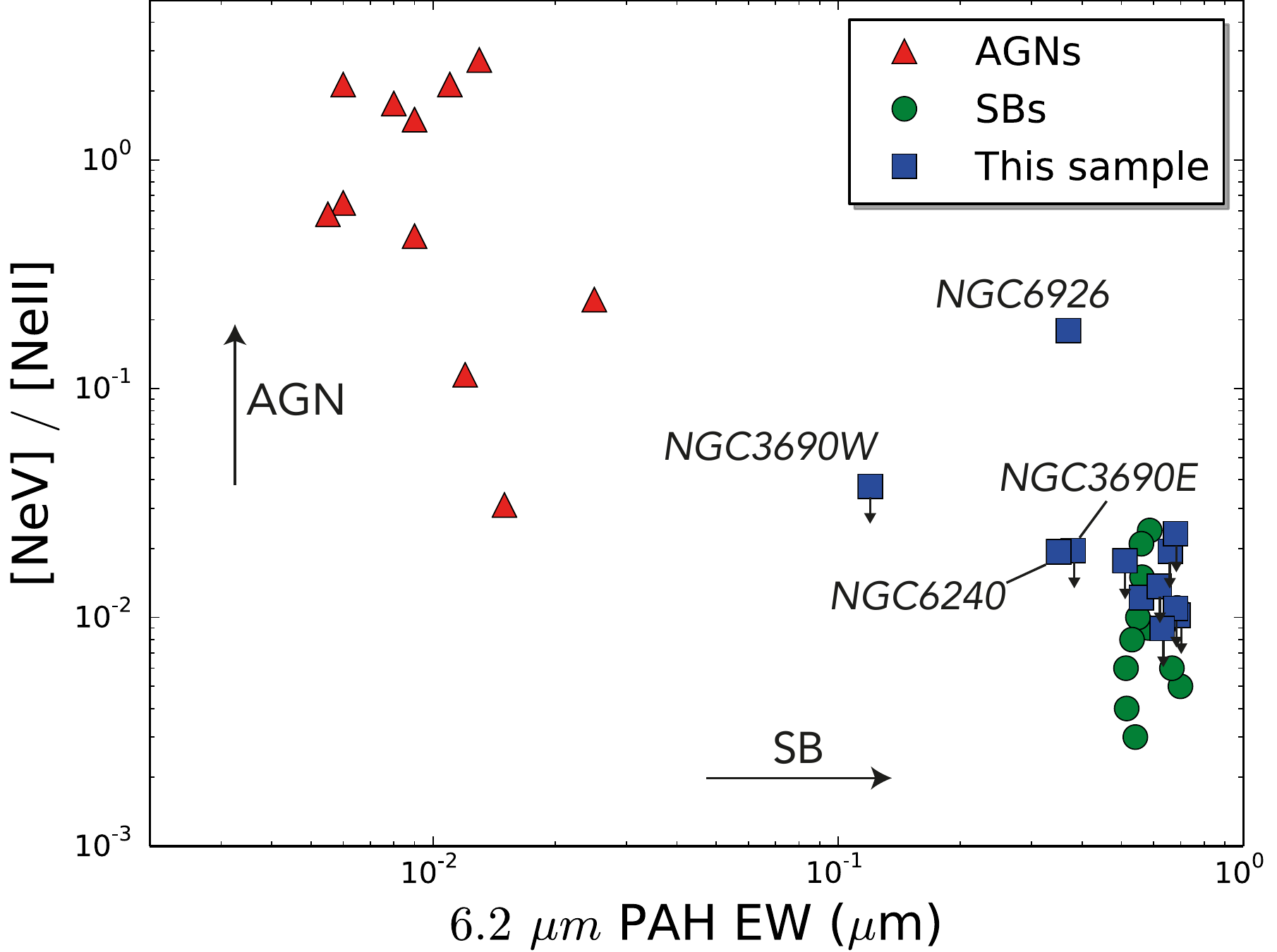}
 \caption{{[}Ne\,{\sc v}{]}/{[}Ne\,{\sc ii}{]} against 6.2\,$\mu$m PAH EW. Our sample is plotted with blue squares. For comparison, we have also plotted a sample of AGN dominated galaxies \citep[red triangles; from][]{weedman05}, and a sample of SB dominated sources \citep[green circles; from][]{brandl06}.}
\label{NeVNeII_PAH.pdf}
\end{figure}

\begin{figure}
\includegraphics[width=\columnwidth]{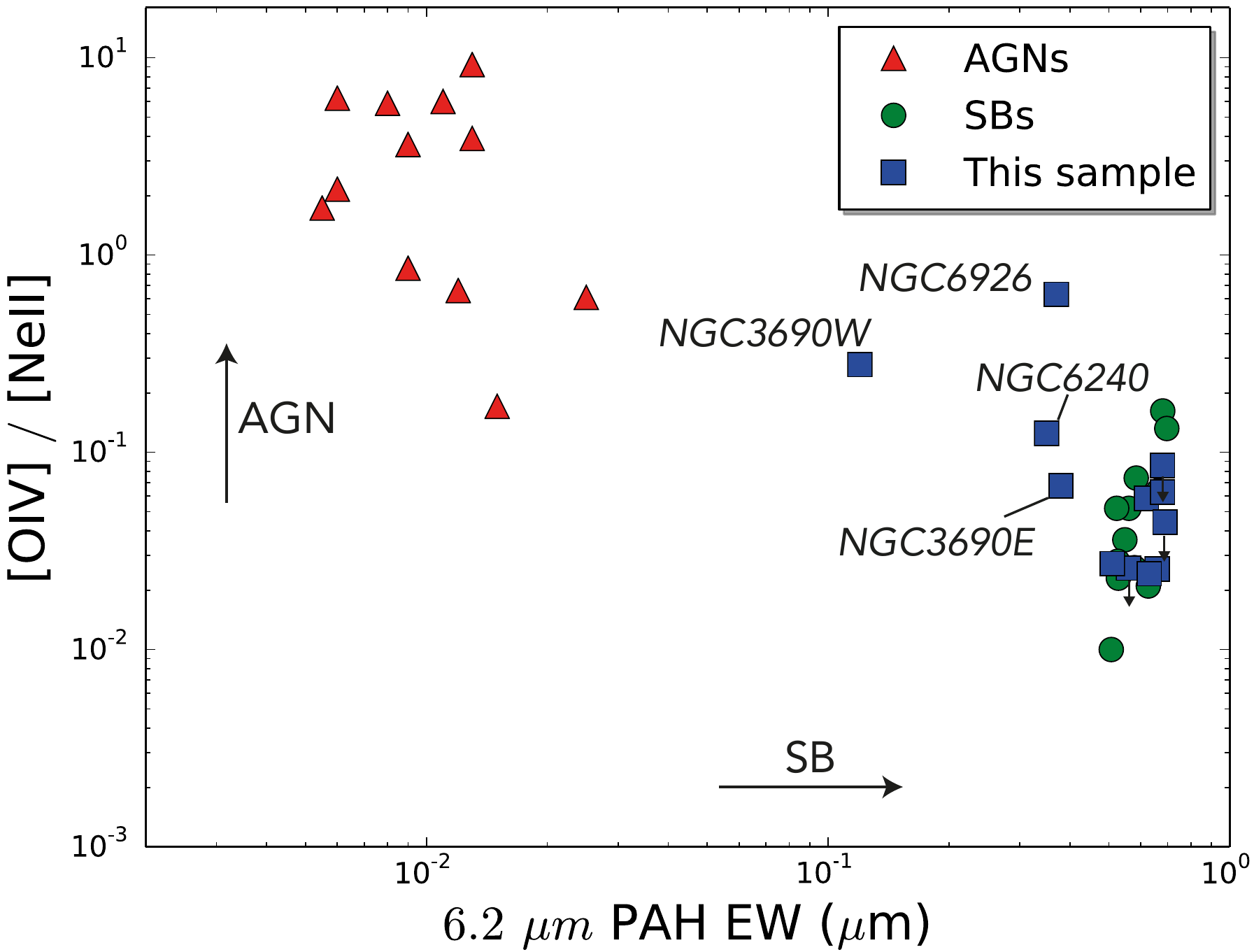}
 \caption{{[}O\,{\sc iv}{]}/{[}Ne\,{\sc ii}{]} against 6.2\,$\mu$m PAH EW. Color codes and samples are the same as in Fig.~\ref{NeVNeII_PAH.pdf}. [O\,{\sc iv}] is detected in a larger number of sources, despite having a lower ionization potential than [Ne\,{\sc v}].}
\label{OIVNeII_PAH.pdf}
\end{figure}

We have plotted these indicators together in Fig.~\ref{NeVNeII_PAH.pdf} and Fig.~\ref{OIVNeII_PAH.pdf}, as done before by other authors \citep{armus07, petric11, vardoulaki14}. In those figures we have plotted the sources from our sample, together with two comparative
 samples: one for AGN dominated systems \citep{weedman05}, and the other formed by SB dominated galaxies \citep{brandl06}.

Two of the galaxies in our sample (NGC\,3690W and NGC\,6926) clearly diverge from the rest of the SB dominated galaxies in the [Ne\,{\sc v}]~/~[Ne\,{\sc ii}] vs PAH EW diagram and two more (NGC\,6240 and NGC\,3690E) show a slight shift (see Fig.~\ref{NeVNeII_PAH.pdf}), which is more clear in the [O\,{\sc iv}]~/~[Ne\,{\sc ii}] vs PAH EW diagram shown in Fig.~\ref{OIVNeII_PAH.pdf}. The results shown by the mid-IR diagnostics are in a relatively good agreement with our model fits. In particular, NGC\,6926 and NGC\,6240 have a very important AGN contribution, although we would also expect IRAS\,17138-1017 to deviate from the SB-dominated loci in both diagnostic plots. With respect to NGC\,3690W, \citet{mattila12} modeled its SED and found an important AGN contribution, compatible with the mid-IR diagnostics and previous observations \citep[e.g.,][]{ballo04}.

A more quantitative comparison can be made by means of the [Ne\,{\sc v}] luminosity, $L_{\mathrm{[Ne\,{\sc v}]}}$. \citet{satyapal07} proposed that $L_{\mathrm{[Ne\,{\sc v}]}}$ is related to the AGN bolometric luminosity through:
\begin{equation}
\log L_\mathrm{AGN}^\mathrm{[Ne\,{\sc v}]} = 0.938 \log L_{\mathrm{[Ne\,{\sc v}]}} +6.317,
\label{eq:nev}
\end{equation}
with luminosities measured in erg\,s$^{-1}$.
Using the [Ne\,{\sc v}] fluxes from \citet{inami13}, we derived the luminosities quoted in Table~\ref{tab:agnindicators}.
{We find that the three sources with a [Ne\,{\sc v}] detection (IC\,883, NGC\,6240, and NGC\,6926) do not follow the above correlation according to our SED-model derived AGN luminosities, albeit NGC\,6240 is compatible within 2$\sigma$. For the remaining sources, there are only upper limits available which, although compatible with the correlation, do not give much information. The relation between [Ne\,{\sc v}] and AGN luminosity arises from the high energetic environment ($\simeq97$\,eV) needed to produce Ne$^{4+}$ \citep{abel08}, typically found around AGN. Nonetheless, these energies can also be reached in young and extreme stellar populations \citep{kewley01}, contaminating this diagnostic tool.}


\subsection{Radio and near-IR comparison}\label{sec:radiovsir}

There is a well-known tight correlation between the total IR luminosity, $L_\mathrm{IR}$, and the radio luminosity at 1.4\,GHz, spanning 5 orders of magnitude, as shown in Figure~\ref{radiovsir.pdf}. In particular, from the sources in our sample, NGC\,6240 is the galaxy with the strongest radio emission and the one that diverges more significantly from the trend. It is also the only source from our sample known to host a dual AGN, in agreement with the overestimated value in the SFR derived from the radio luminosity in section~\ref{sec:sfrindicators}.

The near-IR {band}, although presenting an overall correlation with radio emission (except for some remarkable cases discussed in
this section), traces different processes. Red supergiants, used as indicators of young stellar populations ($\simeq10\,$Myr), radiate most of their energy in the near-IR \citep{oliva95}. Near-IR also includes an important contribution from thermally pulsating asymptotic giant branch (TP-AGB) stars, with young to intermediate populations, up to 2\,Gyr \citep{maraston98}.
 This range of the spectrum can be used to effectively trace star formation in galaxies where the relatively high extinction hinders their study in the optical \citep[e.g.,][]{dametto14}. This makes near-IR a good spectral window in which to study (super) star clusters \citep[e.g.,][]{diaz-santos07, randriamanakoto13a}. On the other hand, the radio emission at 3.6\,cm is dominated by two mechanisms: radio thermal emission (tracing current star formation) and non-thermal synchrotron emission (tracing emission from SNe and SN remnants, i.e., older populations). This explains the general correlation between radio and near-IR, but at the same time, makes the interpretation of the ratio images problematic. In this sense, we find that it is not possible to use the ratios as age maps, unless we are able to, at least, disentangle the thermal and non-thermal radio emission \citep[][]{tabatabaei07, basu13, herrero-illana14}. Furthermore, although near-IR can efficiently penetrate through dust, very dust obscured regions can hinder even near-IR emission, adding an uncertainty to the age interpretation.

Figure~\ref{fig:radioirratios} shows both our radio 3.6\,cm and near-IR $2.2\,\mu$m sub-arcsec resolution images, as well as the flux density ratio between them (reddish for radio dominated regions; blueish for near-IR dominated ones; white implies a flux density ratio of unity). Any pixel that is not above three times the corresponding rms at both 3.6\,cm and 2.2\,$\mu$m is masked out in the ratio maps. For a direct comparison between the radio and the original non-smoothed near-IR images we refer the reader to Figure~\ref{fig:app:radioir}.
We find radio emission to be generally more concentrated in compact nuclear regions. Still, there is not a common trend in all the sources. Indeed, three LIRGs in our sample are completely radio-dominated in their common emission region: IC\,883, CGCG\,049-057, and IRAS\,16516-0948, although near-IR extends further than the radio emission in the latter source.

In the remaining sources, radio and near-IR emission dominate in different regions. We remark, however, the radio silence of bright near-IR knots in three sources: ESO\,440-IG058, IRAS\,18293-3413, and specially IRAS\,16516-0948, which will be discussed in Section~\ref{sec:offnuclear}. The northern source detected in IRAS\,17578-0400 is not in this list, as it is a field star, which does not look like a compact Gaussian emission due to the image convolution with the radio beam. By contrast, IRAS\,18293-3413 has an important compact radio emission $\simeq18^{\prime\prime}$ to the east of the main component that does not show a near-IR counterpart.

A particularly interesting case is NGC\,3690W (the western component of Arp\,299). Its different knots show completely different behaviors regarding their radio and near-IR correlation. While the near-IR dominates the emission of the southern nucleus \citep[designated B in the literature][]{gehrz83}, the northwest component (designated C) has a comparable emission, and the northeast component (C$^\prime$) is very faint at 2.2$\,\mu$m. This near-IR faintness was discussed by \citet{alonso-herrero00}, whose starburst models predict 
a persistence of the emission at 2\,$\mu$m after their assumed Gaussian burst with a FWHM of 5\,Myr. They concluded then that the star formation episode must have been shorter in C$^\prime$. Indeed, \citet{leitherer95} models nicely fit an instantaneous burst, which yields a mass of the SB episode of $3\times10^7M_\odot$, with an age of 4\,Myr, supporting this idea.

{Another interesting case is} NGC\,6240. The nature of the two main compact structures is well determined as a dual AGN \citep{komossa03,gallimore04,risaliti06}. The nature of the fainter northwest radio component, however, is not so clear. This component was first reported by \citet{colbert94} at 3.6\,cm, estimating its $L_\mathrm{IR}$ to be $\simeq2\times10^{10}L_\odot$. They suggested that, considering its steep spectral index ($\alpha=-1.1$), it could be a clump of ejected electrons \citep[a superwind;][]{heckman93}, possibly ejected from a powerful starburst in the nearby compact bright region. \citet{beswick01} found a consistent 1.4\,GHz structure using MERLIN (FWHM: $0.31^{\prime\prime}\times0.15^{\prime\prime}$), but did not detect the source at 5.0\,GHz with the same instrument (FWHM: $0.10^{\prime\prime}\times0.06^{\prime\prime}$), due to the steep spectral index and the limited sensitivity at 5.0\,GHz to detect the diffuse emission. The peak reported by \citet{colbert94} for this relatively faint peak is $1.10\,$mJy/beam, in agreement with our deeper VLA image ($1.06\pm0.23\,$mJy/beam; see Fig.~\ref{fig:radioirratios}).

Finally, we did not find any correlation between {the radio and near-IR total luminosities, nor between} the radio to near-IR ratio maps and the position of the known SSCs in these galaxies \citep{randriamanakoto13a}.

\subsection{An off-nuclear star forming region in IRAS\,16516-0948}\label{sec:offnuclear}

\begin{figure*}
\includegraphics[width=1\textwidth]{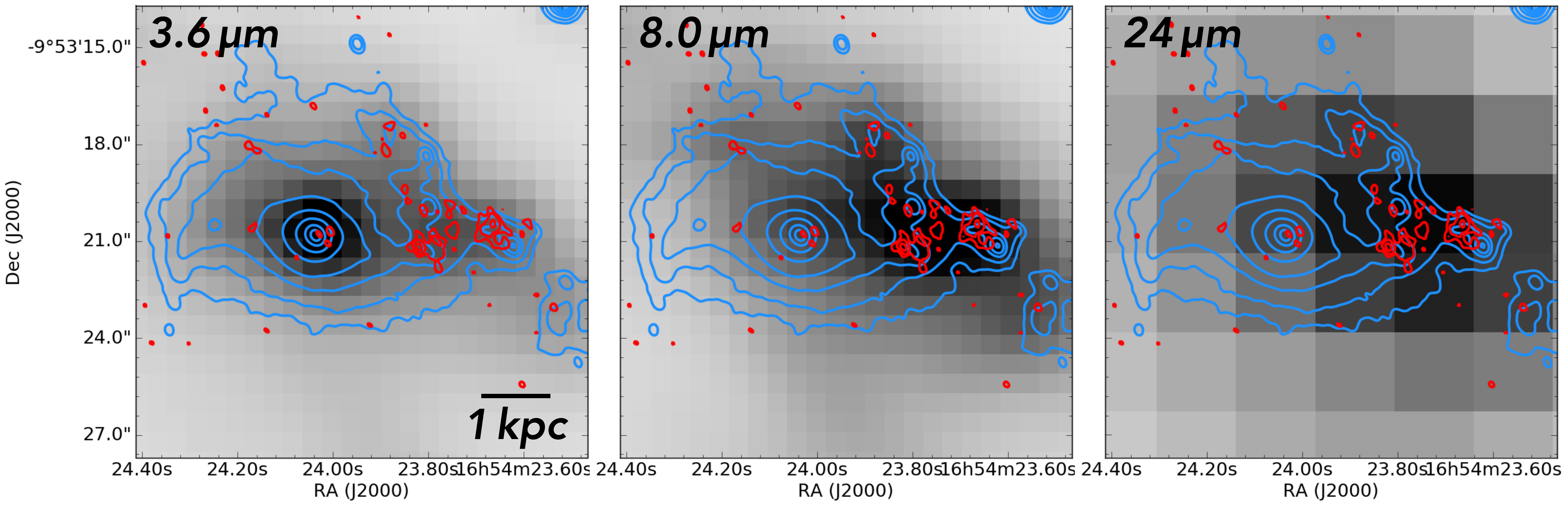}
 \caption{Near-IR and radio contours of IRAS\,16516-0948 overplotted on \emph{Spitzer} images. Left: IRAC $3.6\,\mu$m. Middle: IRAC $8.0\,\mu$m. Right: MIPS $24\,\mu$m. We have overplotted our near-IR K-band (blue) and radio X-band (red) contours. North is up, east is left. Note how the peak IR emission clearly shifts to the west as wavelength increases.}
\label{iras16516_spitzer.pdf}
\end{figure*}

Figure~\ref{fig:radioirratios} shows an unexpected {lack of correlation} between the studied bands for IRAS\,16516-0948. Figure~\ref{iras16516_spitzer.pdf} shows our near-IR and radio data contours overplotted on archival \emph{Spitzer} images of IRAS\,16516-0948. It is clear from the figure that at increasing wavelength the infrared peak is shifting its position towards the west. The surroundings of this galaxy present enough compact sources to ensure unequivocal relative astrometry. The shift in the peak of the 8.0\,$\mu$m with respect to the 3.6\,$\mu$m peak is $\simeq5^{\prime\prime}$. The radio contours correlate well with the peak at $8.0\,\mu$m.

The case of IRAS\,16516-0948 is rare, but not unique among (U)LIRGs: a clear example is the merger II\,Zw\,096 ($\log(L_{\mathrm{IR}}/L_\odot)=11.94$), where $80\%$ of its total IR luminosity comes from a compact optical-invisible source $\simeq5^{\prime\prime}$ away from the optical peak \citep[see Fig.~2 in][]{inami10}, and can have a SFR density up to $780M_\odot$\,yr$^{-1}$\,kpc$^{-2}$. As in IRAS\,16516-0948, radio emission from II\,Zw\,096 correlates well with the MIPS peak, suggesting that, in the region with the strongest mid-IR and radio emission, intense star formation is occurring. Another interesting case is IRAS\,19115-2124, or \emph{The Bird}, the starburst of which is dominated by an off-nucleus minor component of the interaction \citep[][V\"ais\"anen et al., in prep.]{vaisanen08a}. 

\citet{haan11} suggested an explanation for this phenomenon: these galaxies are possibly not yet relaxed with off-nuclear starburst and/or strong shocks, with the off-nuclear emission being associated with spiral arms, a possible secondary nucleus or the region between the merging nuclei.
The sample by \citet{haan11} contains 73\,LIRGs from the GOALS sample, and finds off-nuclear mid-IR emission in six of them ($\simeq9.5\%$). A single case in our sample of 11\,LIRGs agrees with this proportion.

\subsection{Radio spectral behavior} \label{sec:spix}

\begin{figure*}
\includegraphics[width=0.7\textwidth]{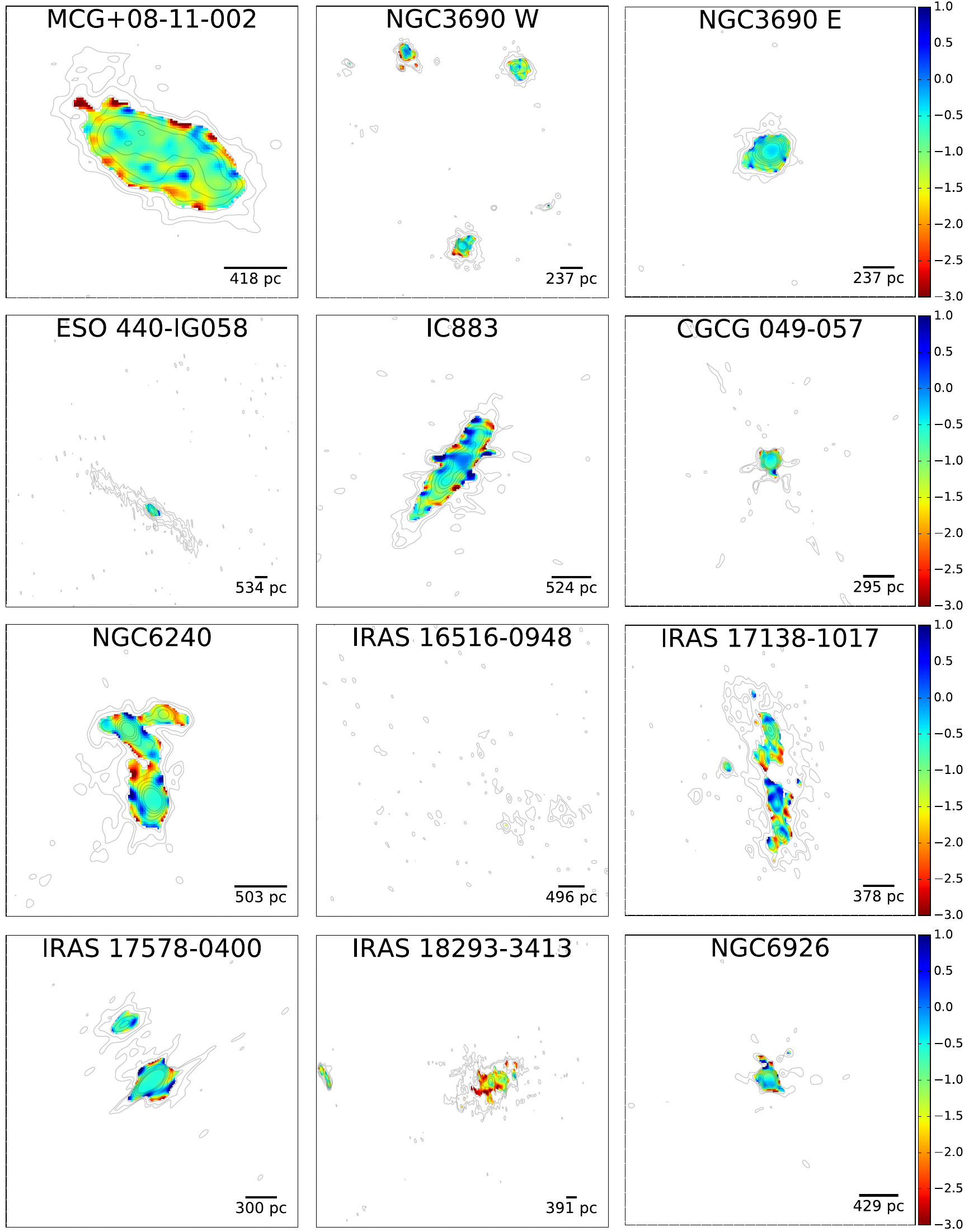}
 \caption{Spectral index maps of our sample. We have masked the maps in any point with a flux density below $10\times\mathrm{rms}$ to ensure the reliability of the spectral indices. Radio contours have been drawn as in Figure~\ref{fig:radioirratios}. All the plots were made with the same color scale. See main text for details on how the maps were obtained.}
\label{spixall2.pdf}
\end{figure*}

Figure~\ref{spixall2.pdf} shows the spectral index maps of the sources. {The most prominent characteristic is the patchy spatial distribution of $\alpha$. This has been reported in the literature \citep[e.g., in NGC\,1596; see][]{lisenfeld04}, and the most likely explanation is that the patches show regions with different ages, therefore the thermal and non-thermal radio emission peaks are not spatially coincident, producing the observed distribution. The regions with the steepest spectrum are dominated by old
radio emitters, while those with flatter spectrum correspond to regions with ongoing activity associated with recently exploded supernovae and/or AGN activity.}

Table~\ref{tab:spix11} quotes the average spectral indices, masked below $10\times\mathrm{rms}$ in the total intensity images to ensure robustness in the spectral index calculation. We note that the signal-to-noise ratio of the radio image of IRAS\,16516-0948 is not good enough to create a reliable spectral index map, since only a few pixels are above the quoted $10\times\mathrm{rms}$ threshold. 

We do not find a correlation between the spectral index maps and the emission ratios shown in Figure~\ref{fig:radioirratios}. The median spectral index of our sample is -0.78, consistent with the values derived for star-forming galaxies at L- ($\simeq18$\,cm) and C-bands($\simeq6$\,cm) \citep[]{leroy11, basu15}, with a single power-law describing the spectral behavior. Thus, we find no significant flattening due to thermal contribution, whose emission has a characteristic spectral index of $\alpha\simeq-0.1$. We note, however, that this invariance in the spectral index does not hold for even lower frequencies, where free-free absorption mechanisms dominate \citep{clemens10}.
This has already been shown for the case of Arp220 with LOFAR observations \citep{varenius16}.

We find that, with the exception of NGC\,3690W and IC\,883, those sources with a steeper spectral index ($\alpha<-0.8$) show extended diffuse emission. \citet{murphy13a} studied a sample of 36 compact starbursts finding the same trend. This can be explained with radio emission arising from radio continuum bridges and tidal tails, result of the interactions, which do not have a direct stellar origin; instead, its emission is due to relativistic electrons cooling down in those regions and producing the steep spectrum.

The steepest spectral index in our sample corresponds to IRAS\,18293-3413, with $\alpha = -1.73$. While this source has the highest SFR from our SED modeling, this is an unexpectedly steep $\alpha$ value for a starburst galaxy. Further observations are needed to obtain its radio spectrum at different bands, allowing us to disentangle the different components of the radio emission and, eventually, understanding the origin of this oddity.

\begin{table}\centering
\caption{Spectral indices.}
\begin{tabular}{rccc}
  	\hline
Name && Average $\alpha$ & Dispersion\\ 
	\hline
MCG+08-11-002&& $-1.09$ & 0.58 \\
NGC\,3690W && $-1.01$ & 0.64 \\
NGC\,3690E  && $-0.65$ & 0.35 \\
ESO\,440-IG058 && $-0.60$ & 0.36 \\
IC\,883&& $-0.68$  & 0.90\\
CGCG\,049-057 & &$-0.78$ & 0.54 \\
NGC\,6240 & &$-0.99$  & 0.76\\
IRAS\,16516-0948 && $\cdots$ & $\cdots$ \\ 
IRAS\,17138-1017&& $-0.85$ & 0.80 \\
IRAS\,17578-0400&& $-0.64$ & 0.70 \\
IRAS\,18293-3413& &$-1.73$  & 0.70\\
NGC\,6926 & & $-0.76$ & 1.00 \\
	\hline
\end{tabular}

\medskip

The quoted dispersions correspond to the standard deviation of the individual (pixel) spectral indices, and are thus a measurement of the uniformity of the spectral index across the region.
\label{tab:spix11}
\end{table}

\section{Summary and conclusions} \label{sec:summary}

We have characterized the star-formation and AGN activity of a sample of 11 local LIRGs imaged 
with subarcsecond angular resolution at radio and near-infrared wavelengths. 
We have performed a complete SED model fit, using a combination of starburst and AGN templates,  aimed at isolating the spheroid, starburst and AGN luminosities as well as to derive the star formation and supernova rates, their age, and their AGN contribution. 
We also compare SFR and AGN luminosities derived using different IR and radio tracers. Finally, we report new radio (3.6\,cm) and near-IR (2.2\,$\mu$m) data of the central kpc region of our LIRG sample.  The main results of our study are the following ones:

\begin{enumerate}

\item 
From our SED modeling, we determine the luminosity contribution of each component, and find that all but one of the sources in our sample are starburst-dominated, with a significant AGN contribution ($\gtrsim20\%$) only in three of them. Only one source (NGC\,6926) is AGN-dominated (64\%). We also derive star formation rates ($40$ to $167\,M_\odot\,\mathrm{yr}^{-1}$), supernova rates (0.4 to $2.0\,\mathrm{SN}\,\mathrm{yr}^{-1}$), and starburst ages (23 to $29\,\mathrm{Myr}$) that are consistent with their LIRG nature.
The only exception is NGC\,6926, the youngest starburst (9\,Myr) of our sample, in which massive stars did not have time to go off supernovae yet.

\item We find an overall consistency among the different (IR and radio) star formation tracers, 
with no star formation tracer under/over estimating systematically the SFR with respect to others. 
The only exception is the overestimated radio-derived SFR in NGC\,6240, likely due to the strong dual AGN influence.

\item We also  find an overall consistency between the common tracers of AGN contribution based on the mid-IR high-ionization line ratios ([Ne\,{\sc v}]~/~[Ne\,{\sc ii}] and [O\,{\sc iv}]~/~[Ne\,{\sc ii}] vs PAH EW) and 
our predicted SED model fitted L$_{\rm AGN}$.
However, the AGN luminosities derived directly from [Ne\,{\sc v}] are not fully compatible with our SED-model derived luminosities, possibly because extreme stellar populations are contributing significantly to the [Ne\,{\sc v}] emission.


\item 
From our sub-arcsecond imaging at radio (8.4\,GHz VLA) and near-IR ($2.2\,\mu$m NACO-VLT and ALTAIR-Gemini) observations, we find that, in general, there is a good spatial correlation of the radio and the near-IR emission, and that both tend to 
be more concentrated in the nuclear regions of  the galaxies.
The lack of a radio counterpart for a few conspicuous near-IR features in some sources points to young starbursts with no supernovae to produce synchrotron radiation.  We note that we clearly identified one off-nuclear starburst in the late merger IRAS\,16516-0948, which may have been likely triggered by the ongoing merging process.

\item  
We {have obtained spectral index maps, showing a patchy spatial distrubtion.} We find an average spectral index of $\alpha \simeq-0.8$ ($S_\nu \propto \nu^\alpha$) from our wide-band VLA observations. This is a typical value for synchrotron-powered LIRGs, and shows that the contribution of thermal emission to the total radio emission at 8.4 GHz is very small. 

\end{enumerate}

\section*{Acknowledgements}

RHI, MAPT, and AA acknowledge support from the Spanish MINECO through grants AYA2012-38491-C02-02 and AYA2015-63939-C2-1-P. PV acknowledges support from the National Research Foundation of South Africa. ZR thanks funding from the South African SKA. {We thank the anonymous referee for his/her useful comments and feedback, which have improved the quality of this work. The authors also thank Cristina Romero-Ca\~nizales and Lorena Hern\'andez-Garc\'ia for the useful discussion.}
The National Radio Astronomy Observatory is a facility of the National Science Foundation operated under cooperative agreement by Associated Universities, Inc. Based in part on observations obtained at the Gemini Observatory (Program IDs: GN-2008A-Q-38, GN-2008B-Q-32, GN-2009A-Q-12, GN-2009B-Q-23, GN-2010A-Q-40, GN-2011A-Q-48, GN-2011B-Q-73, GN-2012A-Q-56, and GS-2015A-Q-7), which is operated by the Association of Universities for Research in Astronomy, Inc., under a cooperative agreement with the NSF on behalf of the Gemini partnership: the National Science Foundation (United States), the National Research Council (Canada), CONICYT (Chile), Ministerio de Ciencia, Tecnolog\'{i}a e Innovaci\'{o}n Productiva (Argentina), and Minist\'{e}rio da Ci\^{e}ncia, Tecnologia e Inova\c{c}\~{a}o (Brazil). And based in part on observations made with the European Southern Observatory Telescopes, at the La Silla Paranal Observatory, Chile, under programs 073.D-0406 (PI: Vaisanen), 087.D-0444 and 089.D-0847 (PI: Mattila). We acknowledge the use of the VizieR photometry tool, developed by Anne-Camille Simon and Thomas Boch.

\bibliographystyle{mn2e}
\small
\bibliography{masterbibdesk}

\begin{thebibliography}{163}
\expandafter\ifx\csname natexlab\endcsname\relax\def\natexlab#1{#1}\fi

\bibitem[{{Abel} \& {Satyapal}(2008)}]{abel08}
{Abel} N.~P., {Satyapal} S., 2008, \apj, 678, 686

\bibitem[{{Ahn} {et~al}\mbox{.}(2012){Ahn}, {Alexandroff}, {Allende Prieto},
  {Anderson}, {Anderton}, {Andrews}, {Aubourg}, {Bailey}, {Balbinot}, {Barnes},
  \& et~al.}]{ahn12}
{Ahn} C.~P. {et~al.}, 2012, \apjs, 203, 21

\bibitem[{{Alonso-Herrero} {et~al}\mbox{.}(2001){Alonso-Herrero},
  {Engelbracht}, {Rieke}, {Rieke}, \& {Quillen}}]{alonso-herrero01}
{Alonso-Herrero} A., {Engelbracht} C.~W., {Rieke} M.~J., {Rieke} G.~H.,
  {Quillen} A.~C., 2001, \apj, 546, 952

\bibitem[{{Alonso-Herrero} {et~al}\mbox{.}(2014){Alonso-Herrero}, {Ramos
  Almeida}, {Esquej}, {Roche}, {Hern{\'a}n-Caballero}, {H{\"o}nig},
  {Gonz{\'a}lez-Mart{\'{\i}}n}, {Aretxaga}, {Mason}, {Packham}, {Levenson},
  {Rodr{\'{\i}}guez Espinosa}, {Siebenmorgen}, {Pereira-Santaella},
  {D{\'{\i}}az-Santos}, {Colina}, {Alvarez}, \& {Telesco}}]{alonso-herrero14}
{Alonso-Herrero} A. {et~al.}, 2014, \mnras, 443, 2766

\bibitem[{{Alonso-Herrero} {et~al}\mbox{.}(2000){Alonso-Herrero}, {Rieke},
  {Rieke}, \& {Scoville}}]{alonso-herrero00}
{Alonso-Herrero} A., {Rieke} G.~H., {Rieke} M.~J., {Scoville} N.~Z., 2000,
  \apj, 532, 845

\bibitem[{{Armus} {et~al}\mbox{.}(2007){Armus}, {Charmandaris},
  {Bernard-Salas}, {Spoon}, {Marshall}, {Higdon}, {Desai}, {Teplitz}, {Hao},
  {Devost}, {Brandl}, {Wu}, {Sloan}, {Soifer}, {Houck}, \& {Herter}}]{armus07}
{Armus} L. {et~al.}, 2007, \apj, 656, 148

\bibitem[{{Armus} {et~al}\mbox{.}(2009){Armus}, {Mazzarella}, {Evans},
  {Surace}, {Sanders}, {Iwasawa}, {Frayer}, {Howell}, {Chan}, {Petric},
  {Vavilkin}, {Kim}, {Haan}, {Inami}, {Murphy}, {Appleton}, {Barnes}, {Bothun},
  {Bridge}, {Charmandaris}, {Jensen}, {Kewley}, {Lord}, {Madore}, {Marshall},
  {Melbourne}, {Rich}, {Satyapal}, {Schulz}, {Spoon}, {Sturm}, {U}, {Veilleux},
  \& {Xu}}]{armus09}
{Armus} L. {et~al.}, 2009, \pasp, 121, 559

\bibitem[{{Asmus} {et~al}\mbox{.}(2014){Asmus}, {H{\"o}nig}, {Gandhi},
  {Smette}, \& {Duschl}}]{asmus14}
{Asmus} D., {H{\"o}nig} S.~F., {Gandhi} P., {Smette} A., {Duschl} W.~J., 2014,
  \mnras, 439, 1648

\bibitem[{{Baan} {et~al}\mbox{.}(1987){Baan}, {Henkel}, \& {Haschick}}]{baan87}
{Baan} W.~A., {Henkel} C., {Haschick} A.~D., 1987, \apj, 320, 154

\bibitem[{{Baan} \& {Kl{\"o}ckner}(2006)}]{baan06}
{Baan} W.~A., {Kl{\"o}ckner} H.-R., 2006, \aap, 449, 559

\bibitem[{{Ballo} {et~al}\mbox{.}(2004){Ballo}, {Braito}, {Della Ceca},
  {Maraschi}, {Tavecchio}, \& {Dadina}}]{ballo04}
{Ballo} L., {Braito} V., {Della Ceca} R., {Maraschi} L., {Tavecchio} F.,
  {Dadina} M., 2004, \apj, 600, 634

\bibitem[{{Bassani} {et~al}\mbox{.}(1999){Bassani}, {Dadina}, {Maiolino},
  {Salvati}, {Risaliti}, {Della Ceca}, {Matt}, \& {Zamorani}}]{bassani99}
{Bassani} L., {Dadina} M., {Maiolino} R., {Salvati} M., {Risaliti} G., {Della
  Ceca} R., {Matt} G., {Zamorani} G., 1999, \apjs, 121, 473

\bibitem[{{Basu} {et~al}\mbox{.}(2015){Basu}, {Beck}, {Schmidt}, \&
  {Roy}}]{basu15}
{Basu} A., {Beck} R., {Schmidt} P., {Roy} S., 2015, \mnras, 449, 3879

\bibitem[{{Basu} \& {Roy}(2013)}]{basu13}
{Basu} A., {Roy} S., 2013, \mnras, 433, 1675

\bibitem[{{Beswick} {et~al}\mbox{.}(2001){Beswick}, {Pedlar}, {Mundell}, \&
  {Gallimore}}]{beswick01}
{Beswick} R.~J., {Pedlar} A., {Mundell} C.~G., {Gallimore} J.~F., 2001, \mnras,
  325, 151

\bibitem[{{Bondi} {et~al}\mbox{.}(2012){Bondi}, {P{\'e}rez-Torres},
  {Herrero-Illana}, \& {Alberdi}}]{bondi12}
{Bondi} M., {P{\'e}rez-Torres} M.~A., {Herrero-Illana} R., {Alberdi} A., 2012,
  \aap, 539, A134

\bibitem[{{Bonzini} {et~al}\mbox{.}(2015){Bonzini}, {Mainieri}, {Padovani},
  {Andreani}, {Berta}, {Bethermin}, {Lutz}, {Rodighiero}, {Rosario}, {Tozzi},
  \& {Vattakunnel}}]{bonzini15}
{Bonzini} M. {et~al.}, 2015, \mnras, 453, 1079

\bibitem[{{Bottinelli} {et~al}\mbox{.}(1986){Bottinelli}, {Gouguenheim}, {Le
  Squeren}, {Martin}, {Dennefeld}, \& {Paturel}}]{bottinelli86}
{Bottinelli} L., {Gouguenheim} L., {Le Squeren} A.~M., {Martin} J.~M.,
  {Dennefeld} M., {Paturel} G., 1986, \iaucirc, 4231, 2

\bibitem[{{Brandl} {et~al}\mbox{.}(2006){Brandl}, {Bernard-Salas}, {Spoon},
  {Devost}, {Sloan}, {Guilles}, {Wu}, {Houck}, {Weedman}, {Armus}, {Appleton},
  {Soifer}, {Charmandaris}, {Hao}, {Higdon}, {Marshall}, \&
  {Herter}}]{brandl06}
{Brandl} B.~R. {et~al.}, 2006, \apj, 653, 1129

\bibitem[{{Bruzual} \& {Charlot}(1993)}]{bruzual93}
{Bruzual} G., {Charlot} S., 1993, \apj, 405, 538

\bibitem[{{Bruzual} \& {Charlot}(2003)}]{bruzual03}
{Bruzual} G., {Charlot} S., 2003, \mnras, 344, 1000

\bibitem[{{Calistro Rivera} {et~al}\mbox{.}(2016){Calistro Rivera}, {Lusso},
  {Hennawi}, \& {Hogg}}]{calistro-rivera16}
{Calistro Rivera} G., {Lusso} E., {Hennawi} J.~F., {Hogg} D.~W., 2016, \apj,
  833, 98

\bibitem[{{Cleary} {et~al}\mbox{.}(2007){Cleary}, {Lawrence}, {Marshall},
  {Hao}, \& {Meier}}]{cleary07}
{Cleary} K., {Lawrence} C.~R., {Marshall} J.~A., {Hao} L., {Meier} D., 2007,
  \apj, 660, 117

\bibitem[{{Clemens} {et~al}\mbox{.}(2010){Clemens}, {Scaife}, {Vega}, \&
  {Bressan}}]{clemens10}
{Clemens} M.~S., {Scaife} A., {Vega} O., {Bressan} A., 2010, \mnras, 405, 887

\bibitem[{{Colbert} {et~al}\mbox{.}(1994){Colbert}, {Wilson}, \&
  {Bland-Hawthorn}}]{colbert94}
{Colbert} E.~J.~M., {Wilson} A.~S., {Bland-Hawthorn} J., 1994, \apj, 436, 89

\bibitem[{{Condon}(1992)}]{condon92}
{Condon} J.~J., 1992, \araa, 30, 575

\bibitem[{{Condon} \& {Broderick}(1991)}]{condon91b}
{Condon} J.~J., {Broderick} J.~J., 1991, \aj, 102, 1663

\bibitem[{{Condon} {et~al}\mbox{.}(1998){Condon}, {Cotton}, {Greisen}, {Yin},
  {Perley}, {Taylor}, \& {Broderick}}]{condon98}
{Condon} J.~J., {Cotton} W.~D., {Greisen} E.~W., {Yin} Q.~F., {Perley} R.~A.,
  {Taylor} G.~B., {Broderick} J.~J., 1998, \aj, 115, 1693

\bibitem[{{Condon} \& {Yin}(1990)}]{condon90}
{Condon} J.~J., {Yin} Q.~F., 1990, \apj, 357, 97

\bibitem[{{Corbett} {et~al}\mbox{.}(2003){Corbett}, {Kewley}, {Appleton},
  {Charmandaris}, {Dopita}, {Heisler}, {Norris}, {Zezas}, \&
  {Marston}}]{corbett03}
{Corbett} E.~A. {et~al.}, 2003, \apj, 583, 670

\bibitem[{{Corbett} {et~al}\mbox{.}(2002){Corbett}, {Norris}, {Heisler},
  {Dopita}, {Appleton}, {Struck}, {Murphy}, {Marston}, {Charmandaris},
  {Kewley}, \& {Zezas}}]{corbett02}
{Corbett} E.~A. {et~al.}, 2002, \apj, 564, 650

\bibitem[{{Corwin}(2004)}]{corwin04}
{Corwin} H.~G., 2004, VizieR Online Data Catalog, 7239, 0

\bibitem[{{Costagliola} {et~al}\mbox{.}(2016){Costagliola}, {Herrero-Illana},
  {Lohfink}, {P{\'e}rez-Torres}, {Aalto}, {Muller}, \&
  {Alberdi}}]{costagliola16}
{Costagliola} F., {Herrero-Illana} R., {Lohfink} A., {P{\'e}rez-Torres} M.,
  {Aalto} S., {Muller} S., {Alberdi} A., 2016, ArXiv e-prints

\bibitem[{{Dametto} {et~al}\mbox{.}(2014){Dametto}, {Riffel}, {Pastoriza},
  {Rodr{\'{\i}}guez-Ardila}, {Hernandez-Jimenez}, \& {Carvalho}}]{dametto14}
{Dametto} N.~Z., {Riffel} R., {Pastoriza} M.~G., {Rodr{\'{\i}}guez-Ardila} A.,
  {Hernandez-Jimenez} J.~A., {Carvalho} E.~A., 2014, \mnras, 443, 1754

\bibitem[{{Davies} {et~al}\mbox{.}(2016){Davies}, {Medling}, {U}, {Max},
  {Sanders}, \& {Kewley}}]{davies16}
{Davies} R.~L., {Medling} A.~M., {U} V., {Max} C.~E., {Sanders} D., {Kewley}
  L.~J., 2016, \mnras, 458, 158

\bibitem[{{de Jong} {et~al}\mbox{.}(1985){de Jong}, {Klein}, {Wielebinski}, \&
  {Wunderlich}}]{dejong85}
{de Jong} T., {Klein} U., {Wielebinski} R., {Wunderlich} E., 1985, \aap, 147,
  L6

\bibitem[{{Del Moro} {et~al}\mbox{.}(2013){Del Moro}, {Alexander}, {Mullaney},
  {Daddi}, {Pannella}, {Bauer}, {Pope}, {Dickinson}, {Elbaz}, {Barthel},
  {Garrett}, {Brandt}, {Charmandaris}, {Chary}, {Dasyra}, {Gilli}, {Hickox},
  {Hwang}, {Ivison}, {Juneau}, {Le Floc'h}, {Luo}, {Morrison}, {Rovilos},
  {Sargent}, \& {Xue}}]{delmoro13}
{Del Moro} A. {et~al.}, 2013, \aap, 549, A59

\bibitem[{{Della Ceca} {et~al}\mbox{.}(2002){Della Ceca}, {Ballo}, {Tavecchio},
  {Maraschi}, {Petrucci}, {Bassani}, {Cappi}, {Dadina}, {Franceschini},
  {Malaguti}, {Palumbo}, \& {Persic}}]{dellaceca02}
{Della Ceca} R. {et~al.}, 2002, \apjl, 581, L9

\bibitem[{{Depoy} {et~al}\mbox{.}(1988){Depoy}, {Wynn-Williams}, {Hill}, \&
  {Becklin}}]{depoy88}
{Depoy} D.~L., {Wynn-Williams} C.~G., {Hill} G.~J., {Becklin} E.~E., 1988, \aj,
  95, 398

\bibitem[{{Diamond-Stanic} {et~al}\mbox{.}(2009){Diamond-Stanic}, {Rieke}, \&
  {Rigby}}]{diamond-stanic09}
{Diamond-Stanic} A.~M., {Rieke} G.~H., {Rigby} J.~R., 2009, \apj, 698, 623

\bibitem[{{D{\'{\i}}az-Santos} {et~al}\mbox{.}(2007){D{\'{\i}}az-Santos},
  {Alonso-Herrero}, {Colina}, {Ryder}, \& {Knapen}}]{diaz-santos07}
{D{\'{\i}}az-Santos} T., {Alonso-Herrero} A., {Colina} L., {Ryder} S.~D.,
  {Knapen} J.~H., 2007, \apj, 661, 149

\bibitem[{{D{\'{\i}}az-Santos} {et~al}\mbox{.}(2011){D{\'{\i}}az-Santos},
  {Charmandaris}, {Armus}, {Stierwalt}, {Haan}, {Mazzarella}, {Howell},
  {Veilleux}, {Murphy}, {Petric}, {Appleton}, {Evans}, {Sanders}, \&
  {Surace}}]{diaz-santos11}
{D{\'{\i}}az-Santos} T. {et~al.}, 2011, \apj, 741, 32

\bibitem[{{Dixon} \& {Joseph}(2011)}]{dixon11}
{Dixon} T.~G., {Joseph} R.~D., 2011, \apj, 740, 99

\bibitem[{{Donley} {et~al}\mbox{.}(2005){Donley}, {Rieke}, {Rigby}, \&
  {P{\'e}rez-Gonz{\'a}lez}}]{donley05}
{Donley} J.~L., {Rieke} G.~H., {Rigby} J.~R., {P{\'e}rez-Gonz{\'a}lez} P.~G.,
  2005, \apj, 634, 169

\bibitem[{{Efstathiou}(2006)}]{efstathiou06}
{Efstathiou} A., 2006, \mnras, 371, L70

\bibitem[{{Efstathiou} {et~al}\mbox{.}(2013){Efstathiou}, {Christopher},
  {Verma}, \& {Siebenmorgen}}]{efstathiou13}
{Efstathiou} A., {Christopher} N., {Verma} A., {Siebenmorgen} R., 2013, \mnras,
  436, 1873

\bibitem[{{Efstathiou} \& {Rowan-Robinson}(1995)}]{efstathiou95}
{Efstathiou} A., {Rowan-Robinson} M., 1995, \mnras, 273, 649

\bibitem[{{Efstathiou} \& {Rowan-Robinson}(2003)}]{efstathiou03}
{Efstathiou} A., {Rowan-Robinson} M., 2003, \mnras, 343, 322

\bibitem[{{Efstathiou} {et~al}\mbox{.}(2000){Efstathiou}, {Rowan-Robinson}, \&
  {Siebenmorgen}}]{efstathiou00}
{Efstathiou} A., {Rowan-Robinson} M., {Siebenmorgen} R., 2000, \mnras, 313, 734

\bibitem[{{Efstathiou} \& {Siebenmorgen}(2009)}]{efstathiou09}
{Efstathiou} A., {Siebenmorgen} R., 2009, \aap, 502, 541

\bibitem[{{Engelbracht} {et~al}\mbox{.}(1996){Engelbracht}, {Rieke}, {Rieke},
  \& {Latter}}]{engelbracht96}
{Engelbracht} C.~W., {Rieke} M.~J., {Rieke} G.~H., {Latter} W.~B., 1996, \apj,
  467, 227

\bibitem[{{Farrah} {et~al}\mbox{.}(2007){Farrah}, {Bernard-Salas}, {Spoon},
  {Soifer}, {Armus}, {Brandl}, {Charmandaris}, {Desai}, {Higdon}, {Devost}, \&
  {Houck}}]{farrah07}
{Farrah} D. {et~al.}, 2007, \apj, 667, 149

\bibitem[{{Gallimore} \& {Beswick}(2004)}]{gallimore04}
{Gallimore} J.~F., {Beswick} R., 2004, \aj, 127, 239

\bibitem[{{Gehrz} {et~al}\mbox{.}(1983){Gehrz}, {Sramek}, \&
  {Weedman}}]{gehrz83}
{Gehrz} R.~D., {Sramek} R.~A., {Weedman} D.~W., 1983, \apj, 267, 551

\bibitem[{{Genzel} {et~al}\mbox{.}(1998){Genzel}, {Lutz}, {Sturm}, {Egami},
  {Kunze}, {Moorwood}, {Rigopoulou}, {Spoon}, {Sternberg}, {Tacconi-Garman},
  {Tacconi}, \& {Thatte}}]{genzel98}
{Genzel} R. {et~al.}, 1998, \apj, 498, 579

\bibitem[{{Greenhill} {et~al}\mbox{.}(2003){Greenhill}, {Kondratko}, {Lovell},
  {Kuiper}, {Moran}, {Jauncey}, \& {Baines}}]{greenhill03}
{Greenhill} L.~J., {Kondratko} P.~T., {Lovell} J.~E.~J., {Kuiper} T.~B.~H.,
  {Moran} J.~M., {Jauncey} D.~L., {Baines} G.~P., 2003, \apjl, 582, L11

\bibitem[{{Haan} {et~al}\mbox{.}(2011){Haan}, {Surace}, {Armus}, {Evans},
  {Howell}, {Mazzarella}, {Kim}, {Vavilkin}, {Inami}, {Sanders}, {Petric},
  {Bridge}, {Melbourne}, {Charmandaris}, {Diaz-Santos}, {Murphy}, {U},
  {Stierwalt}, \& {Marshall}}]{haan11}
{Haan} S. {et~al.}, 2011, \aj, 141, 100

\bibitem[{{Heckman} {et~al}\mbox{.}(1993){Heckman}, {Lehnert}, \&
  {Armus}}]{heckman93}
{Heckman} T.~M., {Lehnert} M.~D., {Armus} L., 1993, in Astrophysics and Space
  Science Library, Vol. 188, The Environment and Evolution of Galaxies, {Shull}
  J.~M., {Thronson} H.~A., eds., p. 455

\bibitem[{{Heckman} {et~al}\mbox{.}(2005){Heckman}, {Ptak}, {Hornschemeier}, \&
  {Kauffmann}}]{heckman05}
{Heckman} T.~M., {Ptak} A., {Hornschemeier} A., {Kauffmann} G., 2005, \apj,
  634, 161

\bibitem[{{Helou} {et~al}\mbox{.}(1985){Helou}, {Soifer}, \&
  {Rowan-Robinson}}]{helou85}
{Helou} G., {Soifer} B.~T., {Rowan-Robinson} M., 1985, \apjl, 298, L7

\bibitem[{{Herrero-Illana} {et~al}\mbox{.}(2012{\natexlab{a}}){Herrero-Illana},
  {P{\'e}rez-Torres}, \& {Alberdi}}]{herrero-illana12a}
{Herrero-Illana} R., {P{\'e}rez-Torres} M.~{\'A}., {Alberdi} A.,
  2012{\natexlab{a}}, \aap, 540, L5

\bibitem[{{Herrero-Illana} {et~al}\mbox{.}(2014){Herrero-Illana},
  {P{\'e}rez-Torres}, {Alonso-Herrero}, {Alberdi}, {Colina}, {Efstathiou},
  {Hern{\'a}ndez-Garc{\'{\i}}a}, {Miralles-Caballero}, {V{\"a}is{\"a}nen},
  {Packham}, {Rajpaul}, \& {Zijlstra}}]{herrero-illana14}
{Herrero-Illana} R. {et~al.}, 2014, \apj, 786, 156

\bibitem[{{Herrero-Illana} {et~al}\mbox{.}(2012{\natexlab{b}}){Herrero-Illana},
  {Romero-Canizales}, {Perez-Torres}, {Alberdi}, {Kankare}, {Mattila}, \&
  {Ryder}}]{herrero-illana12b}
{Herrero-Illana} R., {Romero-Canizales} C., {Perez-Torres} M.~A., {Alberdi} A.,
  {Kankare} E., {Mattila} S., {Ryder} S.~D., 2012{\natexlab{b}}, The
  Astronomer's Telegram, 4432, 1

\bibitem[{{Inami} {et~al}\mbox{.}(2013){Inami}, {Armus}, {Charmandaris},
  {Groves}, {Kewley}, {Petric}, {Stierwalt}, {D{\'{\i}}az-Santos}, {Surace},
  {Rich}, {Haan}, {Howell}, {Evans}, {Mazzarella}, {Marshall}, {Appleton},
  {Lord}, {Spoon}, {Frayer}, {Matsuhara}, \& {Veilleux}}]{inami13}
{Inami} H. {et~al.}, 2013, \apj, 777, 156

\bibitem[{{Inami} {et~al}\mbox{.}(2010){Inami}, {Armus}, {Surace},
  {Mazzarella}, {Evans}, {Sanders}, {Howell}, {Petric}, {Vavilkin}, {Iwasawa},
  {Haan}, {Murphy}, {Stierwalt}, {Appleton}, {Barnes}, {Bothun}, {Bridge},
  {Chan}, {Charmandaris}, {Frayer}, {Kewley}, {Kim}, {Lord}, {Madore},
  {Marshall}, {Matsuhara}, {Melbourne}, {Rich}, {Schulz}, {Spoon}, {Sturm},
  {U}, {Veilleux}, \& {Xu}}]{inami10}
{Inami} H. {et~al.}, 2010, \aj, 140, 63

\bibitem[{{Ivison} {et~al}\mbox{.}(2010){Ivison}, {Magnelli}, {Ibar},
  {Andreani}, {Elbaz}, {Altieri}, {Amblard}, {Arumugam}, {Auld}, {Aussel},
  {Babbedge}, {Berta}, {Blain}, {Bock}, {Bongiovanni}, {Boselli}, {Buat},
  {Burgarella}, {Castro-Rodr{\'{\i}}guez}, {Cava}, {Cepa}, {Chanial},
  {Cimatti}, {Cirasuolo}, {Clements}, {Conley}, {Conversi}, {Cooray}, {Daddi},
  {Dominguez}, {Dowell}, {Dwek}, {Eales}, {Farrah}, {F{\"o}rster Schreiber},
  {Fox}, {Franceschini}, {Gear}, {Genzel}, {Glenn}, {Griffin}, {Gruppioni},
  {Halpern}, {Hatziminaoglou}, {Isaak}, {Lagache}, {Levenson}, {Lu}, {Lutz},
  {Madden}, {Maffei}, {Magdis}, {Mainetti}, {Maiolino}, {Marchetti},
  {Morrison}, {Mortier}, {Nguyen}, {Nordon}, {O'Halloran}, {Oliver}, {Omont},
  {Owen}, {Page}, {Panuzzo}, {Papageorgiou}, {Pearson}, {P{\'e}rez-Fournon},
  {P{\'e}rez Garc{\'{\i}}a}, {Poglitsch}, {Pohlen}, {Popesso}, {Pozzi},
  {Rawlings}, {Raymond}, {Rigopoulou}, {Riguccini}, {Rizzo}, {Rodighiero},
  {Roseboom}, {Rowan-Robinson}, {Saintonge}, {Sanchez Portal}, {Santini},
  {Schulz}, {Scott}, {Seymour}, {Shao}, {Shupe}, {Smith}, {Stevens}, {Sturm},
  {Symeonidis}, {Tacconi}, {Trichas}, {Tugwell}, {Vaccari}, {Valtchanov},
  {Vieira}, {Vigroux}, {Wang}, {Ward}, {Wright}, {Xu}, \& {Zemcov}}]{ivison10}
{Ivison} R.~J. {et~al.}, 2010, \aap, 518, L31

\bibitem[{{Johnson} {et~al}\mbox{.}(2013){Johnson}, {Wilson}, {Tang}, \&
  {Scott}}]{johnson13}
{Johnson} S.~P., {Wilson} G.~W., {Tang} Y., {Scott} K.~S., 2013, \mnras, 436,
  2535

\bibitem[{{Kankare} {et~al}\mbox{.}(2008{\natexlab{a}}){Kankare}, {Mattila},
  {Ryder}, {Alonso-Herrero}, {Diaz Santos}, {Colina}, {Kotilainen}, {Lehto},
  {Perez-Torres}, {Romero-Canizales}, {Alberdi}, {Vaisanen}, \&
  {Efstathiou}}]{kankare08a}
{Kankare} E. {et~al.}, 2008{\natexlab{a}}, Central Bureau Electronic Telegrams,
  1569

\bibitem[{{Kankare} {et~al}\mbox{.}(2014){Kankare}, {Mattila}, {Ryder},
  {Fraser}, {Pastorello}, {Elias-Rosa}, {Romero-Ca{\~n}izales}, {Alberdi},
  {Hentunen}, {Herrero-Illana}, {Kotilainen}, {P{\'e}rez-Torres}, \&
  {V{\"a}is{\"a}nen}}]{kankare14}
{Kankare} E. {et~al.}, 2014, \mnras, 440, 1052

\bibitem[{{Kankare} {et~al}\mbox{.}(2008{\natexlab{b}}){Kankare}, {Mattila},
  {Ryder}, {P{\'e}rez-Torres}, {Alberdi}, {Romero-Canizales},
  {D{\'{\i}}az-Santos}, {V{\"a}is{\"a}nen}, {Efstathiou}, {Alonso-Herrero},
  {Colina}, \& {Kotilainen}}]{kankare08b}
{Kankare} E. {et~al.}, 2008{\natexlab{b}}, \apjl, 689, L97

\bibitem[{{Kankare} {et~al}\mbox{.}(2012){Kankare}, {Mattila}, {Ryder},
  {V{\"a}is{\"a}nen}, {Alberdi}, {Alonso-Herrero}, {Colina}, {Efstathiou},
  {Kotilainen}, {Melinder}, {P{\'e}rez-Torres}, {Romero-Ca{\~n}izales}, \&
  {Takalo}}]{kankare12}
{Kankare} E. {et~al.}, 2012, \apjl, 744, L19

\bibitem[{{Keel} \& {Wu}(1995)}]{keel95}
{Keel} W.~C., {Wu} W., 1995, \aj, 110, 129

\bibitem[{{Kennicutt}(1998)}]{kennicutt98}
{Kennicutt}, Jr. R.~C., 1998, \araa, 36, 189

\bibitem[{{Kewley} {et~al}\mbox{.}(2001){Kewley}, {Dopita}, {Sutherland},
  {Heisler}, \& {Trevena}}]{kewley01}
{Kewley} L.~J., {Dopita} M.~A., {Sutherland} R.~S., {Heisler} C.~A., {Trevena}
  J., 2001, \apj, 556, 121

\bibitem[{{Komossa} {et~al}\mbox{.}(2003){Komossa}, {Burwitz}, {Hasinger},
  {Predehl}, {Kaastra}, \& {Ikebe}}]{komossa03}
{Komossa} S., {Burwitz} V., {Hasinger} G., {Predehl} P., {Kaastra} J.~S.,
  {Ikebe} Y., 2003, \apjl, 582, L15

\bibitem[{{Koss} {et~al}\mbox{.}(2013){Koss}, {Mushotzky}, {Baumgartner},
  {Veilleux}, {Tueller}, {Markwardt}, \& {Casey}}]{koss13}
{Koss} M., {Mushotzky} R., {Baumgartner} W., {Veilleux} S., {Tueller} J.,
  {Markwardt} C., {Casey} C.~M., 2013, \apjl, 765, L26

\bibitem[{{Lacki} \& {Thompson}(2010)}]{lacki10b}
{Lacki} B.~C., {Thompson} T.~A., 2010, \apj, 717, 196

\bibitem[{{Lacki} {et~al}\mbox{.}(2010){Lacki}, {Thompson}, \&
  {Quataert}}]{lacki10a}
{Lacki} B.~C., {Thompson} T.~A., {Quataert} E., 2010, \apj, 717, 1

\bibitem[{{Larson} {et~al}\mbox{.}(2016){Larson}, {Sanders}, {Barnes},
  {Ishida}, {Evans}, {U}, {Mazzarella}, {Kim}, {Privon}, {Mirabel}, \&
  {Flewelling}}]{larson16}
{Larson} K.~L. {et~al.}, 2016, \apj, 825, 128

\bibitem[{{Lehmer} {et~al}\mbox{.}(2010){Lehmer}, {Alexander}, {Bauer},
  {Brandt}, {Goulding}, {Jenkins}, {Ptak}, \& {Roberts}}]{lehmer10}
{Lehmer} B.~D., {Alexander} D.~M., {Bauer} F.~E., {Brandt} W.~N., {Goulding}
  A.~D., {Jenkins} L.~P., {Ptak} A., {Roberts} T.~P., 2010, \apj, 724, 559

\bibitem[{{Leitherer} \& {Heckman}(1995)}]{leitherer95}
{Leitherer} C., {Heckman} T.~M., 1995, \apjs, 96, 9

\bibitem[{{Leroy} {et~al}\mbox{.}(2011){Leroy}, {Evans}, {Momjian}, {Murphy},
  {Ott}, {Armus}, {Condon}, {Haan}, {Mazzarella}, {Meier}, {Privon},
  {Schinnerer}, {Surace}, \& {Walter}}]{leroy11}
{Leroy} A.~K. {et~al.}, 2011, \apjl, 739, L25

\bibitem[{{Li}(2002)}]{li02}
{Li} W.~D., 2002, \iaucirc, 7864, 2

\bibitem[{{Lisenfeld} {et~al}\mbox{.}(1996){Lisenfeld}, {Voelk}, \&
  {Xu}}]{lisenfeld96}
{Lisenfeld} U., {Voelk} H.~J., {Xu} C., 1996, \aap, 314, 745

\bibitem[{{Lisenfeld} {et~al}\mbox{.}(2004){Lisenfeld}, {Wilding}, {Pooley}, \&
  {Alexander}}]{lisenfeld04}
{Lisenfeld} U., {Wilding} T.~W., {Pooley} G.~G., {Alexander} P., 2004, \mnras,
  349, 1335

\bibitem[{{Lu} {et~al}\mbox{.}(2014){Lu}, {Zhao}, {Xu}, {Gao}, {Armus},
  {Mazzarella}, {Isaak}, {Petric}, {Charmandaris}, {D{\'{\i}}az-Santos},
  {Evans}, {Howell}, {Appleton}, {Inami}, {Iwasawa}, {Leech}, {Lord},
  {Sanders}, {Schulz}, {Surace}, \& {van der Werf}}]{lu14}
{Lu} N. {et~al.}, 2014, \apjl, 787, L23

\bibitem[{{Maraston}(1998)}]{maraston98}
{Maraston} C., 1998, \mnras, 300, 872

\bibitem[{{Masini} {et~al}\mbox{.}(2016){Masini}, {Comastri}, {Balokovi{\'c}},
  {Zaw}, {Puccetti}, {Ballantyne}, {Bauer}, {Boggs}, {Brandt}, {Brightman},
  {Christensen}, {Craig}, {Gandhi}, {Hailey}, {Harrison}, {Koss}, {Madejski},
  {Ricci}, {Rivers}, {Stern}, \& {Zhang}}]{masini16}
{Masini} A. {et~al.}, 2016, \aap, 589, A59

\bibitem[{{Matheson} {et~al}\mbox{.}(2002){Matheson}, {Jha}, {Challis},
  {Kirshner}, \& {Calkins}}]{matheson02}
{Matheson} T., {Jha} S., {Challis} P., {Kirshner} R., {Calkins} M., 2002,
  \iaucirc, 7872, 2

\bibitem[{{Mattila} {et~al}\mbox{.}(2012){Mattila}, {Dahlen}, {Efstathiou},
  {Kankare}, {Melinder}, {Alonso-Herrero}, {P{\'e}rez-Torres}, {Ryder},
  {V{\"a}is{\"a}nen}, \& {{\"O}stlin}}]{mattila12}
{Mattila} S. {et~al.}, 2012, \apj, 756, 111

\bibitem[{{Mattila} {et~al}\mbox{.}(2010){Mattila}, {Kankare}, {Datson}, \&
  {Pastorello}}]{mattila10b}
{Mattila} S., {Kankare} E., {Datson} J., {Pastorello} A., 2010, Central Bureau
  Electronic Telegrams, 2149, 1

\bibitem[{{Mattila} \& {Meikle}(2001)}]{mattila01}
{Mattila} S., {Meikle} W.~P.~S., 2001, \mnras, 324, 325

\bibitem[{{Mattila} {et~al}\mbox{.}(2007{\natexlab{a}}){Mattila},
  {V{\"a}is{\"a}nen}, {Farrah}, {Efstathiou}, {Meikle}, {Dahlen}, {Fransson},
  {Lira}, {Lundqvist}, {{\"O}stlin}, {Ryder}, \& {Sollerman}}]{mattila07b}
{Mattila} S. {et~al.}, 2007{\natexlab{a}}, \apjl, 659, L9

\bibitem[{{Mattila} {et~al}\mbox{.}(2007{\natexlab{b}}){Mattila}, {Vaisanen},
  {Meikle}, {Dahlen}, {Efstathiou}, {Farrah}, {Fransson}, {Lira}, {Lundqvist},
  {Ostlin}, {Ryder}, \& {Sollerman}}]{mattila07a}
{Mattila} S. {et~al.}, 2007{\natexlab{b}}, Central Bureau Electronic Telegrams,
  858, 1

\bibitem[{{Mauch} \& {Sadler}(2007)}]{mauch07}
{Mauch} T., {Sadler} E.~M., 2007, \mnras, 375, 931

\bibitem[{{McMullin} {et~al}\mbox{.}(2007){McMullin}, {Waters}, {Schiebel},
  {Young}, \& {Golap}}]{mcmullin07}
{McMullin} J.~P., {Waters} B., {Schiebel} D., {Young} W., {Golap} K., 2007, in
  Astronomical Society of the Pacific Conference Series, Vol. 376, Astronomical
  Data Analysis Software and Systems XVI, {Shaw} R.~A., {Hill} F., {Bell}
  D.~J., eds., p. 127

\bibitem[{{Miluzio} {et~al}\mbox{.}(2013){Miluzio}, {Cappellaro}, {Botticella},
  {Cresci}, {Greggio}, {Mannucci}, {Benetti}, {Bufano}, {Elias-Rosa},
  {Pastorello}, {Turatto}, \& {Zampieri}}]{miluzio13}
{Miluzio} M. {et~al.}, 2013, \aap, 554, A127

\bibitem[{{Modica} {et~al}\mbox{.}(2012){Modica}, {Vavilkin}, {Evans}, {Kim},
  {Mazzarella}, {Iwasawa}, {Petric}, {Howell}, {Surace}, {Armus}, {Spoon},
  {Sanders}, {Wong}, \& {Barnes}}]{modica12}
{Modica} F. {et~al.}, 2012, \aj, 143, 16

\bibitem[{{Monreal-Ibero} {et~al}\mbox{.}(2010){Monreal-Ibero}, {Arribas},
  {Colina}, {Rodr{\'{\i}}guez-Zaur{\'{\i}}n}, {Alonso-Herrero}, \&
  {Garc{\'{\i}}a-Mar{\'{\i}}n}}]{monreal-ibero10}
{Monreal-Ibero} A., {Arribas} S., {Colina} L., {Rodr{\'{\i}}guez-Zaur{\'{\i}}n}
  J., {Alonso-Herrero} A., {Garc{\'{\i}}a-Mar{\'{\i}}n} M., 2010, \aap, 517,
  A28

\bibitem[{{Mori{\'c}} {et~al}\mbox{.}(2010){Mori{\'c}}, {Smol{\v c}i{\'c}},
  {Kimball}, {Riechers}, {Ivezi{\'c}}, \& {Scoville}}]{moric10}
{Mori{\'c}} I., {Smol{\v c}i{\'c}} V., {Kimball} A., {Riechers} D.~A.,
  {Ivezi{\'c}} {\v Z}., {Scoville} N., 2010, \apj, 724, 779

\bibitem[{{Mullaney} {et~al}\mbox{.}(2011){Mullaney}, {Alexander}, {Goulding},
  \& {Hickox}}]{mullaney11}
{Mullaney} J.~R., {Alexander} D.~M., {Goulding} A.~D., {Hickox} R.~C., 2011,
  \mnras, 414, 1082

\bibitem[{{Murphy} {et~al}\mbox{.}(2011){Murphy}, {Condon}, {Schinnerer},
  {Kennicutt}, {Calzetti}, {Armus}, {Helou}, {Turner}, {Aniano}, {Beir{\~a}o},
  {Bolatto}, {Brandl}, {Croxall}, {Dale}, {Donovan Meyer}, {Draine},
  {Engelbracht}, {Hunt}, {Hao}, {Koda}, {Roussel}, {Skibba}, \&
  {Smith}}]{murphy11}
{Murphy} E.~J. {et~al.}, 2011, \apj, 737, 67

\bibitem[{{Murphy} {et~al}\mbox{.}(2013){Murphy}, {Stierwalt}, {Armus},
  {Condon}, \& {Evans}}]{murphy13a}
{Murphy} E.~J., {Stierwalt} S., {Armus} L., {Condon} J.~J., {Evans} A.~S.,
  2013, \apj, 768, 2

\bibitem[{{Nakai} {et~al}\mbox{.}(2002){Nakai}, {Sato}, \&
  {Yamauchi}}]{nakai02}
{Nakai} N., {Sato} N., {Yamauchi} A., 2002, \pasj, 54, L27

\bibitem[{{Neff} {et~al}\mbox{.}(2004){Neff}, {Ulvestad}, \& {Teng}}]{neff04}
{Neff} S.~G., {Ulvestad} J.~S., {Teng} S.~H., 2004, \apj, 611, 186

\bibitem[{{Netzer} {et~al}\mbox{.}(2007){Netzer}, {Lutz}, {Schweitzer},
  {Contursi}, {Sturm}, {Tacconi}, {Veilleux}, {Kim}, {Rupke}, {Baker},
  {Dasyra}, {Mazzarella}, \& {Lord}}]{netzer07}
{Netzer} H. {et~al.}, 2007, \apj, 666, 806

\bibitem[{{Oliva} {et~al}\mbox{.}(1999){Oliva}, {Moorwood}, {Drapatz}, {Lutz},
  \& {Sturm}}]{oliva99}
{Oliva} E., {Moorwood} A.~F.~M., {Drapatz} S., {Lutz} D., {Sturm} E., 1999,
  \aap, 343, 943

\bibitem[{{Oliva} {et~al}\mbox{.}(1995){Oliva}, {Origlia}, {Kotilainen}, \&
  {Moorwood}}]{oliva95}
{Oliva} E., {Origlia} L., {Kotilainen} J.~K., {Moorwood} A.~F.~M., 1995, \aap,
  301, 55

\bibitem[{{Padovani} {et~al}\mbox{.}(2011){Padovani}, {Miller}, {Kellermann},
  {Mainieri}, {Rosati}, \& {Tozzi}}]{padovani11}
{Padovani} P., {Miller} N., {Kellermann} K.~I., {Mainieri} V., {Rosati} P.,
  {Tozzi} P., 2011, \apj, 740, 20

\bibitem[{{Pannella} {et~al}\mbox{.}(2015){Pannella}, {Elbaz}, {Daddi},
  {Dickinson}, {Hwang}, {Schreiber}, {Strazzullo}, {Aussel}, {Bethermin},
  {Buat}, {Charmandaris}, {Cibinel}, {Juneau}, {Ivison}, {Le Borgne}, {Le
  Floc'h}, {Leiton}, {Lin}, {Magdis}, {Morrison}, {Mullaney}, {Onodera},
  {Renzini}, {Salim}, {Sargent}, {Scott}, {Shu}, \& {Wang}}]{pannella15}
{Pannella} M. {et~al.}, 2015, \apj, 807, 141

\bibitem[{{Parra} {et~al}\mbox{.}(2007){Parra}, {Conway}, {Diamond}, {Thrall},
  {Lonsdale}, {Lonsdale}, \& {Smith}}]{parra07}
{Parra} R., {Conway} J.~E., {Diamond} P.~J., {Thrall} H., {Lonsdale} C.~J.,
  {Lonsdale} C.~J., {Smith} H.~E., 2007, \apj, 659, 314

\bibitem[{{Pereira-Santaella} {et~al}\mbox{.}(2010){Pereira-Santaella},
  {Diamond-Stanic}, {Alonso-Herrero}, \& {Rieke}}]{pereira-santaella10}
{Pereira-Santaella} M., {Diamond-Stanic} A.~M., {Alonso-Herrero} A., {Rieke}
  G.~H., 2010, \apj, 725, 2270

\bibitem[{{P{\'e}rez-Torres} {et~al}\mbox{.}(2010){P{\'e}rez-Torres},
  {Alberdi}, {Romero-Ca{\~n}izales}, \& {Bondi}}]{perez-torres10}
{P{\'e}rez-Torres} M.~A., {Alberdi} A., {Romero-Ca{\~n}izales} C., {Bondi} M.,
  2010, \aap, 519, L5+

\bibitem[{{P{\'e}rez-Torres} {et~al}\mbox{.}(2007){P{\'e}rez-Torres},
  {Mattila}, {Alberdi}, {Colina}, {Torrelles}, {V{\"a}is{\"a}nen}, {Ryder},
  {Panagia}, \& {Wilson}}]{perez-torres07}
{P{\'e}rez-Torres} M.~A. {et~al.}, 2007, \apjl, 671, L21

\bibitem[{{P{\'e}rez-Torres} {et~al}\mbox{.}(2009){P{\'e}rez-Torres},
  {Romero-Ca{\~n}izales}, {Alberdi}, \& {Polatidis}}]{perez-torres09b}
{P{\'e}rez-Torres} M.~A., {Romero-Ca{\~n}izales} C., {Alberdi} A., {Polatidis}
  A., 2009, \aap, 507, L17

\bibitem[{{Petric} {et~al}\mbox{.}(2011){Petric}, {Armus}, {Howell}, {Chan},
  {Mazzarella}, {Evans}, {Surace}, {Sanders}, {Appleton}, {Charmandaris},
  {D{\'{\i}}az-Santos}, {Frayer}, {Haan}, {Inami}, {Iwasawa}, {Kim}, {Madore},
  {Marshall}, {Spoon}, {Stierwalt}, {Sturm}, {U}, {Vavilkin}, \&
  {Veilleux}}]{petric11}
{Petric} A.~O. {et~al.}, 2011, \apj, 730, 28

\bibitem[{{Portegies Zwart} {et~al}\mbox{.}(2010){Portegies Zwart}, {McMillan},
  \& {Gieles}}]{portegies-zwart10}
{Portegies Zwart} S.~F., {McMillan} S.~L.~W., {Gieles} M., 2010, \araa, 48, 431

\bibitem[{{Randriamanakoto}
  {et~al}\mbox{.}(2013{\natexlab{a}}){Randriamanakoto}, {Escala},
  {V{\"a}is{\"a}nen}, {Kankare}, {Kotilainen}, {Mattila}, \&
  {Ryder}}]{randriamanakoto13b}
{Randriamanakoto} Z., {Escala} A., {V{\"a}is{\"a}nen} P., {Kankare} E.,
  {Kotilainen} J., {Mattila} S., {Ryder} S., 2013{\natexlab{a}}, \apjl, 775,
  L38

\bibitem[{{Randriamanakoto}
  {et~al}\mbox{.}(2013{\natexlab{b}}){Randriamanakoto}, {V{\"a}is{\"a}nen},
  {Ryder}, {Kankare}, {Kotilainen}, \& {Mattila}}]{randriamanakoto13a}
{Randriamanakoto} Z., {V{\"a}is{\"a}nen} P., {Ryder} S., {Kankare} E.,
  {Kotilainen} J., {Mattila} S., 2013{\natexlab{b}}, \mnras, 431, 554

\bibitem[{{Rau} \& {Cornwell}(2011)}]{rau11}
{Rau} U., {Cornwell} T.~J., 2011, \aap, 532, A71

\bibitem[{{Rich} {et~al}\mbox{.}(2015){Rich}, {Kewley}, \& {Dopita}}]{rich15}
{Rich} J.~A., {Kewley} L.~J., {Dopita} M.~A., 2015, \apjs, 221, 28

\bibitem[{{Rieke} {et~al}\mbox{.}(1993){Rieke}, {Loken}, {Rieke}, \&
  {Tamblyn}}]{rieke93}
{Rieke} G.~H., {Loken} K., {Rieke} M.~J., {Tamblyn} P., 1993, \apj, 412, 99

\bibitem[{{Risaliti} {et~al}\mbox{.}(2000){Risaliti}, {Gilli}, {Maiolino}, \&
  {Salvati}}]{risaliti00}
{Risaliti} G., {Gilli} R., {Maiolino} R., {Salvati} M., 2000, \aap, 357, 13

\bibitem[{{Risaliti} {et~al}\mbox{.}(2006){Risaliti}, {Sani}, {Maiolino},
  {Marconi}, {Berta}, {Braito}, {Della Ceca}, {Franceschini}, \&
  {Salvati}}]{risaliti06}
{Risaliti} G. {et~al.}, 2006, \apjl, 637, L17

\bibitem[{{Rodr{\'{\i}}guez-Zaur{\'{\i}}n}
  {et~al}\mbox{.}(2011){Rodr{\'{\i}}guez-Zaur{\'{\i}}n}, {Arribas},
  {Monreal-Ibero}, {Colina}, {Alonso-Herrero}, \&
  {Alfonso-Garz{\'o}n}}]{rodriguez-zaurin11}
{Rodr{\'{\i}}guez-Zaur{\'{\i}}n} J., {Arribas} S., {Monreal-Ibero} A., {Colina}
  L., {Alonso-Herrero} A., {Alfonso-Garz{\'o}n} J., 2011, \aap, 527, A60

\bibitem[{{Romero-Ca{\~n}izales} {et~al}\mbox{.}(2017){Romero-Ca{\~n}izales},
  {Alberdi}, {Ricci}, {Ar{\'e}valo}, {P{\'e}rez-Torres}, {Conway}, {Beswick},
  {Bondi}, {Muxlow}, {Argo}, {Bauer}, {Efstathiou}, {Herrero-Illana},
  {Mattila}, \& {Ryder}}]{romero-canizales17}
{Romero-Ca{\~n}izales} C. {et~al.}, 2017, \mnras, 467, 2504

\bibitem[{{Romero-Ca{\~n}izales} {et~al}\mbox{.}(2014){Romero-Ca{\~n}izales},
  {Herrero-Illana}, {P{\'e}rez-Torres}, {Alberdi}, {Kankare}, {Bauer}, {Ryder},
  {Mattila}, {Conway}, {Beswick}, \& {Muxlow}}]{romero-canizales14}
{Romero-Ca{\~n}izales} C. {et~al.}, 2014, \mnras, 440, 1067

\bibitem[{{Romero-Ca{\~n}izales} {et~al}\mbox{.}(2011){Romero-Ca{\~n}izales},
  {Mattila}, {Alberdi}, {P{\'e}rez-Torres}, {Kankare}, \&
  {Ryder}}]{romero-canizales11}
{Romero-Ca{\~n}izales} C., {Mattila} S., {Alberdi} A., {P{\'e}rez-Torres}
  M.~A., {Kankare} E., {Ryder} S.~D., 2011, \mnras, 415, 2688

\bibitem[{{Romero-Ca{\~n}izales}
  {et~al}\mbox{.}(2012{\natexlab{a}}){Romero-Ca{\~n}izales},
  {P{\'e}rez-Torres}, \& {Alberdi}}]{romero-canizales12a}
{Romero-Ca{\~n}izales} C., {P{\'e}rez-Torres} M.~{\'A}., {Alberdi} A.,
  2012{\natexlab{a}}, \mnras, 422, 510

\bibitem[{{Romero-Ca{\~n}izales}
  {et~al}\mbox{.}(2012{\natexlab{b}}){Romero-Ca{\~n}izales},
  {P{\'e}rez-Torres}, {Alberdi}, {Argo}, {Beswick}, {Kankare}, {Batejat},
  {Efstathiou}, {Mattila}, {Conway}, {Garrington}, {Muxlow}, {Ryder}, \&
  {V{\"a}is{\"a}nen}}]{romero-canizales12b}
{Romero-Ca{\~n}izales} C. {et~al.}, 2012{\natexlab{b}}, \aap, 543, A72

\bibitem[{{Roy} \& {Norris}(1997)}]{roy97}
{Roy} A.~L., {Norris} R.~P., 1997, \mnras, 289, 824

\bibitem[{{Ryder} {et~al}\mbox{.}(2010){Ryder}, {Mattila}, {Kankare}, \&
  {Perez-Torres}}]{ryder10}
{Ryder} S., {Mattila} S., {Kankare} E., {Perez-Torres} M., 2010, Central Bureau
  Electronic Telegrams, 2189, 1

\bibitem[{{Sales} {et~al}\mbox{.}(2010){Sales}, {Pastoriza}, \&
  {Riffel}}]{sales10}
{Sales} D.~A., {Pastoriza} M.~G., {Riffel} R., 2010, \apj, 725, 605

\bibitem[{{Sanders} {et~al}\mbox{.}(2003){Sanders}, {Mazzarella}, {Kim},
  {Surace}, \& {Soifer}}]{sanders03}
{Sanders} D.~B., {Mazzarella} J.~M., {Kim} D.-C., {Surace} J.~A., {Soifer}
  B.~T., 2003, \aj, 126, 1607

\bibitem[{{Sato} {et~al}\mbox{.}(2005){Sato}, {Yamauchi}, {Ishihara}, {Sorai},
  {Kuno}, {Nakai}, {Balasubramanyam}, \& {Hall}}]{sato05}
{Sato} N., {Yamauchi} A., {Ishihara} Y., {Sorai} K., {Kuno} N., {Nakai} N.,
  {Balasubramanyam} R., {Hall} P., 2005, \pasj, 57, 587

\bibitem[{{Satyapal} {et~al}\mbox{.}(2007){Satyapal}, {Vega}, {Heckman},
  {O'Halloran}, \& {Dudik}}]{satyapal07}
{Satyapal} S., {Vega} D., {Heckman} T., {O'Halloran} B., {Dudik} R., 2007,
  \apjl, 663, L9

\bibitem[{{Scoville} {et~al}\mbox{.}(2000){Scoville}, {Evans}, {Thompson},
  {Rieke}, {Hines}, {Low}, {Dinshaw}, {Surace}, \& {Armus}}]{scoville00}
{Scoville} N.~Z. {et~al.}, 2000, \aj, 119, 991

\bibitem[{{Shao} {et~al}\mbox{.}(2010){Shao}, {Lutz}, {Nordon}, {Maiolino},
  {Alexander}, {Altieri}, {Andreani}, {Aussel}, {Bauer}, {Berta},
  {Bongiovanni}, {Brandt}, {Brusa}, {Cava}, {Cepa}, {Cimatti}, {Daddi},
  {Dominguez-Sanchez}, {Elbaz}, {F{\"o}rster Schreiber}, {Geis}, {Genzel},
  {Grazian}, {Gruppioni}, {Magdis}, {Magnelli}, {Mainieri}, {P{\'e}rez
  Garc{\'{\i}}a}, {Poglitsch}, {Popesso}, {Pozzi}, {Riguccini}, {Rodighiero},
  {Rovilos}, {Saintonge}, {Salvato}, {Sanchez Portal}, {Santini}, {Sturm},
  {Tacconi}, {Valtchanov}, {Wetzstein}, \& {Wieprecht}}]{shao10}
{Shao} L. {et~al.}, 2010, \aap, 518, L26

\bibitem[{{Shipley} {et~al}\mbox{.}(2016){Shipley}, {Papovich}, {Rieke},
  {Brown}, \& {Moustakas}}]{shipley16}
{Shipley} H.~V., {Papovich} C., {Rieke} G.~H., {Brown} M.~J.~I., {Moustakas}
  J., 2016, \apj, 818, 60

\bibitem[{{Smith} {et~al}\mbox{.}(1995){Smith}, {Herter}, {Haynes}, {Beichman},
  \& {Gautier}}]{smith95}
{Smith} D.~A., {Herter} T., {Haynes} M.~P., {Beichman} C.~A., {Gautier}, III
  T.~N., 1995, \apj, 439, 623

\bibitem[{{Stierwalt} {et~al}\mbox{.}(2014){Stierwalt}, {Armus},
  {Charmandaris}, {Diaz-Santos}, {Marshall}, {Evans}, {Haan}, {Howell},
  {Iwasawa}, {Kim}, {Murphy}, {Rich}, {Spoon}, {Inami}, {Petric}, \&
  {U}}]{stierwalt14}
{Stierwalt} S. {et~al.}, 2014, \apj, 790, 124

\bibitem[{{Stierwalt} {et~al}\mbox{.}(2013){Stierwalt}, {Armus}, {Surace},
  {Inami}, {Petric}, {Diaz-Santos}, {Haan}, {Charmandaris}, {Howell}, {Kim},
  {Marshall}, {Mazzarella}, {Spoon}, {Veilleux}, {Evans}, {Sanders},
  {Appleton}, {Bothun}, {Bridge}, {Chan}, {Frayer}, {Iwasawa}, {Kewley},
  {Lord}, {Madore}, {Melbourne}, {Murphy}, {Rich}, {Schulz}, {Sturm},
  {Vavilkin}, \& {Xu}}]{stierwalt13}
{Stierwalt} S. {et~al.}, 2013, \apjs, 206, 1

\bibitem[{{Sturm} {et~al}\mbox{.}(2000){Sturm}, {Lutz}, {Tran}, {Feuchtgruber},
  {Genzel}, {Kunze}, {Moorwood}, \& {Thornley}}]{sturm00}
{Sturm} E., {Lutz} D., {Tran} D., {Feuchtgruber} H., {Genzel} R., {Kunze} D.,
  {Moorwood} A.~F.~M., {Thornley} M.~D., 2000, \aap, 358, 481

\bibitem[{{Tabatabaei} {et~al}\mbox{.}(2007){Tabatabaei}, {Beck}, {Kr{\"u}gel},
  {Krause}, {Berkhuijsen}, {Gordon}, \& {Menten}}]{tabatabaei07}
{Tabatabaei} F.~S., {Beck} R., {Kr{\"u}gel} E., {Krause} M., {Berkhuijsen}
  E.~M., {Gordon} K.~D., {Menten} K.~M., 2007, \aap, 475, 133

\bibitem[{{Tadhunter} {et~al}\mbox{.}(2007){Tadhunter}, {Dicken}, {Holt},
  {Inskip}, {Morganti}, {Axon}, {Buchanan}, {Gonz{\'a}lez Delgado}, {Barthel},
  \& {van Bemmel}}]{tadhunter07}
{Tadhunter} C. {et~al.}, 2007, \apjl, 661, L13

\bibitem[{{Terashima} {et~al}\mbox{.}(2015){Terashima}, {Hirata}, {Awaki},
  {Oyabu}, {Gandhi}, {Toba}, \& {Matsuhara}}]{terashima15}
{Terashima} Y., {Hirata} Y., {Awaki} H., {Oyabu} S., {Gandhi} P., {Toba} Y.,
  {Matsuhara} H., 2015, \apj, 814, 11

\bibitem[{{Treister} {et~al}\mbox{.}(2009){Treister}, {Urry}, \&
  {Virani}}]{treister09}
{Treister} E., {Urry} C.~M., {Virani} S., 2009, \apj, 696, 110

\bibitem[{{Ulvestad}(2009)}]{ulvestad09}
{Ulvestad} J.~S., 2009, \aj, 138, 1529

\bibitem[{{V{\"a}is{\"a}nen}
  {et~al}\mbox{.}(2008{\natexlab{a}}){V{\"a}is{\"a}nen}, {Mattila}, {Kniazev},
  {Adamo}, {Efstathiou}, {Farrah}, {Johansson}, {{\"O}stlin}, {Buckley},
  {Burgh}, {Crause}, {Hashimoto}, {Lira}, {Loaring}, {Nordsieck},
  {Romero-Colmenero}, {Ryder}, {Still}, \& {Zijlstra}}]{vaisanen08a}
{V{\"a}is{\"a}nen} P. {et~al.}, 2008{\natexlab{a}}, \mnras, 384, 886

\bibitem[{{V{\"a}is{\"a}nen} {et~al}\mbox{.}(2012){V{\"a}is{\"a}nen},
  {Rajpaul}, {Zijlstra}, {Reunanen}, \& {Kotilainen}}]{vaisanen12}
{V{\"a}is{\"a}nen} P., {Rajpaul} V., {Zijlstra} A.~A., {Reunanen} J.,
  {Kotilainen} J., 2012, \mnras, 420, 2209

\bibitem[{{V{\"a}is{\"a}nen}
  {et~al}\mbox{.}(2008{\natexlab{b}}){V{\"a}is{\"a}nen}, {Ryder}, {Mattila}, \&
  {Kotilainen}}]{vaisanen08b}
{V{\"a}is{\"a}nen} P., {Ryder} S., {Mattila} S., {Kotilainen} J.,
  2008{\natexlab{b}}, \apjl, 689, L37

\bibitem[{{van der Kruit}(1973)}]{vanderkruit73}
{van der Kruit} P.~C., 1973, \aap, 29, 263

\bibitem[{{Vardoulaki} {et~al}\mbox{.}(2014){Vardoulaki}, {Charmandaris},
  {Murphy}, {Diaz-Santos}, {Armus}, {Evans}, {Mazzarella}, {Privon},
  {Stierwalt}, \& {Barcos-Munoz}}]{vardoulaki14}
{Vardoulaki} E. {et~al.}, 2014, ArXiv e-prints

\bibitem[{{Varenius} {et~al}\mbox{.}(2016){Varenius}, {Conway},
  {Mart{\'{\i}}-Vidal}, {Aalto}, {Barcos-Mu{\~n}oz}, {K{\"o}nig},
  {P{\'e}rez-Torres}, {Deller}, {Mold{\'o}n}, {Gallagher}, {Yoast-Hull},
  {Horellou}, {Morabito}, {Alberdi}, {Jackson}, {Beswick}, {Carozzi},
  {Wucknitz}, \& {Ram{\'{\i}}rez-Olivencia}}]{varenius16}
{Varenius} E. {et~al.}, 2016, ArXiv e-prints

\bibitem[{{Vega} {et~al}\mbox{.}(2008){Vega}, {Clemens}, {Bressan}, {Granato},
  {Silva}, \& {Panuzzo}}]{vega08}
{Vega} O., {Clemens} M.~S., {Bressan} A., {Granato} G.~L., {Silva} L.,
  {Panuzzo} P., 2008, \aap, 484, 631

\bibitem[{{Veilleux} {et~al}\mbox{.}(1995){Veilleux}, {Kim}, {Sanders},
  {Mazzarella}, \& {Soifer}}]{veilleux95}
{Veilleux} S., {Kim} D.-C., {Sanders} D.~B., {Mazzarella} J.~M., {Soifer}
  B.~T., 1995, \apjs, 98, 171

\bibitem[{{Voelk}(1989)}]{voelk89}
{Voelk} H.~J., 1989, \aap, 218, 67

\bibitem[{{Voit}(1992)}]{voit92}
{Voit} G.~M., 1992, \mnras, 258, 841

\bibitem[{{Weedman} {et~al}\mbox{.}(2005){Weedman}, {Hao}, {Higdon}, {Devost},
  {Wu}, {Charmandaris}, {Brandl}, {Bass}, \& {Houck}}]{weedman05}
{Weedman} D.~W. {et~al.}, 2005, \apj, 633, 706

\bibitem[{{Wu} {et~al}\mbox{.}(2009){Wu}, {Charmandaris}, {Huang}, {Spinoglio},
  \& {Tommasin}}]{wu09}
{Wu} Y., {Charmandaris} V., {Huang} J., {Spinoglio} L., {Tommasin} S., 2009,
  \apj, 701, 658

\bibitem[{{Yuan} {et~al}\mbox{.}(2010){Yuan}, {Kewley}, \& {Sanders}}]{yuan10}
{Yuan} T.-T., {Kewley} L.~J., {Sanders} D.~B., 2010, \apj, 709, 884

\bibitem[{{Yun} {et~al}\mbox{.}(2001){Yun}, {Reddy}, \& {Condon}}]{yun01}
{Yun} M.~S., {Reddy} N.~A., {Condon} J.~J., 2001, \apj, 554, 803

\bibitem[{{Zezas} {et~al}\mbox{.}(2003){Zezas}, {Ward}, \& {Murray}}]{zezas03}
{Zezas} A., {Ward} M.~J., {Murray} S.~S., 2003, \apjl, 594, L31

\end{thebibliography}

\appendix
\section{Non-smoothed near-IR images with radio contours}
\begin{figure*}
\includegraphics[width=0.9\textwidth]{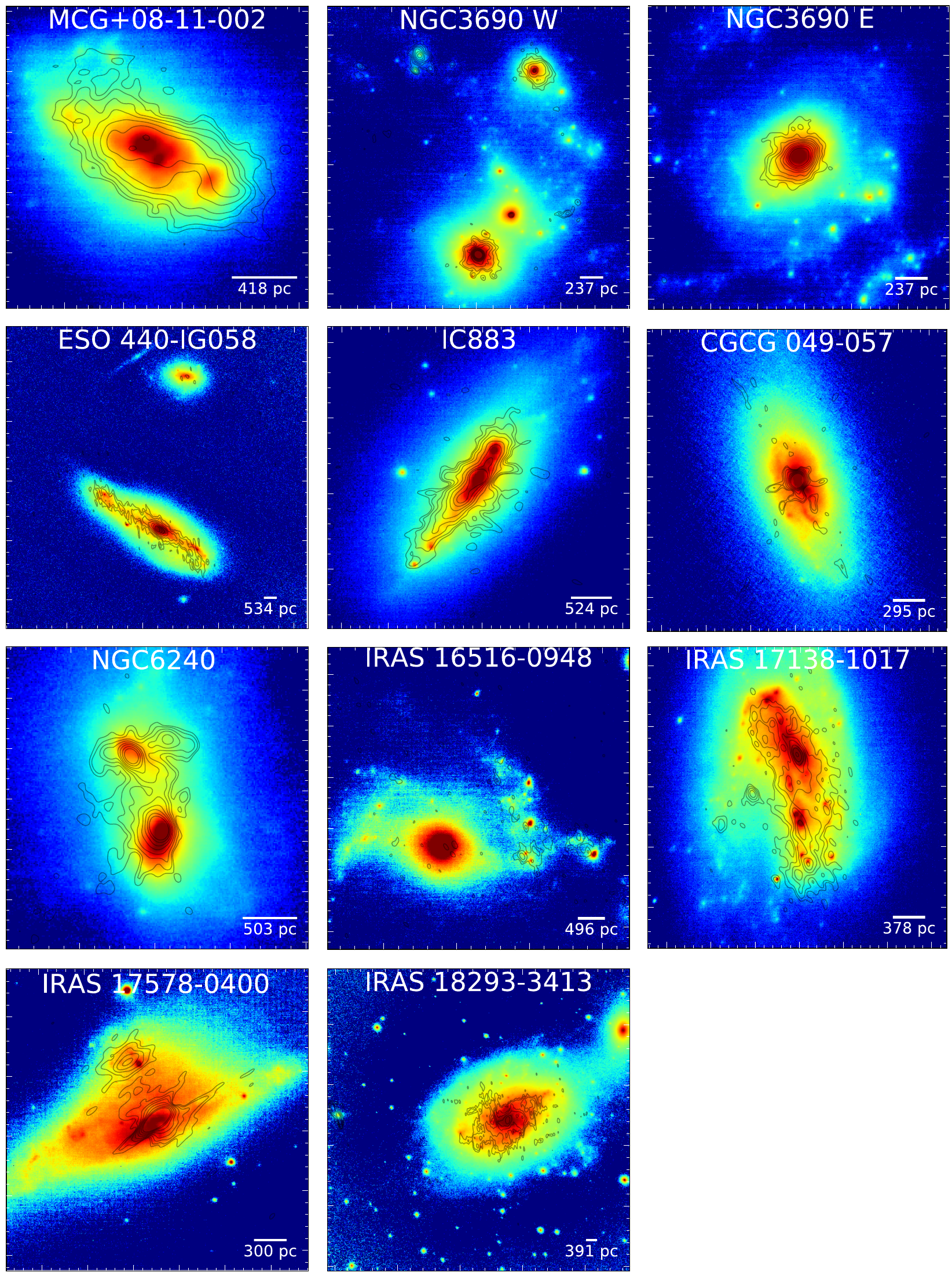}
 \caption{$2.2\mu$m non-smoothed near-IR images with overlaid $3.6\,$cm radio contours. Contours are drawn every $3\left(\sqrt{3}\right)^n\times\mathrm{rms}$ level, with $n=0,1,2,3...$}
\label{fig:app:radioir}
\end{figure*}

We show in Figure~\ref{fig:app:radioir} the original non-smoothed near-IR images with the overlaid contours corresponding to the radio emission.

\bsp	
\label{lastpage}
\end{document}